%% file: NewPhase_Retrievalv23_Arxiv.tex
\documentclass[onecolumn,draftcls]{IEEEtran}

\input{preamble}
\usepackage{xspace}
\usepackage[ colorlinks = true,
             linkcolor = blue,
             urlcolor  = blue,
             citecolor = red,
             anchorcolor = green,
]{hyperref}

\usepackage{cite}
\allowdisplaybreaks[1]
\usepackage{amssymb}
\usepackage{pifont}

\newcommand{\rev}[1]{{\color{black} #1}}

\begin{document}
\title{ \fontsize{23}{25} \selectfont Support Recovery in the Phase Retrieval Model: Information-Theoretic Fundamental Limits} 

\author{Lan V.\ Truong and 
	Jonathan Scarlett,
\thanks{L.\ V .\ Truong is with the Department of Engineering, the University of Cambridge, Cambridge CB2 1PZ UK (e-mail:
	\url{lt407@cam.ac.uk}).} \thanks{J.\ Scarlett is with the Department of Computer Science, School of Computing,
	National University of Singapore (NUS), Singapore 117417, and also with the Department of Mathematics, NUS, Singapore 119076 (e-mail: \url{scarlett@comp.nus.edu.sg}).} 
 \thanks{This work is supported by an NUS Early Career Research Award. This paper was presented in part at the 2019 IEEE Information Theory Workshop.}
\thanks{Copyright \copyright~2017 IEEE. Personal use of this material is permitted.  However, permission to use this material for any other purposes must be obtained from the IEEE by sending a request to \url{pubs-permissions@ieee.org}.}
}

\maketitle
\begin{abstract} 
    The support recovery problem consists of determining a sparse subset of variables that is relevant in generating a set of observations.  In this paper, we study the support recovery problem in the phase retrieval model consisting of noisy phaseless measurements, which arises in a diverse range of settings such as optical detection, X-ray crystallography, electron microscopy, and coherent diffractive imaging.  Our focus is on information-theoretic fundamental limits under an approximate recovery criterion, considering both discrete and Gaussian models for the sparse non-zero entries, along with Gaussian measurement matrices. In both cases, our bounds provide sharp thresholds with near-matching constant factors in several scaling regimes on the sparsity and signal-to-noise ratio.  As a key step towards obtaining these results, we develop new concentration bounds for the conditional information content of log-concave random variables, which may be of independent interest.
\end{abstract}   

%
%
\section{Introduction}\label{intro}

Recently, there has been a growing interest in recovering an unknown signal $\beta \in \bbC^p$ from \rev{a number of phaseless quadratic observations, each taking the form $Y=|\langle \beta, X\rangle|^2 + Z$}, where  $X \in \bbC^{p}$ is a measurement vector, and $Z \in \bbR$ represents measurement noise. Since only the magnitude of $\langle \beta, X \rangle$ is measured, and not the phase (or the sign, in the real case), this problem is referred to as \emph{phase retrieval}. The phase retrieval problem has many applications including optical detection, $X$-ray crystallography, electron microscopy, and coherent diffractive imaging~\cite{Eldar2014}.

\subsection{Sparse Phase Retrieval}

Similarly to the basic linear model, various works have shown that the number of measurements can be reduced significantly if the signal $\beta \in \bbC^p$  is {\em sparse}, i.e., it has at most $k$ non-zero entries for some $k \ll p$.  Here we provide a non-exhaustive list of relevant results from the literature.

It is shown in~\cite{Eldar2014} that for real-valued signals, stable phase retrieval can be achieved with $O(k\log(\frac{p}{k}))$ measurements in the noiseless setting, and with $O(k\log k\log(\frac{p}{k}))$ measurements in the noisy setting under some conditions on the noise distribution.  \rev{The measurements considered in \cite{Eldar2014}  are isotropic and sub-Gaussian, including (real) Gaussian measurements as a special case.  The corresponding results are information-theoretic and are not shown to be attained with any practical algorithm, and to the best of our knowledge, it remains an open problem to attain comparable theoretical results for practical algorithms.  However, numerical evidence has been given for the success of generalized approximate message passing (GAMP) with roughly $2k\log(p/k)$ Gaussian measurements in the noiseless case, and for the robustness of GAMP to noise \cite{schniter2014compressive}.  On the other hand, rigorous results for efficient algorithms with Gaussian measurements typically require significantly more measurements; for instance, an $O(k^2 \log p)$ bound is attained in \cite{LV2013a} via a semidefinite programming (SDP) approach.
}

\rev{
While our focus will be on Gaussian measurements, it is also worth highlighting some works that achieve computationally efficient sparse phase retrieval with a similar number of {\em carefully-designed} non-Gaussian measurements. Cai {\em et al.}  \cite{Cai2014SUPERSS} designed an algorithm that succeeds with $O(k)$ measurements and $O(k \log k)$ decoding time in the noiseless complex-valued setting; the measurement matrix is generated from the structure of a series of bipartite graphs (between signal components and measurements) with various desirable properties.  More recently, in the real-valued setting, Nakos \cite{Nakos2017e} proposed an algorithm that recovers an approximately $k$-sparse vector under the $\ell_2/\ell_2$ guarantee, with $O(k\log p)$ measurements and $O(k^{1+\gamma}{\rm poly}(\log p))$ decoding time for any constant $\gamma > 0$.  See also \cite{YiLi2018e} for additional variants.

In the noisy complex-valued setting, Iwen {\em et al.}~\cite{IWEN2017135} provided a simple two-stage sparse phase retrieval strategy that can stably reconstruct $\beta$ up to a global phase shift using only $O(k\log(\frac{p}{k}))$ measurements, under some bounded noise assumptions.  In addition, Pedarsani {\em et al.}  \cite{Pedarsani2017a} used a sparse-graph coding approach to attain an approximate support recovery guarantee for quantized signals with: (i) $O(k \log p)$ measurements and $O(p \log p)$ decoding time, or (ii) $O(k \log^3 p)$ measurements and $O(k \log^3 p)$ decoding time.  In the noiseless case, these further reduce to $O(k)$, even without the assumption of quantized signals.}

Fourier measurements are also commonly considered, and are particularly relevant in many practical applications.  For example, in this setting, Jaganathan {\em et al.} \cite{Jaganathan2017a} gave guarantees on recovering sparse signals whose support is aperiodic, and proposed an efficient two-stage algorithm is proposed that first identifies the support, and then the sparse signal values.

\subsection{Support Recovery}

A distinct goal that has received less attention in phase retrieval, but considerable attention in other models, is the support recovery problem~\cite{Malyutov2013a, Foucart2013, Miller}, where one wishes to exactly or approximately determine the support $S = {\rm supp}(\beta)$ given a collection of observations $\bY \in \bbR^n$ and the corresponding measurement matrix $\bX \in \bbC^{n\times p}$.\footnote{\rev{As we mention in the notation section below, the boldness of these symbols is used to highlight their association with multiple measurements.  In contrast, while $X \in \bbC^{p}$ represents a vector, it is non-bold because it is only associated with a single measurement. }}  This problem is of direct interest when the goal is to find {\em which} variables influence the output (rather than their associated weights), and may also be used as a first step towards estimating the values of $\beta$ (e.g., see \cite{Wang2016,Jaganathan2017a}).

Under general linear and non-linear models, Scarlett and Cevher~\cite{Scarlet2017a} provided achievability and converse bounds characterizing the trade-off between error probability and number of measurements. They applied their general bounds to the linear, $1$-bit, and group testing models to obtain precise thresholds on the number of measurements required to achieve vanishing decoding error probability in the high-dimensional limit.  Numerous other related works also exist, with the focus being mainly on linear models~\cite{Wainwright2009a, Wang2009a, Rahnama,Tulino2013a,Scarlett2013a}; see \cite{Scarlet2017a} for a more detailed overview.  In particular, {\em approximate recovery} criteria were studied by Reeves and Gastpar \cite{Reeves2012a, Gastpar2013a} in the regime $k = \Theta(p)$, and by Scarlett and Cevher \cite{Scarlet2017a} in the regime $k = o(p)$; we focus on the latter setting.

Although the initial bounds in~\cite{Scarlet2017a} are very general, applying these bounds to new models can still be very challenging, due to the need to establish concentration bounds and mutual information bounds on a case-by-case basis.  In this paper, we use this approach to establish fundamental limits for approximate support recovery in the phase retrieval model, under a log-concavity assumption on the noise distribution.  To achieve this goal, we need to overcome at least two key challenges: establishing concentration bounds for information quantities in the phase retrieval model, and upper and lower bounding key conditional mutual information terms that have no closed form expressions.  For each of these challenges, we develop novel auxiliary results, some of which may be of independent interest. The following subsection lists our specific contributions in more detail.

\subsection{Contributions} 

Our main contributions in this paper are as follows:
\begin{itemize}
    \item We extend the concentration bounds of the unconditional information content of log-concave densities by Fradelizi {\em et al.} \cite[Theorem 3.1]{Fradelizi2016} to conditional versions ({\em cf.} Corollary~\ref{verykey}) in which joint log-concavity does not hold. Due to this extension, we can establish concentration bounds for the conditional information density of $n$-dimensional random variables ({\em cf.} Theorem~\ref{thre1}) and apply these bounds to the phase retrieval model. Because of their generality, our extended concentration bounds might be of independent interest.
    \item Under i.i.d.~complex Gaussian measurement matrices $\bX$, we establish tight upper and lower bounds on the required number of measurements to achieve approximate support recovery (i.e., recovering a given proportion of the support) under both discrete ({\em cf.} Lemma~\ref{main1}) and Gaussian ({\em cf.} Theorem~\ref{Gaussian}) modeling assumptions on the non-zero entries of $\beta$.  In both cases, the upper and lower bounds coincide up to an explicit constant factor in certain sparsity regimes, and this constant factor is often very close to one (e.g., when the signal-to-noise ratio is sufficiently high).
\end{itemize}

\subsection{Notation} 

We use the similar notation to \cite{Scarlet2017a}. We use upper-case letters for random variables, and lower cases for their realizations. A non-bold character may be a scalar or a vector, whereas a bold character refers to a collection of $n$ scalars (e.g., $\bY \in \bbR^n$) or vectors (e.g., $\bX \in \bbR^{n\times p}$), where $n$ is the number of measurements. We write $\beta_S$ to denote the subvector of $\beta$ at the columns indexed by $S$, and $\bX_{S}$ to denote the submatrix of $\bX$ containing the columns indexed by $S$. The complement with respect to $\{1,2,\ldots,p\}$ is denoted by $(\cdot)^c$. 

The symbol $\sim$ means ``distributed as". For a given joint probability density distribution $f_{XY}$, the corresponding marginal distributions are denoted by $f_X$ and $f_Y$, and similarly for conditional probability density marginals (e.g., $f_{Y|X}$).  The notation $f_{XY}^n$, $f_X^n$, etc.~denotes the corresponding i.i.d.~distribution in which each term is distributed as $f_{XY}$, $f_X$, etc.  We write $\bbP[\cdot]$ for probabilities, $\bbE[\cdot]$ for expectations, and $\var[\cdot]$ for variances. 

We use usual notations for the differential entropy (e.g., $h(X)$)  and mutual information (e.g., $I(X;Y)$), and their conditional counterparts (e.g., $h(X|Z), I(X;Y|Z)$).  We use the notation $\calN(\mu,\sigma^2)$ for real Gaussian random variables, $\calC\calN(\mu,\sigma^2)$ for complex Gaussians (with variance $\frac{\sigma^2}{2}$ in each of the real and imaginary parts), and $\chi_k^2$ for the central chi squared distribution with $k$ degrees of freedom.
 
We make use of the standard asymptotic notations $O(\cdot), o(\cdot), \Theta(\cdot), \Omega(\cdot)$ and $\omega(\cdot)$. We define the function $[\cdot]^+=\max\{0,\cdot\}$ and write the floor and ceiling functions as $\lfloor \cdot \rfloor$ and $\lfloor \cdot \rfloor$, respectively. The function $\log$ has base $e$, and all information quantities are measured in nats.

Throughout the paper, we frequently make use of integrals written as $\int (\,\dotsc) \mu(dx)$,  $\int (\,\dotsc) \mu(dx \times dy)$, etc., where $\mu(\cdot)$ denotes a suitable measure that can typically be taken to be the Lebesgue measure.  For $t > 0$, we say that a function $f(\bx)$ on $\bbR^n$ is in $L^t(\bbR^n)$ is $|f(\bx)|^t$ is integrable.

\subsection{Structure of the Paper}

In Section~\ref{model}, we formally introduce the problem setup and overview our main results.  In Section \ref{sec:aux}, we provide the main auxiliary results on log-concavity, concentration of measure, and mutual information bounds.  Sections~\ref{sec:discrete} and~\ref{sec:Gaussian} provide the proofs of our main support recovery results. Conclusions are drawn in Section~\ref{sec:conclu}.

%
%
\section{Problem Setup and Main Results}\label{model}

\subsection{Model and Assumptions}\label{subsecA}

Let $p$ denote the ambient dimension, $k$ the sparsity level, and $n$ the number of measurements. We let $\calS$ be the set of subsets of $\{1,2,\ldots,p\}$ having cardinality $k$. The key random variables in the support retrieval problem are the support set $S \in \calS$, the unknown signal $\beta \in \bbC^p$, the measurement matrix $\bX \in \bbC^{n \times p}$, and the observation vector $\bY \in \bbR^n$.

The support set $S$ is assumed to be equiprobable on the ${p\choose k}$ subsets within $\calS$. Given $S$, the entries of $\beta_{S^c}$ are deterministically set to zero, and the remaining entries are generated according to some distribution $\beta_{S} \sim f_{\beta_{S}}$.\footnote{We allow for both discrete and continuous distributions on $\beta_S$, meaning that in some cases $f_{\beta_{S}}$ represents a probability mass function rather than a density function.}  We assume that these non-zero entries follows the same distribution for all the ${p\choose k}$ possible realizations of $S$, and that this distribution is permutation-invariant.  

We consider the setting of (complex) Gaussian measurements, in which the measurement matrix takes i.i.d.~values on $\mathcal{CN}(0,1)$, whose density is denoted by $f_X$.  We write $f_X^{n\times p}$, to denote the corresponding i.i.d. distribution for matrices, and we write $f_X^k$ as a shorthand for $f_X^{k \times 1}$. Given $S=s$, each entry of the observation vector $\bY$ is generated in a conditionally independent manner according to the following model:
\begin{align}
    \label{acqui}
    Y=|\langle X_s,\beta_s\rangle|^2 +Z,
\end{align}
where $X_s \sim f_X^k$, $\beta_s \in \bbC^k$, and $Z \sim f_Z$, with $f_Z$ being an arbitrary {\em log-concave} density function.  This log-concavity assumption is made for mathematical convenience, but also captures a wide range of noise distributions, including Gaussian.  We note that the permutation-invariance of $Y$, $X_S$ and $\beta_S$ with respect to $S$ allows us to condition on a fixed $S=s$ \rev{of cardinality $k$} throughout the analysis (e.g. $s = \{1,\dotsc,k\}$) without loss of generality; such conditioning should henceforth be assumed unless explicitly stated otherwise.

The relation~\eqref{acqui} induces the following conditional joint distribution of $(\beta_s,X_s,Y)$ (given $S=s$):
\begin{align}
    f_{\beta_s X_s Y}(b_s,x_s,y)&=f_{\beta_s}(b_s) f_X^k(x_s) f_{Y|X_s\beta_s}(y|x_s,b_s)\\
    &=f_{\beta_s}(b_s) f_X^k(x_s) f_Z(y-|\langle x_s, b_s \rangle|^2),
\end{align} 
and its multiple-observation counterpart
\begin{align}
    \label{jointdis}
    f_{\beta_s\bX_s\bY}(b_s,\bx_s,\by)=f_{\beta_s}(b_s) f_{X}^{n \times k}(\bx_s) f_{Y|X_s\beta_s}^n(\by|\bx_s,b_s),
\end{align}
where $f_{Y|X_s\beta_s}^n(\by|\bx_s,b_s)$ is the $n$-fold product of $f_{Y|X_s\beta_s}(\cdot|\cdot,b_s)$. The remaining entries of the measurement matrix are distributed as $\bX_{s^c}\sim f_X^{n\times (p-k)}$. 

Given $\bX$ and $\bY$, a decoder forms an estimate $\hatS$ of $S$. Like previous works studying the information-theoretic limits of support recovery (e.g.,~\cite{Scarlet2017a,Wainwright2009a}), we assume that the decoder knows the system model, including $f_{Y|X_s\beta_s}$ and $f_{\beta_s}$.  We focus on the {\em approximate recovery criterion}, only requiring that at least $k-\lfloor \alpha^* k\rfloor+1$ entries of $S$ are successfully identified (approximate recovery) for some $\alpha^*\in (0,1)$. Following~\cite{Reeves2012a,Scarlet2017a}, the error probability is given by
\begin{align}
    \rvP_{\rme}(\alpha^*):=\bbP\Big[\{|S\setminus \hatS|\geq \lfloor \alpha^* k\rfloor\}\cup \{|\hatS \setminus S|\geq \lfloor \alpha^* k\rfloor\}\Big].
\end{align}
Note that if both $S$ and $\hatS$ have cardinality $k$ with probability one, then the two events in the union are identical, and hence either of the two can be removed.  A more stringent performance criterion also considered in literature is the exact support recovery problem, where the error probability is given by $\rvP_{\rme}(0)$, but our techniques currently appear to be less suited to that setting.

Our main goal is to derive necessary and sufficient conditions on $n$ (as a function of $k$ and $p$) such that $\rvP_{\rme}(\alpha^*)$ vanishes as $p\to \infty$. Moreover, when considering converse results, we will not only be interested in conditions under which $\rvP_{\rme}(\alpha^*) \not \to 0$, but also conditions under which the stronger statement $\rvP_{\rme}(\alpha^*) \to 1$ holds.

\subsection{Overview of Main Results}

Here we state and discuss the two main results of this paper.  Both of the theorems concern the information-theoretic limits of support recovery in the phase retrieval as described above, but with two different models of interest for the non-zero entries $\beta_s$.  \rev{We note that our results are asymptotic as $p \to \infty$ and $k \to \infty$, and we seek explicit constant factors in the leading term, but leave higher-order terms unspecified.  Sharper (e.g., non-asymptotic) characterizations appear to be much more challenging, and are beyond the scope of this work.  In addition, we emphasize that our achievability result is based on a computationally intractable information-theoretic decoder, and approaching the fundamental limits with practical decoding techniques remains an interesting direction for future studies.}

{\bf Discrete setting.} The first result concerns a discrete distribution on $\beta_s$, namely, $\beta_s$ is a uniformly random permutation of a fixed complex vector $(b_1,\dotsc,b_k)$.  We let $(b'_1,\dotsc,b'_k)$ be the sorted version of $(b_1,\dotsc,b_k)$ such that $|b'_1| \leq |b'_2|\leq \cdots |b'_k|$, and define the following mutual information quantities:
\begin{align}
    \label{defI1}
    I_1(\alpha,k)&:=\frac{1}{2}\log \bigg[\bigg(\frac{4}{\exp(2h(Z))}\bigg)\bigg(\sum_{i=1}^{\lfloor \alpha k \rfloor} |b'_i|^2\bigg)^2+1\bigg],\\
    \label{defI2}
    I_2(\alpha,k)&:=\frac{1}{2}\log\bigg[\bigg(\frac{2\pi e}{\exp(2h(Z))}\bigg)\bigg(\sum_{i=1}^{\lfloor \alpha k \rfloor} |b'_i|^2\bigg)^2+1\bigg]\nonumber\\
    &\quad +\frac{1}{2}\log\bigg[1+\frac{\big(\sum_{i=1}^{\lfloor \alpha k \rfloor} |b'_i|^2\big)\big(\sum_{i=\lfloor \alpha k \rfloor +1}^k |b'_i|^2\big)} {\big(\sum_{i=1}^{\lfloor \alpha k \rfloor} |b'_i|^2\big)^2+\frac{\exp(2h(Z))}{2\pi e}}\bigg]+\frac{1}{2}\log \bigg(\frac{\pi e}{2}\bigg),
\end{align}
where $h(Z)$ is the differential entropy of $Z$.

\begin{theorem}\label{cormain1} 
    Consider the phase retrieval setup in Section~\ref{model}, with $\beta_s$ being a uniformly random permutation of a fixed complex vector $(b_1,b_2,\ldots,b_k)$.   Let $|b_{\rm{min}}|=\min\{|b_i|: i\in \{1,\cdots,k\}\}$ and $|b_{\rm{max}}|=\max\{|b_i|: i\in \{1,\cdots,k\}\}$, and assume that $|b_{\rm{min}}|=\Theta(|b_{\rm{max}}|)$, and that $k \to \infty$ with $\|b_s\|_2=\Theta(1)$ as $p \to \infty$.  In addition, assume that there are $m_{\beta} \in \{1,\dotsc,k\}$ distinct elements in $(b_1,\ldots,b_k)$.  

    We have $\rvP_{\rme}(\alpha^*) \to 0$ as $p\to \infty$ provided that
    \begin{align}
        \label{eq3main}
        n \geq \max_{\alpha \in [\alpha^*,1]}\frac{\alpha k\log (\frac{p}{k})}{I_1(\alpha,k)}(1+\eta)
    \end{align} 
    for arbitrarily small $\eta > 0$ if either of the following additional conditions hold: (i) $m_{\beta}=\Theta(1)$ and $k=o(p)$, or (ii) $\log k = o(\log p)$ (and $m_{\beta}$ is arbitrary).
    
    Conversely, under the general scaling $k = o(p)$ and arbitrary $m_{\beta}$, we have $\rvP_{\rme}(\alpha^*) \to 1$ as $p\to \infty$ whenever 
    \begin{align}
        \label{eq4main}
        n\leq \max_{\alpha \in [\alpha^*,1]}\frac{(\alpha-\alpha^*)k\log (\frac{p}{k})}{I_2(\alpha,k)} (1-\eta),
    \end{align} 
    for arbitrarily small $\eta > 0$.
\end{theorem}
\begin{IEEEproof}
    See Section \ref{sec:discrete}.
\end{IEEEproof}

We observe that the upper and lower bounds are nearly in closed form, other than the optimization over a single scalar $\alpha$.  Moreover, the two have a very similar form, with the main difference being the appearance of $\alpha$ vs.~$(\alpha-\alpha^*)$ in the numerator, and $I_1$ vs.~$I_2$ in the denominator.  The bounds hold for an arbitrary log-concave noise distribution $f_Z$.

Since the noise variance $\sigma^2$ is fixed and the measurement matrix has normalized $\mathcal{CN}(0,1)$ entries, the assumption  $\|b\|_2 = \Theta(1)$ corresponds to the case that the signal-to-noise ratio (SNR) is constant.  We observe that under this assumption, the upper and lower bounds provide matching $\Theta\big(k \log\frac{p}{k}\big)$ behavior.  Perhaps more significantly, in the high-SNR limit (i.e., $\|b\|_2 \to \infty$), we obtain nearly identical {\em constant factors}.  To see this, it suffices to crudely lower bound $I_1(\alpha,k)$ by $\frac{1}{2}\log \big[\big(\frac{4}{\exp(2h(Z))}\big)\big(\lfloor \alpha k\rfloor |b_{\rm{min}}|^2\big)^2+1\big]$, and upper bound $I_2(\alpha,k)$ by $\frac{1}{2}\log\big[\big(\frac{2\pi e}{\exp(2h(Z))}\big)\big(\lfloor \alpha k\rfloor |b_{\rm{max}}|^2\big)^2+1\big]+\frac{1}{2}\log\big[1+\frac{\lfloor \alpha k\rfloor  k |b_{\rm{max}}|^4 } {\lfloor \alpha k \rfloor^2 |b_{\rm{min}}|^4}\big]+\frac{1}{2}\log \big(\frac{\pi e}{2}\big)$.  For any $\alpha$ bounded away from zero, since $|b_{\rm{min}}|=\Theta(|b_{\rm{max}}|)$, these both behave as $\log( k |b_{\rm{min}}|^2 ) (1+o(1))$ as $\|b\|_2 \to \infty$ (or equivalently $k |b_{\rm{min}}|^2 \to \infty$), which implies that the maxima in \eqref{eq3main} and \eqref{eq4main} are attained by $\alpha = 1$ in this limit, and the upper and lower bounds coincide up to a factor of $\frac{1}{1-\alpha^*}$.

We believe that the additional assumptions on $m_{\beta}$ and $k$ in the achievability part are an artifact of our analysis, and note that similar assumptions were made for the linear model in \cite{Scarlet2017a}.  The conditions in Theorem \ref{cormain1} are less restrictive than those in \cite{Scarlet2017a} since we are considering approximate recovery instead of exact recovery.

{\bf Gaussian setting.} We now turn to a (complex) Gaussian model on the non-zero entries in which $\beta_s \sim \mathcal{CN}(0,\bI_k \sigma_{\beta}^2)$, $\sigma_{\beta}^2=\frac{c_{\beta}}{k}$ for some $c_{\beta}>0$. This is analogous to a model considered for the linear setting in \cite{Reeves2012a,Scarlet2017a}.  Our result is stated in terms of the mutual information quantities 
\begin{align}
    \label{defbarI1}
    \barI_1(\alpha)&:=\frac{1}{2}\log\bigg(1+ 4\bigg(\frac{c_{\beta} g(\alpha)}{ \sigma\sqrt{2\pi e}}\bigg)^2\bigg),\\
    \label{defbarI2k}
    \barI_2(\alpha)&:=\frac{1}{2}\log\bigg(1+ \bigg(\frac{ c_{\beta} g(\alpha)}{\sigma}\bigg)^2\bigg) + \frac{1}{2}\log\bigg(1+\frac{c_{\beta}^2 g(\alpha) (1-g(\alpha)) }{g^2(\alpha) c_{\beta}^2 + \sigma^2}  \bigg)+\frac{1}{2}\log \bigg(\frac{\pi e}{2}\bigg),
\end{align}
where $g(\cdot)$ is defined as
\begin{align}
    \label{defgalpha}
    g(\alpha):=\int_0^{\infty}[\alpha-F_{1}(u)]^+ du,
\end{align} 
with $F_{1}$ denoting the cumulative distribution function of a  $|\mathcal{CN}(0,1)|^2$ random variable.

\begin{theorem}\label{Gaussian}
    Consider the phase retrieval setup in Section~\ref{model} where $Z \sim \calN(0,\sigma^2)$, and $\beta_s \sim \mathcal{CN}(0,\bI_k \sigma_{\beta}^2)$ with $\sigma_{\beta}^2=\frac{c_{\beta}}{k}$ for some constant $c_{\beta} > 0$.
    If $k \to \infty$ with $\log k = o(\log p)$, then we have $\rvP_{\rme}(\alpha^*)\to 0$ as $p \to \infty$ provided that
    \begin{align}
        \label{facto2000}
        n\geq \max_{\alpha \in [\alpha^*,1]} \frac{\alpha k\log \frac{p}{k}}{\barI_1(\alpha)} (1+\eta),
    \end{align} 
    for arbitrarily small $\eta > 0$. 

    Conversely, under the broader scaling regime $k \to \infty$ with $k = o(p)$, we have $\rvP_{\rme}(\alpha^*)\to 1$ as $p\to \infty$ whenever
    \begin{align}
        \label{facto2001}
        n\leq \max_{\alpha \in [\alpha^*,1]}  \frac{(\alpha-\alpha^*)k\log \frac{p}{k}}{\barI_2(\alpha)} (1-\eta)
    \end{align} 
    for arbitrarily small $\eta > 0$. 
\end{theorem}

The assumption $\log k = o(\log p)$ in the achievability part (which holds, for example, when $k = O( (\log p)^c )$ for some $c > 0$) is rather restrictive compared to the general $k = o(p)$ scaling in the converse part.  The former arises from a significant technical challenge (see Proposition \ref{gaprof} below), and we expect that the requirement is merely an artifact of our analysis.\footnote{In fact, extending our analysis to the broader scaling regime $k = O(p^{1-\epsilon})$ (for some $\epsilon > 0$) leads to the correct scaling $n = O(k \log p)$, but unfortunately the resulting constant factors are quite loose compared to Theorem \ref{Gaussian}.} In addition, we note that while we allowed an arbitrary log-concave distribution in the discrete setting, here we have focused on $Z \sim \calN(0,\sigma^2)$ to simplify the analysis.  Despite this restriction, we believe that Gaussian noise still captures the essential features of the phase retrieval problem.

Once again, the scaling  $\sigma_{\beta}^2=\frac{c_{\beta}}{k}$ amounts to a fixed SNR.  As mentioned in \cite{Reeves2012a}, exact recovery is not possible for Gaussian $\beta_s$ when the SNR is constant, and may even need a huge number of measurements when the SNR increases with $p$.  This motivates the consideration of approximate recovery in this setting.

The differences between the upper and lower bounds are similar to the discrete case.  In particular, although the constants differ, the bounds are similar, and always have the same scaling laws. In the limit $c_{\beta} \to \infty$, we have $\barI_1(\alpha)=\log(c_{\beta})(1+o(1))$ and $\barI_2(\alpha)=\log(c_{\beta})(1+o(1))$; in this case, the maxima in~\eqref{facto2000}-\eqref{facto2001} are both achieved with $\alpha \to 1$, and hence, the two bounds coincide to within a multiplicative factor of $\frac{1}{1-\alpha^*}$.

\begin{figure}
	\centering
		\includegraphics[width=0.60\textwidth]{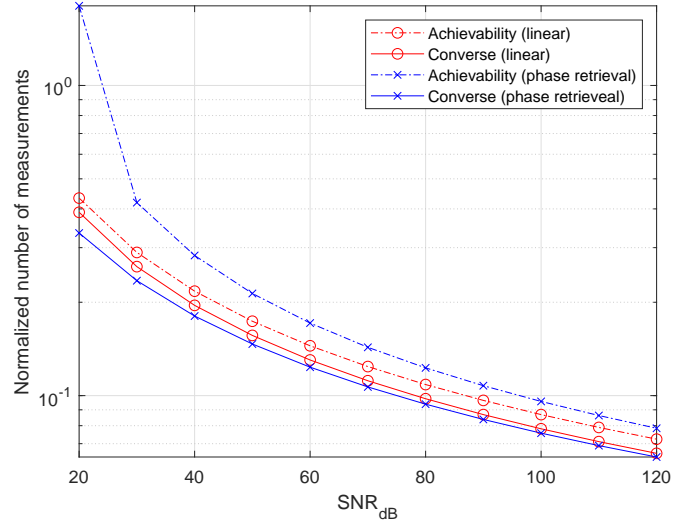}
    \caption{(Discrete case) Asymptotic thresholds on the number of measurements required for approximate support recovery for the  linear model \cite{Scarlet2017a}  and phase retrieval model (discrete case) with $\calN(0,1)$ Gaussian noise, and with distortion level $\alpha^*=0.1$ and non-zero entries $b_1=\cdots=b_k=\sqrt{\frac{c_{\beta}}{k}}$.  The asymptotic number of measurements is normalized by $k \log(\frac{p}{k})$, and ${\rm{SNR}}_{\rm{dB}}:=10 \log\big(\frac{2c_{\beta}^2}{\sigma^2}\big)=10 \log (2 c_{\beta}^2)$ (with $\sigma^2 = 1$).} 
	\label{fig:EvalPic0}
\end{figure}

\begin{figure}
	\centering
	\includegraphics[width=0.60\textwidth]{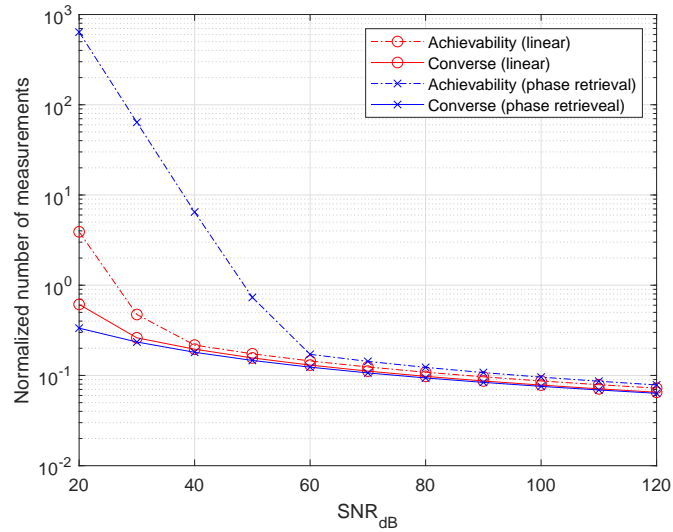}
	\caption{(Gaussian case) Asymptotic thresholds on the number of measurements required for approximate support recovery for the linear model~\cite{Scarlet2017a} and the phase retrieval model with $\calN(0,1)$ Gaussian noise, and with distortion level $\alpha^*=0.1$ and $\mathcal{CN}\big(0,\frac{c_{\beta}}{k}\big)$ non-zero entries in $\beta_s$. The asymptotic number of measurements is normalized by $k \log(\frac{p}{k})$, and ${\rm{SNR}}_{\rm{dB}}:=10 \log\big(2\frac{(k\sigma_{\beta}^2)^2}{\sigma^2}\big)=10 \log (2 c_{\beta}^2$) (with $\sigma^2 = 1$).} 
	\label{fig:EvalPic1}
\end{figure}

{\bf Comparison to the linear model.} In Figures \ref{fig:EvalPic0} and \ref{fig:EvalPic1}, we plot the upper and lower bounds of Theorem~\ref{cormain1} and Theorem \ref{Gaussian} for $\alpha^* = 0.1$ under various signal-to-noise ratios (SNRs), along with the counterparts for the linear model in \cite{Scarlet2017a}.\footnote{The approximate recovery result for the discrete case was not explicitly stated in \cite{Scarlet2017a}, but it is easily inferred from the analysis, and amounts to a much simpler version of the analysis of the present paper.}  For the discrete model, we focus on the simple case that $Z \sim \mathcal{N}(0,\sigma^2)$ and
\begin{equation}    
    b_1=\cdots=b_k=\sqrt{\frac{c_{\beta}}{k}}
\end{equation}
for some $c_{\beta} > 0$, corresponding to $m_{\beta} = 1$ in Theorem \ref{cormain1}.  In Appendix \ref{app:SNR}, we describe how we equate the SNR in the linear and phase retrieval models, and also how to evaluate the bounds of Theorem \ref{cormain1}  when $m_{\beta} = 1$.

As predicted by the discussion following Theorems~\ref{cormain1} and \ref{Gaussian}, the upper and lower bounds are close (though still with a constant gap) when the SNR is sufficiently high.  In addition, in this regime the information-theoretic limits of the phase retrieval model and the linear model are very similar, especially in the Gaussian case.

However, at lower SNR, the gap for the phase retrieval model can widen significantly more than that of the linear model.  This appears to be because the key mutual information quantities arising in the analysis can only be expressed in closed form in the linear model, while requiring possibly-loose bounds in the phase retrieval model.  However, all that is needed to close these gaps (at least partially) is to deduce improved mutual information bounds for the phase retrieval setting ({\em cf.}, Section \ref{sec:mut}).

%
%
\section{Auxiliary Results} \label{sec:aux}

In this section, we introduce the main auxiliary results needed to prove Theorems \ref{cormain1} and \ref{Gaussian}.  We first introduce some notation and recall the initial bounds for general observation models from \cite{Scarlet2017a}, and then present the relevant log-concavity properties, mutual information bounds, and concentration bounds.

\subsection{Information-Theoretic Definitions} 

We first outline some information theoretic definitions from~\cite{Scarlet2017a}, recalling that we are conditioning on a fixed $S=s$ throughout. We consider partitions of the support set $s\in \calS$ into two disjoint sets $s_{\rm{dif}}\neq \emptyset$ and $s_{\rm{eq}}$, where $s_{\rm{eq}}$ will typically correspond to an overlap between $s$ and some other set $\bars$ (i.e., $s\cap \bars$, the ``equal'' part), and $s_{\rm{dif}}$ will correspond to the indices in one set but not in the other (i.e., $s\setminus \bars$, the ``differing'' part).

For fixed $s \in \calS$ and a corresponding pair $(s_{\rm{dif}},s_{\rm{eq}})$, we introduce the notation
\begin{align}
    \label{defcon1}
    f_{\bY|\bX_{s_{\rm{dif}}}\bX_{s_{\rm{eq}}}}(\by|\bx_{s_{\rm{dif}}},\bx_{s_{\rm{eq}}})&:=f_{\bY|\bX_s}(\by|\bx_s),\\
    \label{defcon2}
    f_{Y|X_{s_{\rm{dif}}}X_{s_{\rm{eq}}}}(y|x_{s_{\rm{dif}}},x_{s_{\rm{eq}}},b_s)&:=f_{Y|X_s \beta_s}(y|x_s,\beta_s),
\end{align} 
where $f_{\bY|\bX_s}$ is the marginal distribution of~\eqref{jointdis}. While the left-hand sides of~\eqref{defcon1} and~\eqref{defcon2} represent the same quantities for any pair $(s_{\rm{dif}},s_{\rm{eq}})$, it will still prove convenient to work with these in place of the right-hand sides. In particular, this allows us to introduce the marginal distributions
\begin{align}
    f_{\bY|\bX_{s_{\rm{eq}}}}(\by|\bx_{s_{\rm{eq}}}) &:= \sum_{\bx_{s_{\rm{dif}}}} f_X^{n\times \ell}(\bx_{s_{\rm{dif}}})f_{\bY|\bX_{s_{\rm{dif}}}\bX_{s_{\rm{eq}}}}(\by|\bx_{s_{\rm{dif}}},\bx_{s_{\rm{eq}}}),\\
    f_{Y|X_{s_{\rm{eq}}}}(y|x_{s_{\rm{eq}}}) &:= \sum_{x_{s_{\rm{dif}}}} f_X^{\ell}(x_{s_{\rm{dif}}})f_{Y|X_{s_{\rm{dif}}}X_{s_{\rm{eq}}}}(y|x_{s_{\rm{dif}}},x_{s_{\rm{eq}}}),
\end{align} 
where $\ell := |s_{\rm{dif}}|$. Using the preceding definitions, we introduce two information densities (in the terminology of the information theory literature, e.g., \cite{PPV10}). The first contains probabilities averaged over $\beta_s$, 
\begin{align}
    i(\bx_{s_{\rm{dif}}};\by|\bx_{s_{\rm{eq}}}):=\log \frac{f_{\bY|\bX_{s_{\rm{dif}}}\bX_{s_{\rm{eq}}}}(\by|\bx_{s_{\rm{dif}}},\bx_{s_{\rm{eq}}})}{f_{\bY|\bX_{s_{\rm{eq}}}}(\by|\bx_{s_{\rm{eq}}})},
\end{align}
whereas the second conditions on $\beta_s=b_s$:
\begin{align}
    i^n(\bx_{s_{\rm{dif}}};\by|\bx_{s_{\rm{eq}}},b_s):=\sum_{i=1}^n i(x_{s_{\rm{dif}}}^{(i)};y^{(i)}|x_{s_{\rm{eq}}}^{(i)},b_s),
\end{align} 
where $(x^{(i)},y^{(i)})$ is the $i$-th measurement, and the single-letter information density is
\begin{align}
    \label{eqteso}
    i(x_{s_{\rm{dif}}};y|x_{s_{\rm{eq}}},b_s):=\log \frac{f_{Y|X_{s_{\rm{dif}}}X_{s_{\rm{eq}}}\beta_s}(y|x_{s_{\rm{dif}}},x_{s_{\rm{eq}}},b_s)}{f_{Y|X_{s_{\rm{eq}}}\beta_s}(y|x_{s_{\rm{eq}}},b_s)}.
\end{align}
Averaging~\eqref{eqteso} with respect to the distribution in~\eqref{defcon2} conditioned on $\beta_s=b_s$ yields a conditional mutual information quantity, which is denoted by
\begin{align}
    \label{defI}
    I_{s_{\rm{dif}},s_{\rm{eq}}}(b_s):=I(X_{s_{\rm{dif}}};Y|X_{s_{\rm{eq}}}, \beta_s=b_s).
\end{align}

\subsection{General Achievability and Converse Bounds} 

For the general support recovery problem with probabilistic models, the following achievability and converse bounds are given in~\cite{Scarlet2017a}.  While these are stated for the real-valued setting in \cite{Scarlet2017a}, the proofs apply verbatim to the complex-valued setting.

\begin{theorem}~\cite[Theorem 5]{Scarlet2017a}\label{scarletthm1} 
    Fix any constants $\delta_1>0, \delta_2\in (0,1)$, and $\gamma > 0$, and functions $\{\psi_{\ell}\}_{\ell=\lfloor \alpha^* k\rfloor}^k (\psi_{\ell}: \bbZ \times \bbR \to \bbR)$ such that the following holds:
    \begin{gather}
        \label{condA1}
        \bbP\Big[i^n(\bX_{s_{\rm{dif}}};\bY|\bX_{s_{\rm{eq}}},\beta_s)\leq n(1-\delta_2)I_{s_{\rm{dif}},s_{\rm{eq}}}(b_s) \,\big|\, \beta_s=b_s\Big] \le \psi_{|s_{\rm{dif}}|}(n,\delta_2),\\
        \label{condA2}
        n\geq \frac{\log{p-k \choose |s_{\rm{dif}}|}+\log\Big(\frac{k^2}{\delta_1^2}{k \choose |s_{\rm{dif}}|}^2 \Big)+\gamma}{I_{s_{\rm{dif}},s_{\rm{eq}}}(b_s)(1-\delta_2)},
    \end{gather} 
    for all $(s_{\rm{dif}},s_{\rm{eq}})$ with $\lfloor \alpha^* k\rfloor \leq |s_{\rm{dif}}|\leq k$ and for all $b_s$ in some (typical) set $\calT_{\beta}$. Then we have
    \begin{align}
        \label{achie1}
        \rvP_{\rme}(\alpha^*)\leq  \sum_{l=\lfloor \alpha^* k \rfloor}^k {k \choose \ell} \psi_{\ell}(n,\delta_2)+P_0(\gamma)+2 \delta_1 + \bbP[\beta_s \notin \calT_{\beta}],
    \end{align}
    where
    \begin{align}
        \label{defP0}
        P_0(\gamma):=\bbP\bigg[\log \frac{f_{\bY|\bX_s \beta_s}(\bY|\bX_s,\beta_s)}{f_{\bY|\bX_s}(\bY|\bX_s)}>\gamma\bigg].
    \end{align}
\end{theorem}

\begin{theorem}~\cite[Theorem 6]{Scarlet2017a} \label{scarletthm2} 
    Fix any constants $\delta_1>0, \delta_2>0$, and functions $\{\psi'_{\ell}\}_{\ell=\lfloor \alpha^* k\rfloor}^k (\psi'_{\ell}: \bbZ \times \bbR \to \bbR)$ such that the following holds:
    \begin{gather}
        \label{condA3}
        \bbP\Big[i^n(\bX_{s_{\rm{dif}}};\bY|\bX_{s_{\rm{eq}}},\beta_s)\leq n(1+\delta_2)I_{s_{\rm{dif}},s_{\rm{eq}}}(b_s) \,\big|\, \beta_s=b_s\Big] \ge 1-\psi'_{|s_{\rm{dif}}|}(n,\delta_2),\\
        \label{condA4}
        n \leq \frac{\log{p-k+ |s_{\rm{dif}}|\choose |s_{\rm{dif}}|}-\log \big(\sum_{d=0}^{\lfloor \alpha^* k\rfloor}{p-k\choose d}{|s_{\rm{dif}}|\choose d}\big)-\log \delta_1}{I_{s_{\rm{dif}},s_{\rm{eq}}}(b_s)(1+\delta_2)},
    \end{gather} 
    for all $(s_{\rm{dif}},s_{\rm{eq}})$ with $|s_{\rm{dif}}| \in [\lfloor \alpha^* k\rfloor, k]$, and for all $b_s$ in some (typical) set $\calT_{\beta}$.  Then we have
    \begin{align}
        \label{conv1}
        \rvP_{\rme}(\alpha^*)\geq \bbP[\beta_s \in \calT_{\beta}] \Big( 1-\max_{\ell=\lfloor\alpha^* k \rfloor,\dotsc,k}\psi'_{\ell}(n,\delta_2) \Big) -\delta_1.
    \end{align} 
\end{theorem}

\noindent The steps for applying and simplifying these bounds are as follows:
\begin{enumerate}
    \item Establish an explicit characterization of each mutual information term $I_{s_{\rm{dif}},s_{\rm{eq}}}(b_s)$ (e.g., upper and lower bounds);
    \item Use concentration of measure to find expressions for each function $\psi_{\ell}$ and $\psi'_{\ell}$ in Theorems \ref{scarletthm1} and \ref{scarletthm2}, i.e., functions satisfying \eqref{condA1} and \eqref{condA3};
     \item According to the specific model on the non-zero entries $\beta_s$ under consideration, choose a suitable typical set $\calT_{\beta}$, and also a value of $\gamma$, so that both $\bbP[\beta_s \notin \calT_{\beta}]$ and $P_0(\gamma)$ can be proved to be vanishing as $p \to \infty$;
    \item Combine and simplify the preceding steps to deduce the final sample complexity bound.
\end{enumerate}
These steps turn out to be highly non-trivial in the phase retrieval setting.  In the following subsections, we provide general-purpose tools for Steps 1 and 2; we defer Steps 3 and 4 to Section \ref{sec:discrete} for discrete $\beta_s$, and to Section \ref{sec:Gaussian} for Gaussian $\beta_s$.

\subsection{Log-Concavity Properties}  \label{sec:log_conc}

Both our mutual information bounds and concentration bounds will crucially rely on the log-concavity properties stated in the following lemma.

\begin{lemma} \label{lem:log_conc}
    Under the  phase retrieval setup in Section~\ref{model}, we have the following:
    \begin{enumerate}
        \item Given $S=s$ and $\beta_s=b_s$, the conditional marginal density of $Y$ is log-concave;
        \item Given $S=s$, $\beta_s=b_s$, and $X_{s_{\rm{eq}}}= x_{s_{\rm{eq}}}$ for some $s_{\rm eq} \subset s$, the conditional marginal density of $Y$ is log-concave.
    \end{enumerate}
\end{lemma}
\begin{IEEEproof}
    Recall that $Z$ is log-concave by assumption, and $Y = |\langle X_s, b_s \rangle|^2 + Z$ with $X_s$ having i.i.d.~$\calC\calN(0,1)$ entries.  In other words, $Y=U+Z$, where $U$ is the squared magnitude of a $\mathcal{CN}(0,\|b_s\|_2^2)$ random variable.  We observe that $Y$ is log-concave, since the $\chi^2$ distribution with two degrees of freedom is log-concave~\cite{Yu2011} and the convolution of two log-concave functions is log-concave \cite{Saumard2014}.   
    
    In addition, given $S=s$, $\beta_s=b_s$, and $X_{s_{\rm{eq}}}=x_{s_{\rm{eq}}}$, we have $Y=U+Z$, where $U$ is the squared magnitude of a $\mathcal{CN}(\langle b_{s_{\rm{eq}}}, x_{s_{\rm eq}}\rangle, \|b_{s_{\rm{dif}}}\|_2^2)$ random variable.  This distribution on $Y$ is also log-concave by a similar argument, and the fact that the {\em non-central} $\chi^2$ distribution with two degrees of freedom is log-concave \cite{Yu2011}.
\end{IEEEproof}

\subsection{Mutual Information Bounds} \label{sec:mut}

While an exact expression for the mutual information $I_{s_{\rm{dif}},s_{\rm{eq}}}(b)$ does not appear to be possible, the following theorem states closed-form upper and lower bounds.  While there is a gap between the two in general, the asymptotic behavior is similar when $v_{\rm{dif}}=\sum_{i \in s_{\rm{dif}}}|b_i|^2$ grows large; this fact ultimately leads to tight sample complexity bounds in the high-SNR setting.

\begin{theorem}\label{mutulthm} 
    For the phase retrieval setup in Section~\ref{model}, the following holds for $I_{s_{\rm{dif}},s_{\rm{eq}}}(b_s)$ defined in~\eqref{defI}:
    \begin{align}
        \label{boundmut}
        &\frac{1}{2}\log \bigg[\bigg(\frac{4}{\exp(2h(Z))}\bigg)v_{\rm{dif}}^2+1\bigg] \leq I_{s_{\rm{dif}},s_{\rm{eq}}}(b_s)\nonumber\\
        &\qquad \leq \frac{1}{2}\log \bigg(\frac{\pi e}{2}\bigg)+\frac{1}{2}\log\bigg[\bigg(\frac{2\pi e}{\exp(2h(Z))}\bigg)v_{\rm{dif}}^2+1\bigg]+\frac{1}{2}\log\bigg(1+\frac{v_{\rm{dif}}v_{\rm{eq}}}{v_{\rm{dif}}^2+\frac{\exp(2h(Z))}{2\pi e}}\bigg),
    \end{align} 
    where $v_{\rm{eq}}=\sum_{i \in s_{\rm{eq}}} |b_i|^2$ and  $v_{\rm{dif}}=\sum_{i \in s_{\rm{dif}}}|b_i|^2$.
\end{theorem}
\begin{IEEEproof} 
    The upper bound is based on the entropy power inequality and the maximum entropy property of the Gaussian distribution, and the lower bound is based on (known) results that give nearly-matching lower bounds for log-concave random variables.  The details are given in Appendix \ref{app:mutualinfo}.
\end{IEEEproof}

\subsection{Concentration Bounds} \label{sec:conce}

Perhaps the most technically challenging part of our analysis is to establish concentration bounds amounting to explicit expressions for $\psi_{\ell}$ and $\psi'_{\ell}$ in Theorems \ref{scarletthm1} and \ref{scarletthm2}.   

Before stating the final concentration bounds, we provide a general result that may be of independent interest, giving a concentration bound on {\em conditional} information random variables of the form $\tilde{h}(\bY|\bX) = - \log f_{\bY|\bX}(\bY|\bX)$ (in generic notation) under certain log-concavity assumptions.  Such a result is provided as a corollary of the following, which considers generic random variables $(X,Y)$ that need not be associated with the phase retrieval problem at this point.

\begin{proposition} \label{keyext} 
    Suppose that $(X,Y) \in \bbR^{2k} \times \bbR$ with joint density function $f_{X Y}$. For each $t \in \bbR^+$, define
    \begin{align}
        \label{defL}
       L(t)&:=\int_{\bbR^{2k} \times \bbR} f_{Y|X}^t (y|x) f_{X}(x)\mu(dx \times dy),
    \end{align}  
    and assume that
    \begin{align}
        \label{condB10}
        L(t) &<\infty
    \end{align} 
    for all $t \in \bbR^+$.  Moreover, for an arbitrary positive number $\barQ > 0$ (to be chosen later), define
    \begin{align}
        \label{defKK}
        K_1&:=\max\bigg\{\sup_{t \in (0,1]} t^2 (t L(t))'',\sup_{t \in (1,\infty)}t^2 (\barQ^{1-t} L(t))''\bigg\},
   \end{align} 
    and assume that
    \begin{align}
	\label{condB0}
   	K_1 <\infty.
	\end{align}
	Then, the following holds:
    \begin{align}
        \label{keyfact}
        \bbE[\exp(\mu(\tilh(Y|X)-h(Y|X))] \leq \exp((K_1+1) r(-\mu)), \quad \forall \mu \in \bbR,
		\end{align} 
    where
    \begin{align}
        \tilh(Y|X)&:=-\log f_{Y|X}(Y|X),\\
				\label{deftilh}
        h(Y|X)&:=\bbE[\tilh(Y|X)]=-\int_{\bbR^{k} \times \bbR} f_{X,Y}(x,y) \log f_{Y|X}(y|x) \mu(dx\times dy),\\
        \label{defru}
        r(u)&=\begin{cases} u-\log(1+u)&\quad \mbox{for}\quad -1 <u < \infty \\  +\infty \quad \mbox{otherwise}. \end{cases}.
		\end{align}
\end{proposition}
\begin{IEEEproof}
     We follow the general approach of \cite{Fradelizi2016}, which considers the unconditional information variable $\tilde{h}(x) = -\log f_{X}(x)$; however, many of the details differ significantly.  The reader is referred to Appendix \ref{app:conc_general}.
\end{IEEEproof}

From this, we immediately deduce a similar result for i.i.d.~product distributions.

\begin{corollary} \label{extthmnew} 
    Let $k \in \bbZ^+$. Suppose that $(\bX,\bY) \in \bbR^{2kn} \times \bbR^n$ with distribution $f_{X Y}^n$ (i.e., i.i.d.~on $f_{XY}$), where $f_{XY}$ satisfies \eqref{condB10} and \eqref{condB0}. Then, the following holds:
    \begin{align}
        \label{keyfactnew}
        \bbE[\exp(\mu(\tilh(\bY|\bX)-h(\bY|\bX))] \leq \exp(n(K_1+1) r(-\mu)), \quad \forall \mu \in \bbR,
		\end{align} 
    where
    \begin{align}
		\label{deftilhnew}
        \tilh(\bY|\bX)&:=-\log f_{\bY|\bX}(\bY|\bX),\\
        h(\bX,\bY)&:=\bbE[\tilh(\bY|\bX)]=-\int_{\bbR^{2kn} \times \bbR^n} f_{\bX,\bY}(\bx,\by) \log f_{\bY|\bX}(\by|\bx) \mu(d\bx\times d\by),
		\end{align} 
        and $K_1$ is defined in~\eqref{defKK}.
\end{corollary}
\begin{IEEEproof}
    The i.i.d.~assumption readily yields $h(\bY|\bX) = n h(Y|X)$ and $\bbE[\exp(\mu \tilh(\bY|\bX))] = \big(\bbE[\exp(\mu \tilh(Y|X))]\big)^n$, where $(X,Y) \sim f_{XY}$.  Hence,
    \begin{align}
        \bbE[\exp(\mu(\tilh(\bY|\bX)-h(\bY|\bX))] =\Big(\bbE[\exp(\mu(\tilh(Y|X)-h(Y|X)))]\Big)^n,
    \end{align}
    and the corollary follows by bounding the expectation via Proposition~\ref{keyext}.
\end{IEEEproof}

We are now ready to state a general result on the concentration of conditional information variables.

\begin{corollary}\label{verykey} 
    Let $(\bX,\bY)\sim f_{XY}^n$ with $X \in \bbR^{2k}, Y\in \bbR$. Then, under conditions~\eqref{condB10} and \eqref{condB0} of Proposition~\ref{keyext}, the following holds for any $\mu>0$: 
    \begin{align}
    \bbP[\tilh(\bY|\bX)-nh(Y|X) \geq n(K_1+1) \mu] \leq \exp(-n (K_1+1) r(\mu)),\\
    \bbP[\tilh(\bY|\bX)-nh(Y|X) \leq -n(K_1+1) \mu] \leq \exp(-n (K_1+1) r(-\mu)),
    \end{align}
    where $\tilh(\bY|\bX)$ is defined in~\eqref{deftilh}, $K_1$ in \eqref{defKK}, and $r(\mu)$ in~\eqref{defru}. 
\end{corollary}
\begin{IEEEproof} 
    With Corollary \ref{extthmnew} in place, this is a fairly straightforward application of the Chernoff bound.  The details are given in Appendix \ref{proof:corkey}.
\end{IEEEproof}

\begin{remark} Some remarks are in order. 
\begin{itemize}
    \item In~\cite[Theorem 3.1]{Fradelizi2016}, the authors showed that $\bbE[\exp(\mu(\tilh(Y)-h(Y)))] \leq \exp(K r(-\mu))$ for any $\mu\in \bbR$ and any random vector $Y \in \bbR$ such that $f_{Y} \in L^{t}(\bbR)$ for all $t > 0$ (i.e., $|f_{Y}(y)|^t$ is absolutely integrable). Theorem~\ref{keyext} shows that this fact can be extended to conditional distributions under some assumptions on the joint distribution $f_{XY}$.
    \item When $\bX$ and $\bY$ are independent, Theorem~\ref{keyext} is very similar to~\cite[Theorem 3.1]{Fradelizi2016}.  
    \item When we apply Corollary \ref{verykey} to the phase retrieval problem, we will bound $K_1$ using the log-concavity properties in Lemma \ref{lem:log_conc}.
    \item If $\bX$ and $\bY$ were jointly log-concave, a variant of \eqref{keyfact} with an alternative definition for $K_1$ could be used based on \cite[Theorem 3.1]{Fradelizi2016} and the union bound, since $\tilh(\bY|\bX)=\tilh(\bX,\bY)-\tilh(\bX)$ and $h(\bY|\bX)=h(\bX,\bY)-h(\bX)$. However, such a bound is not suitable for out purposes, since the measurement variables and outputs in the phase retrieval problem are not jointly log-concave.
    \item Alternatively, using only the fact that $f_{\bY|\bX}(\cdot|\bx)$ is log-concave for all $\bx$, \cite[Theorem 3.1]{Fradelizi2016} gives for suitably-defined $K$ that
    \begin{align}
        \label{eqfrade}
        \bbE[\exp(\mu(\tilh(\bY|\bX=\bx)-h(\bY|\bX=\bx))] \leq \exp(K r(-\mu)).
    \end{align} However, \eqref{keyfactnew} does not appear to follow from~\eqref{eqfrade}. 
\end{itemize}
\end{remark}

Although the preceding results are general, finding an explicit expression or upper bound for $K_1$ in \eqref{defKK} is non-trivial.   With some technical effort, we are able to attain such a bound for the phase retrieval model and deduce the following key concentration result used in the proofs of Theorems \ref{cormain1} and \ref{Gaussian}.

\begin{theorem} \label{thre1} 
    Under the phase retrieval setup in Section~\ref{model}, the following bounds hold:
    \begin{align}
        \label{tofact1}
        \bbP\big[i^n(\bX_{s_{\rm{dif}}};\bY|\bX_{s_{\rm{eq}}},\beta_s=b_s)-n I_{s_{\rm{dif}},s_{\rm{eq}}}(b_s) \leq -2n C(b_s) \mu\big] &\leq \exp(-n C(b_s) r(\mu)) + \exp(-nC(b_s)  r(-\mu)),\\
        \label{tofact2}
        \bbP\big[i^n(\bX_{s_{\rm{dif}}};\bY|\bX_{s_{\rm{eq}}},\beta_s=b_s)-n I_{s_{\rm{dif}},s_{\rm{eq}}}(b_s)\geq 2nC(b_s)  \mu\big] &\leq \exp(-nC(b_s) r(\mu)) + \exp(-nC(b_s) r(-\mu)),
    \end{align} 
    for all $\mu>0$, where $I_{s_{\rm{dif}},s_{\rm{eq}}}(b_s)$ is defined in~\eqref{defI}, $C(b_s)$ is a constant depending on $b_s\in \bbC^k$,\footnote{The definition is given in~\eqref{defCb} in Lemma~\ref{bigband2018}.} and $r(\mu)$ is defined in~\eqref{defru}.
\end{theorem}
\begin{IEEEproof}
    See Appendix \ref{app:conc_phase}.
\end{IEEEproof}

It turns out that the constant $C(b_s)$ behaves as $\Theta(1)$ whenever $\|b_s\|_2 = \Theta(1)$, leading to the following corollary.
    
\begin{corollary}\label{trivicor} 
    For the complex phase retrieval problem in~\eqref{acqui}, equations~\eqref{tofact1} and~\eqref{tofact2} hold with $C(b_s)$ replaced by some constant $C=\Theta(1)$ under the condition $\|b_s\|_2=\Theta(1)$.
\end{corollary}
\begin{IEEEproof}
    See Appendix \ref{app:conc_phase}.
\end{IEEEproof}

%
%
\section{Proof of Theorem \ref{cormain1} (Discrete $\beta_s$)}\label{sec:discrete}

As a stepping stone to proving Theorem \ref{main1}, we state the following lemma, which can be thought of as a version of that theorem before applying the suitable mutual information bounds and asymptotic simplifications.  Recall that  $I_{s_{\rm{dif}},s_{\rm{eq}}}(b_s)$ is defined in~\eqref{defI}.

\begin{lemma}\label{main1}
     Consider the setup of Theorem \ref{cormain1} with $\beta_s$ being a uniformly random permutation of $b_s=(b_1,b_2,\ldots,b_k)$ satisfying $|b_{\rm{min}}|=\Theta(|b_{\rm{max}}|)$, and $\|b_s\|_2=\Theta(1)$, and $k \to \infty$, and $m_{\beta}$ distinct elements in $(b_1,b_2,\ldots,b_k)$.  

    We have $\rvP_{\rme}(\alpha^*) \to 0$ as $p\to \infty$ provided that
    \begin{align}
        \label{eq1main1}
        n\geq \max_{(s_{\rm{dif}},s_{\rm{eq}}) \,:\, \lfloor \alpha^* k \rfloor \leq |s_{\rm{dif}}| \leq k} \frac{\log{p-k \choose |s_{\rm{dif}}|}}{I_{s_{\rm{dif}},s_{\rm{eq}}}(b_s)}(1+\eta),
    \end{align} 
    for arbitrarily small $\eta > 0$ if either of the following additional conditions holds: (i) $m_{\beta}=\Theta(1)$ and $k=o(p)$; or (ii) $\log k = o(\log p)$.
    
    Conversely, for general $m_{\beta}$ and $k = o(p)$, we have $\rvP_{\rme}(\alpha^*) \to 1$ as $p\to \infty$ provided that
    \begin{align}
        \label{eq2main1}
        n\leq \max_{(s_{\rm{dif}},s_{\rm{eq}}) \,:\, \lfloor \alpha^* k \rfloor \leq |s_{\rm{dif}}| \leq k} \frac{\log{p-k+ |s_{\rm{dif}}|\choose |s_{\rm{dif}}|}-\log \big(\sum_{d=0}^{\lfloor \alpha^* k\rfloor}{p-k\choose d}{|s_{\rm{dif}}|\choose d}\big)}{I_{s_{\rm{dif}},s_{\rm{eq}}}(b_s)}(1-\eta)
    \end{align} 
    for  arbitrarily small $\eta > 0$ . 
\end{lemma}

\subsection{Proof of Lemma~\ref{main1}}

We apply Theorem \ref{scarletthm1} in several steps as follows.

\emph{Step 1: Choose the typical set.} Let $\calT_{\beta}$ be the set of all permutations of the fixed complex vector $b_s = (b_1,b_2,\ldots,b_k)$. Under the conditions $|b_{\rm{min}}|=\Theta(|b_{\rm{max}}|)$ and $\|b_s\|_2=\Theta(1)$, we observe that the quantity $v_{\rm{dif}}=\sum_{i \in s_{\rm{dif}}}|b_i|^2$ also behaves as $\Theta(1)$, while $v_{\rm{eq}}=\sum_{i \in s_{\rm{eq}}} |b_i|^2$ behaves as $O(1)$ (note that we only consider $s_{\rm{dif}}$ with cardinality $\Theta(k)$, a constant fraction of the total $k$).  Hence, we find from~\eqref{boundmut} of Theorem~\ref{mutulthm} that
\begin{align}
    \label{goodth}
    I_{s_{\rm{dif}},s_{\rm{eq}}}(b_s)= \Theta(1).
\end{align}
In addition, since there are at most $m_{\beta}^k$ possible random permutations by the definition of $m_{\beta}$, choosing $\gamma=\log \frac{1}{\min_{b_s}f_{\beta_s}(b_s)}\leq k \log m_{\beta}$ gives $P_0(\gamma)=0$; this immediately follows by writing $f_{\bY|\bX_s}(\by|\bx_s) = \sum_{b_s} f_{\beta_s}(b_s)f_{\bY|\bX_s \beta_s}(\by|\bx_s,b_s)$ in \eqref{defP0}.

\emph{Step 2: Bound the information density tail probabilities.} Fix $\delta_2>0$ (to be chosen later), and define
\begin{align}
    \label{158facto}
    \mu_{|s_{\rm{dif}}|}&:=\frac{\delta_2 I_{s_{\rm{dif}},s_{\rm{seq}}}(b_s)}{2 C}
\end{align} 
for each $|s_{\rm dif}|$, where $C$ is defined in Corollary~\ref{trivicor}.

Now, for each integer $\ell$ representing $|s_{\rm{dif}}|$, set
\begin{align}
    \label{defsdif}
    \psi_{\ell}(n,\delta_2) := \psi'_{\ell}(n,\delta_2) &:=\exp(-n C r(\mu_{\ell})) + \exp(-n C r(-\mu_{\ell}))\\
    \label{defsdifphay}
    &\leq 2 \exp(-\min\{r(\mu_{\ell}),r(-\mu_{\ell})\}nC ).
\end{align}
By setting $\mu=\mu_{|s_{\rm{dif}}|}$ in~\eqref{tofact1}, and applying Corollary~\ref{trivicor}, we have
\begin{align}
    \label{res1}
    \bbP[i^n(\bX_{s_{\rm{dif}}};\bY|\bX_{s_{\rm{eq}}},\beta_s=b_s)\leq n (1-\delta_2) I_{s_{\rm{dif}},s_{\rm{eq}}}(b_s)] &\leq \psi_{|s_{\rm{dif}}|}(n,\delta_2).
\end{align}
Similarly, we obtain from~\eqref{tofact2} that
\begin{align}
    \label{res2}
    \bbP[i^n(\bX_{s_{\rm{dif}}};\bY|\bX_{s_{\rm{eq}}},\beta_s=b_s)\leq n (1+\delta_2) I_{s_{\rm{dif}},s_{\rm{eq}}}(b_s)] &\geq 1-\psi'_{|s_{\rm{dif}}|}(n,\delta_2).
\end{align}
This means that the conditions~\eqref{condA1} and~\eqref{condA3} are satisfied with $\psi_{\ell}$ and $\psi'_{\ell}$ defined in~\eqref{defsdif}, respectively. 

\emph{Step 3: Control the remainder terms.} We first consider the remainder term $\psi'_{\ell}$ in the converse bound \eqref{conv1} resulting from \eqref{res2}.  Since $|s_{\rm{dif}}| \in [\lfloor \alpha^* k\rfloor, k]$, we have $|s_{\rm{dif}}|=\lfloor \alpha k \rfloor$ for some $ \alpha \in [\alpha^*,1]$.  Since $I_{s_{\rm{dif}},s_{\rm{eq}}}(b_s)= \Theta(1)$ as stated in \eqref{goodth}, we deduce from \eqref{158facto} that
\begin{align}
    \label{cho1}
    \mu_{\lfloor \alpha k\rfloor}= \Theta(\delta_2),
\end{align}
so that \eqref{defsdifphay} yields
\begin{align}
    \label{hl0b}
    \max_{l \in [\lfloor \alpha^* k\rfloor, k]} \psi'_{l} (n,\delta_2) \le 2\exp\Big(- n r\big(\Theta( \delta_2 ) \big) \Big).
\end{align}
We choose $\delta_2$ to be a slowly vanishing function of $p$, so that a simple Taylor expansion in the definition of $r(\cdot)$ in \eqref{defru} yields $r\big(\Theta( \delta_2 ) \big) = \Theta(\delta_2^2)$.  Therefore, \eqref{hl0b} vanishes as $p \to \infty$ if $n=\omega\big(\frac{1}{\delta^2_2}\big)$.

We now turn to the achievability part.  First observe that the term $\sum_{\ell=\lfloor \alpha^* k\rfloor}^k {k \choose \ell} \psi_{\ell}(n,\delta_2)$ in \eqref{achie1} vanishes as $p \to \infty$ provided that $(1-\alpha^*) k \max_{\ell\in [\lfloor \alpha^* k\rfloor, k]} {k \choose \ell}\psi_{\ell}(n,\delta_2) \to 0$. Since $\psi_{\ell}(n,\delta_2)>0$, this is equivalent to
\begin{align}
    \label{embed3}
    -\log k -\log {k \choose \ell} -\log[\psi_{\ell}(n,\delta_2)]  \to \infty
\end{align} 
for all $\ell \in [\lfloor \alpha^* k\rfloor,k]$.  From~\eqref{defsdif} and~\eqref{embed3}, we find that $\sum_{\ell=\lfloor \alpha^* k\rfloor}^k {k \choose \ell} \psi_{\ell}(n,\delta_2) \to 0$ provided that
\begin{align}
    \label{extrac}
    -\log k -\ell \log \frac{ke}{\ell} +\min\{r(\mu_{\ell}),r(-\mu_{\ell})\} n  \to \infty
\end{align} 
as $p \to \infty$ for all $\ell \in [\lfloor \alpha^* k\rfloor, k]$, where we have used $\log{k \choose \ell} \le \ell\log\frac{ke}{\ell}$.

Since $\ell=\lfloor \alpha k \rfloor$ for some $\alpha \in [\alpha^*,1]$, \eqref{extrac} holds provided that
\begin{align}
    \label{ghost1}
    n\geq  \max_{\alpha \in [\alpha^*,1]}\frac{\log k + \alpha k\log\frac{e}{\alpha}}{\min\{r(\mu_{\lfloor \alpha k\rfloor}),r(-\mu_{\lfloor \alpha k\rfloor})\}}(1+\eta)
\end{align} 
for arbitrarily small $\eta > 0$.
Again using $\min\{r(\mu_{\ell}),r(-\mu_{\ell})\} = \Theta(\delta_2^2)$ for slowly vanishing $\delta_2$ (as established in the above converse part), we find that this condition simplifies to $n=\Omega\big(\frac{k}{\delta_2^2}\big)$.

We also need to consider the effect of the term $\gamma$ in Theorem \ref{scarletthm1}, recalling that we already established that $P_0(\gamma) = 0$ with $\gamma \le k \log m_{\beta}$. For the first case in Lemma~\ref{main1}, i.e., $m_{\beta}=\Theta(1)$ and $k=o(p)$, we have $\gamma = O(k)$.  In the second case, i.e. $m_{\beta}=O(k)$ and $\log k=o(\log p)$, we have $\gamma = O( k \log k) = o\big(k \log \frac{p}{k}\big)$.  Hence, in both cases, we have $\gamma = o\big(k \log \frac{p}{k}\big)$.

\emph{Step 4: Combine and simplify.} For the converse part, since \eqref{hl0b} vanishes when $n = \omega\big(\frac{1}{\delta_2^2}\big)$, we deduce from Theorem~\ref{scarletthm2} (with $\delta_1\to 0$ and $\delta_2 \to 0$ sufficiently slowly) that $\rvP_{\rme}(\alpha^*) \to 1$ when~\eqref{eq2main1} holds, as required.  Specifically, \eqref{eq2main1} is merely an asymptotic simplification of \eqref{condA4}.

For the achievability part, by choosing $\delta_1 \to 0$ and $\delta_2 \to 0$ sufficiently slowly in Theorem~\ref{scarletthm1}, we find that the condition \eqref{condA2} reduces to
\begin{align}
    \label{testingkey}
    n\geq \max_{(s_{\rm{dif}},s_{\rm{eq}}) \,:\, \lfloor \alpha^* k \rfloor\leq |s_{\rm{dif}}|\leq k} \frac{\log {p-k \choose |s_{\rm{dif}}|}+2 \log(k {p-k \choose |s_{\rm{dif}}|})-2\log \delta_1+\gamma}{I_{s_{\rm{dif}},s_{\rm{eq}}}(b_s)(1-\delta_2)}(1+\eta)
\end{align} 
for arbitrarily small $\eta > 0$.  Since $k=o(p)$ and $|s_{\rm{dif}}|=\lfloor \alpha k \rfloor$ for some $\alpha \in [\alpha^*,1]$, the first term in the numerator of~\eqref{testingkey} behaves as $\Theta(\alpha k\log (\frac{p}{k}))$, and the second term behaves as $\Theta(\log k+|s_{\rm{dif}}|\log\frac{k}{|s_{\rm{dif}}|})=\Theta(\alpha k)$. Since for both cases (i) and (ii) of Lemma~\ref{main1}, we have $\gamma=o(k \log(\frac{p}{k}))$, it immediately follows that the numerator in~\eqref{testingkey} is dominated by the first term and the others can be factored into the remainder term $\eta > 0$. Moreover, the condition
$n=\Omega\big(\frac{k}{\delta_2^2}\big)$ stated following \eqref{ghost1} behaves as $o\big(\alpha k\log \frac{p}{k}\big)$ when $\delta_2 \to 0$ sufficiently slowly (e.g., $\delta_2=\Theta(\frac{1}{\log(\log k)})$).  Combining these observations, we deduce that we only require \eqref{testingkey}, with the first term alone kept in the numerator, and the rest factored into $\eta$ in \eqref{eq1main1}.

\subsection{Proof of Theorem~\ref{cormain1}} \label{discproof}

Recall the definitions $v_{\rm{dif}}=\sum_{i \in s_{\rm{dif}}}|b_i|^2$ and $v_{\rm{eq}}=\sum_{i \in s_{\rm{eq}}}|b_i|^2$.  Since  $|s_{\rm{dif}}| \in [\lfloor \alpha^* k\rfloor, k]$, we have $|s_{\rm{dif}}|=\lfloor \alpha k \rfloor$ for some $ \alpha \in [\alpha^*,1]$. 

For the achievability part, we use the lower bound in \eqref{boundmut} of Theorem \ref{mutulthm}.  Since this lower bound is increasing in $v_{\rm{dif}}$ and does not depend on $v_{\rm{eq}}$, we have the following whenever $|s_{\rm{dif}}|=\lfloor \alpha k \rfloor$:
\begin{equation}
    \label{mutbound1}
    I_{s_{\rm{dif}},s_{\rm{eq}}}(b_s) \ge I_1(\alpha,k),
\end{equation}
recalling that $I_1(\alpha,k)$ defined in \eqref{defI1} replaces $v_{\rm dif}$ by the value corresponding to the lowest-magnitude entries of $b_s$.  Hence,~\eqref{eq3main} of Theorem~\ref{cormain1} follows from~\eqref{eq1main1} of Lemma~\ref{main1} by observing that the numerator of~\eqref{eq1main1} behaves as $\big(\alpha k \log \frac{p}{k}\big)(1+o(1))$ and the denominator is lower bounded by $I_1(\alpha,k)$ via~\eqref{mutbound1}. 

For the converse part, we use the upper bound in \eqref{boundmut} of Theorem \ref{mutulthm}.  While this bound depends on $v_{\rm{dif}}$ and $v_{\rm{eq}}$ in a more complicated fashion, the converse bound \eqref{eq2main1} remains valid when we replace the maximum over $(s_{\rm{dif}}, s_{\rm{eq}})$ by {\em any} fixed choice.  Under the choice in which $s_{\rm{dif}}$ contains the indices corresponding to the $\lfloor \alpha k \rfloor$ entries of $b_s$ with the smallest magnitude, \eqref{boundmut} yields
\begin{align}
    \label{mutbound2}
    I_{s_{\rm{dif}},s_{\rm{eq}}}(b_s) \le I_2(\alpha,k),
\end{align}
where $I_2(\alpha,k)$ is defined in \eqref{defI1}.  

Regarding the numerator in \eqref{eq2main1}, it was shown in \cite[Proof of Cor. 2]{Scarlet2017a} via simple asymptotic expansions that the term $\log \big(\sum_{d=0}^{\lfloor \alpha^* k\rfloor}{p-k\choose d}{|s_{\rm{dif}}|\choose d}\big)$ is dominated by $\alpha^* k \log \frac{p}{k}$ as $p \to \infty$ with $k = o(p)$, and that the overall numerator in~\eqref{eq2main1} simplifies to $(\alpha-\alpha^*) k \log (\frac{p}{k}) (1+o(1))$ if $|s_{\rm{dif}}|=\lfloor \alpha k \rfloor$. Combining this fact with \eqref{mutbound2}, we have that $\rvP_{\rme}(\alpha^*) \to 1$ as $p\to \infty$ if
\begin{align}
    \label{gfact1}
    n\leq \max_{\alpha \in [\alpha^*,1]}\frac{(\alpha-\alpha^*) k \log (\frac{p}{k})}{I_2(\alpha,k)}(1-\eta)
\end{align} 
for some $\eta > 0$. This yields equation~\eqref{eq4main} of Theorem~\ref{cormain1}. 


%
%
\section{Proof of Theorem \ref{Gaussian} (Gaussian $\beta_s$)}\label{sec:Gaussian}

One of the key challenges in the Gaussian setting compared to the discrete setting is bounding the quantity $P_0(\gamma)$ appearing in Theorem \ref{scarletthm1}.  As noted in \cite{Scarlet2017a}, this roughly amounts to bounding the mutual information quantity $I(\beta_s;\bY|\bX_s)$, for which the approaches proposed in \cite{Scarlet2017a} appear to be insufficient.  The following proposition states a bound on  $P_0(\gamma)$ resulting from a novel approach.

\begin{proposition} \label{gaprof} 
    Under the phase retrieval setup in Section~\ref{model} with $Z \sim \calN(0,\sigma^2)$ for some $\sigma \in \bbR^+$, $\beta_s \sim \mathcal{CN}(0, \bI_k \sigma_{\beta}^2)$ for some $\sigma_{\beta}^2 = \Theta\big(\frac{1}{k}\big)$, and $k \to \infty$ with $n = \Omega(k)$, the following holds:
    \begin{align}
        \label{defI00}
        P_0(\gamma) \leq \frac{O(k \log n)}{\gamma}+o(1)
    \end{align}
    for any $\gamma>0$, where $P_0(\gamma)$ is defined in~\eqref{defP0} of Theorem~\ref{scarletthm1}, i.e., $P_0(\gamma):=\bbP\big[\log\frac{f_{\bY|\bX_s \beta_s}(\bY|\bX_s,\beta)}{f_{\bY|\bX_s}(\bY|\bX_s)}>\gamma \big]$.
\end{proposition}
\begin{IEEEproof} 
    See Section \ref{proof:gaprof}.
\end{IEEEproof}

The following proposition characterizes the behavior of the $\lfloor \alpha k\rfloor$ entries of $\beta_s$ having the smallest magnitude for fixed $\alpha$. For the real linear model in~\cite{Scarlet2017a}, $(\beta_s)_i^2$ follows a chi-squared distribution with one degree of freedom for all $i=1,\cdots, k$. However, for our phase retrieval model (cf.~Section~\ref{model}), $|(\beta_s)_i|^2$ follows a chi-squared distribution with two degrees of freedom.  This difference only amounts to a minor change in the definition of $g(\alpha)$ in~\eqref{defgalpha}, and \cite[Prop.~3]{Scarlet2017a} extends immediately to the following.

\begin{proposition}\cite[Prop.~3]{Scarlet2017a} \label{scarprop3} 
    For $\beta_s$ i.i.d.~on $\calC\calN\big(0,\frac{\sigma_{\beta}^2}{k})$ for fixed $\sigma_{\beta}^2$, we have with probability one that the following holds for all $\alpha \in [0,1]$:
    \begin{align}
    \label{goodpoint}
    \lim_{k\to \infty} \frac{1}{k \sigma_{\beta}^2} \sum_{i=1}^{\lfloor \alpha k\rfloor} |(\beta_s')_i|^2 = g(\alpha),
    \end{align}
    where $\beta_s'$ is the permutation of $\beta_s$ whose entries are listed in increasing order of magnitude, and $g(\alpha)$ is defined in~\eqref{defgalpha}.
\end{proposition}

Note that this result is essentially an application of the Glivenko-Cantelli theorem \cite[Thm.~19.1]{Van00}, stating uniform convergence of the empirical cumulative distribution function (CDF) to the true CDF.

\subsection{Proof of  Theorem~\ref{Gaussian}}

    The proof of Theorem~\ref{Gaussian} follows the same high-level steps as those for the discrete case.

    \emph{Step 1: Choose a typical set.} Based on the result in Proposition~\ref{scarprop3}, we set $\calT_{\beta}$ to be the set of vectors $b_s$ such that $\max_{\alpha \in [0,1]} \big|\frac{1}{k \sigma_{\beta}^2} \sum_{i=1}^{\lfloor \alpha k\rfloor} |(b_s')_i|^2 - g(\alpha)\big|\leq \eps$ as $k \to \infty$, where $\eps$ is chosen to decay sufficiently slowly so that $\bbP[\calT_{\beta}] \to 1$.  We therefore have
	\begin{align}
        \label{lastingsup1}
	   \frac{1}{k\sigma_{\beta}^2} \sum_{i=1}^{\lfloor \alpha k \rfloor}|b'_i|^2 \to g(\alpha) 
	\end{align} 
    for all $b_s \in \calT_{\beta}$, and in particular $\frac{1}{k\sigma_{\beta}^2}\|b_s\|_2^2\to 1$ by setting $\alpha = 1$.  In addition, we obtain
	\begin{align}
        \label{eesup}
    	\bigg(\sum_{i=1}^{\lfloor \alpha k \rfloor}|b'_i|^2 \bigg)\bigg(\|b_s\|_2^2-\sum_{i=1}^{\lfloor \alpha k \rfloor}|b'_i|^2\bigg)		
    	\to c_{\beta}^2 g(\alpha) (1-g(\alpha))
	\end{align} 
     by using $c_{\beta}=k \sigma_{\beta}^2$ and $\frac{1}{k\sigma_{\beta}^2}\|b_s\|_2^2 \to 1$.

    We proceed similarly to Section \ref{discproof} for the discrete setting, recalling that $v_{\rm{dif}}=\sum_{i \in s_{\rm{dif}}}|b_i|^2$.  For the achievability part, \eqref{lastingsup1} and the mutual information lower bound in \eqref{boundmut} (with $Z\sim \calN(0,\sigma^2)$) imply (within the typical set) that for any $(s_{\rm{dif}},s_{\rm{eq}})$ with $|s_{\rm{dif}}|=\lfloor \alpha k\rfloor$, we have
    \begin{align}
        \label{tfact1}
        I_{s_{\rm{dif}},s_{\rm{eq}}}(b_s) \ge \barI_1(\alpha) (1+o(1)),
    \end{align}
    where $\barI_1(\alpha)$ is defined in \eqref{defbarI1}.

    For the converse part, we do not need to consider all pairs $(s_{\rm{dif}},s_{\rm{eq}})$, since any fixed choice still provides a valid converse.  Hence, for a given cardinality $|s_{\rm{dif}}|=\lfloor \alpha k\rfloor$, we only consider the choice such that $s_{\rm{dif}}$ contains the indices corresponding to the $\lfloor \alpha k \rfloor$ entries of $b_s$ with the smallest magnitude.  Under this choice, we have from \eqref{lastingsup1}--\eqref{eesup} and the upper bound in  \eqref{boundmut} (with $Z\sim \calN(0,\sigma^2)$) that
    \begin{align}
        \label{tfact2}
        I_{s_{\rm{dif}},s_{\rm{eq}}}(b_s) \le \barI_2(\alpha) (1+o(1)),
    \end{align}
    where $\barI_2(\alpha)$ is defined in \eqref{defbarI2k}.
    
    \emph{Step 2: Bound the information density tail probabilities.} We again make use of Theorem~\ref{thre1} and its subsequent expression for $\psi_{\ell}$ and $\psi_{\ell}'$ in \eqref{defsdifphay}.
   
    \emph{Step 3: Control the remainder terms.} Recall that $P_0(\gamma)$ is defined in~\eqref{defP0} of Theorem~\ref{scarletthm1}.  By Proposition~\ref{gaprof}, we have $P_0(\gamma) \to 0$ under any choice of $\gamma$ satisfying $\gamma =\vartheta_p k \log n$ for some $\vartheta_p$ growing to $\infty$ arbitrarily slowly.  When this growth is sufficiently slow and $n = O\big(k \log \frac{p}{k}\big)$, we have
    \begin{align}
        \gamma = o\Big( k \log \frac{p}{k} \Big) \label{beahat}
    \end{align}
    due to the assumption $\log k = o(\log p)$.  Note that $n = O\big(k \log \frac{p}{k}\big)$ holds trivially under the condition \eqref{facto2001} in the converse, whereas for the achievability we can assume without loss of generality that \eqref{facto2000} holds with equality, since additional measurements can only improve the information-theoretic performance.


   By our choice of $\calT_{\beta}$, we may focus on realizations $b_s$ of $\beta_s$ satisfying~\eqref{goodpoint}. For such realizations, we have for all $s_{\rm{dif}}$ with $|s_{\rm{dif}}|=\Theta(k)$ that $v_{\rm{dif}}=\sum_{i\in s_{\rm{dif}}}|b_i|^2=\Theta(1)$ by~\eqref{goodpoint} and the assumption that $\sigma_{\beta}^2=\frac{c_{\beta}}{k}$. Hence, by by~\eqref{tfact1}--\eqref{tfact2} and the fact that $|s_{\rm{dif}}|=\lfloor \alpha k\rfloor$ for some $\alpha \in [\alpha^*,1]$, we have $I_{s_{\rm{dif}},s_{\rm{eq}}}(b_s)=\Theta(1)$ as $k\to \infty$. By using the same arguments as~\eqref{cho1} and \eqref{hl0b}, we deduce that the remainder term $\psi'_{\ell}$ resulting from \eqref{res2} in the converse bound vanishes asymptotically if $n=\omega\big(\frac{1}{\delta^2_2}\big)$. 

    For the achievability part, we have $\sum_{\ell=\lfloor \alpha^* k\rfloor}^k {k \choose \ell} \psi_{\ell}(n,\delta_2) \to 0$ as $k \to \infty$ if $n=\Omega\big(\frac{k}{\delta_2^2}\big)$ by using the same arguments as~\eqref{embed3}--\eqref{ghost1}.  Recalling that we also established above \eqref{beahat} that $P_0(\gamma) \to 0$, we deduce that $\rvP_{\rme}(\alpha^*) \to 0$ since $\rvP_{\rme}(\alpha^*) \le \sum_{\ell=\lfloor \alpha^* k\rfloor}^k {k \choose \ell} \psi_{\ell}(n,\delta_2)+ P_0(\gamma)+2\delta_1$ by Theorem~\ref{scarletthm1}.

     \emph{Step 4: Combine and simplify.} The condition~\eqref{facto2000} is obtained from~\eqref{condA2} of Theorem~\ref{scarletthm1} and~\eqref{tfact1}. By the assumption $k=o(p)$ and~\eqref{beahat}, the numerator in~\eqref{condA2} of Theorem~\ref{scarletthm1} is dominated by ${p-k \choose \lfloor \alpha k\rfloor}$, which behaves as $\big(\alpha k \log \frac{p}{k}\big)(1+o(1))$. The factor $\gamma=o\big(k \log \frac{p}{k}\big)$ (see \eqref{beahat}) and the factor $\log(\frac{k^2}{\delta_1^2}{k\choose |s_{\rm{dif}}|}^2)$ in~\eqref{condA2} have been factored into $\eta$; note that the latter term behaves as $O(k)$ when $\delta_1 \to 0$ sufficiently slowly.
    
    The converse bound in~\eqref{facto2001} is obtained similarly by using~\eqref{condA4} of Theorem~\ref{scarletthm2} and~\eqref{tfact2}. Note that by~\cite[Proof of Cor. 2]{Scarlet2017a}, we have that $\log \big(\sum_{d=0}^{\lfloor \alpha^* k\rfloor}{p-k\choose d}{|s_{\rm{dif}}|\choose d}\big)$ simplifies to $\big(\alpha^* k \log \frac{p}{k}\big)(1+o(1))$.  Combining this fact with the assumption that $k=o(p)$, and observing that $|s_{\rm{dif}}|=\lfloor \alpha k \rfloor$ for some $\alpha \in [\alpha^*,1]$, the numerator in~\eqref{condA4} of Theorem~\ref{scarletthm2} simplifies to $(\alpha-\alpha^*) k \log (\frac{p}{k})(1+o(1))$, thus yielding \eqref{facto2001}.

\subsection{Proof of Proposition \ref{gaprof}} \label{proof:gaprof}

{\bf Overview.} We first outline the intuition behind the proof.  In \cite{Scarlet2017a}, the method for controlling $P_0(\gamma)$ was upper bounding $I(\beta_s;\bY|\bX_s)$ via the expansion $I(\beta_s;\bY|\bX_s) = h(\bY|\bX_s) - h(\bY|\bX_s,\beta_s)$.  Our analysis is instead based on the expansion $I(\beta_s;\bY|\bX_s) = h(\beta_s) - h(\beta_s|\bX_s,\bY)$ (note that $\beta_s$ is independent of $\bX_s$).  However, a difficulty with this expansion is in showing that $h(\beta_s|\bX_s,\bY)$ is not too negative, and we overcome this difficulty as follows:
\begin{itemize}
    \item Carefully choose a typical set in which the triplet $(\beta_s,\bX_s,\bY)$ lies with high probability;
    \item Show that a quantity similar to $h(\beta_s|\bX_s,\bY)$, but with conditioning on lying in the typical set, cannot be too negative by showing that given $(\bX_s,\bY)$, the most probable $\beta_s^*$ also has a surrounding region of $\beta_s$ vectors with a similar conditional density value.  This limits how high the conditional density of $\beta_s$ can be, and hence how negative the differential entropy can be.
\end{itemize}
We proceed in several steps.

{\bf Defining a typical set.} Let 
\begin{align}
    \label{defsetA}
    \calA:=\bigg\{(b_s,\bx_s,\by)\in \bbC^k \times \bbC^{kn} \times \bbR^n: \{\|\bx_s\|_{\rmF} \leq C\} \cap \{\|b_s\|_2 \leq C'\} \cap \{\|\bz_b\|_2 \leq C''\}\bigg\},
\end{align} 
with $C=\sqrt{2 k n}$, $C'= \sqrt{ k\sigma_{\beta}^2 \log n}$, and $C''= \sqrt{2 n \sigma^2}$, where $\|\bx_s\|_{\rmF}$ is the Frobenius norm, and
\begin{align}
    \bz_b:=\Big[y^{(1)}-|\langle x_s^{(1)},b_s\rangle|^2,y^{(2)}-|\langle x_s^{(2)},b_s\rangle|^2,\cdots,y^{(n)}-|\langle x_s^{(n)},b_s\rangle|^2\Big]^T. 
\end{align}
By the union bound, we have
\begin{align}
    \bbP[\calA]
    \ge 1-\bigg(\bbP\bigg[\frac{\|\bX_s\|_{\rmF}^2}{{kn}} > {2} \bigg]+\bbP\bigg[\frac{\|\beta_s\|^2_2}{{k\sigma_{\beta}^2} } > \log n \bigg] +\bbP\bigg[\frac{\|\bZ\|_2^2}{{n\sigma^2}} > {2} \bigg]\bigg).
\end{align}
Recall that $\bX_s$, $\beta_s$, and $\bZ$ are i.i.d.~Gaussian vectors with variances $1$, $\sigma_{\beta}^2$, and $\sigma^2$ respectively. Applying the weak law of large numbers to the first and third probabilities, and Markov's inequality to the middle one, we deduce that $\bbP[\calA] \to 1$ as $p \to \infty$ (with $k \to \infty$ and $n \to \infty$ simultaneously).


{\bf Useful properties within the typical set.} Fix $(b_s,\bx_s,\by) \in \calA$, as well as some $\tilde{b}_s \in \bbC^k$ satisfying
\begin{align}
    \|b_s-\tilde{b}_s\|_2 \leq \eps, \label{b_eps}
\end{align}
for some $\eps>0$ to be chosen later. From $\|b_s-\tilde{b}_s\|_2 \leq \eps$ and $\|b_s\|_2 \leq C'$, we have
\begin{align}
    \|\tilde{b}_s\|_2^2 &=\|b_s+(\tilde{b}_s-b_s)\|_2^2\\
    &\leq (\|b_s\|+\|\tilde{b}_s-b_s\|_2)^2\\
    &= \|b_s\|_2^2 + \|\tilde{b}_s-b_s\|_2^2 + 2\|b_s\| \cdot \|\tilde{b}_s-b_s\|\\
    \label{etric3}
    &\leq \|b_s\|_2^2+ \eps^2 + 2 C' \eps,
\end{align}
and hence
\begin{align}
    \label{etric5}
    \|\tilde{b}_s\|_2^2-\|b_s\|_2^2 \leq \eps^2 + 2 C' \eps.
\end{align}

On the other hand, we also have
\begin{align}
\|\tilde{b}_s\|_2 &\leq \|b_s\|_2 + \|\tilde{b}_s-b_s\|_2\\
&\leq C' +\eps
\end{align}
by the assumptions $\|b_s\|_2 \leq C'$ and $\|b_s-\tilde{b}_s\|_2 \leq \eps$.  It follows that
\begin{align}
\label{la0bug}
\max\{\|b_s\|_2,\|\tilde{b}_s\|_2\} \leq C'+\eps.
\end{align}

Now, we have for each $i\in \{1,2,\dotsc,n\}$ that
\begin{align}
    \label{norm1}
    |\langle x_s^{(i)}, b_s\rangle|^2&= |\langle x_s^{(i)}, \tilde{b}_s\rangle +\langle x_s^{(i)}, b_s-\tilde{b}_s\rangle  |^2 \\
    &\leq  \big(|\langle x_s^{(i)}, \tilde{b}_s\rangle| +|\langle x_s^{(i)}, b_s-\tilde{b}_s\rangle|\big)^2 \label{norm1a} \\
    &=  |\langle x_s^{(i)}, \tilde{b}_s\rangle|^2 +|\langle x_s^{(i)}, b_s-\tilde{b}_s\rangle  |^2+ 2 |\langle x_s^{(i)}, \tilde{b}_s\rangle| \cdot  |\langle x_s^{(i)}, b_s-\tilde{b}_s\rangle|  \label{normb}\\
    &\leq  |\langle x_s^{(i)}, \tilde{b}_s\rangle|^2 +\|x_s^{(i)}\|_2^2 \cdot  \|b_s-\tilde{b}_s\|_2^2+2\|x_s^{(i)}\|_2^2 \cdot \|\tilde{b}_s\| \cdot  \|b_s-\tilde{b}_s\| \label{normb2}\\
    \label{norm3}
    &\leq  |\langle x_s^{(i)}, \tilde{b}_s\rangle|^2 + \|x_s^{(i)}\|_2^2 \eps^2 + 2 \|x_s^{(i)}\|_2^2 (C'+\eps) \eps,
\end{align} 
where \eqref{norm1a} applies the triangle inequality, \eqref{normb2} is by Cauchy-Schwartz, and  \eqref{norm3} applies \eqref{b_eps} and~\eqref{la0bug}.

It follows from~\eqref{norm3} that
\begin{align}
    \label{tricky1a}
    |\langle x_s^{(i)}, b_s\rangle|^2-|\langle x_s^{(i)}, \tilde{b}_s\rangle|^2\leq  \|x_s^{(i)}\|_2^2 \eps^2 + 2 \|x_s^{(i)}\|_2^2 (C'+\eps) \eps,
\end{align}
and by interchanging the roles of $b_s$ and $\tilde{b}_s$ (and noting that~\eqref{la0bug} holds), we obtain
\begin{align}
    \label{kytric1b}
    \big||\langle x_s^{(i)}, b_s\rangle|^2-|\langle x_s^{(i)}, \tilde{b}_s\rangle|^2\big|\leq  \|x_s^{(i)}\|_2^2 \eps^2 + 2 \|x_s^{(i)}\|_2^2 (C'+\eps) \eps.
\end{align}
Summing over $i$, we obtain
\begin{align}
    \sum_{i=1}^n\Big| |\langle x_s^{(i)}, b_s\rangle|^2-|\langle x_s^{(i)}, \tilde{b}_s\rangle|^2\Big| 
    &\leq \sum_{i=1}^n \big( \|x_s^{(i)}\|_2^2 \eps^2 + 2 \|x_s^{(i)}\|_2^2 (C'+\eps) \eps \big)\\
    &= \|\bx_s\|_F^2 \eps^2+ 2 \|\bx_s\|_F^2 (C'+\eps) \eps\\
    \label{kytric1}
    &\leq  C^2 \eps^2 + 2 C^2 (C'+\eps) \eps 
\end{align}
by the condition $\|\bx_s\|_F \le C$ in $\calA$.

Similarly, from~\eqref{kytric1b}, we have
\begin{align}
    \label{kytric2b}
    \big|\langle x_s^{(i)}, b_s\rangle|^2-|\langle x_s^{(i)}, \tilde{b}_s\rangle|^2\big|^2 &\leq  \|x_s^{(i)}\|_2^4 (\eps^2 + 2(C'+\eps)\eps)^2,
\end{align}
and summing over $i$, we obtain
\begin{align}
    \sum_{i=1}^n\Big| |\langle x_s^{(i)}, b_s\rangle|^2-|\langle x_s^{(i)}, \tilde{b}_s\rangle|^2\Big|^2 
    &\leq \sum_{i=1}^n \|x_s^{(i)}\|_2^4 (\eps^2 + 2(C'+\eps)\eps)^2\\
    &\leq \bigg(\sum_{i=1}^n \|x_s^{(i)}\|_2^2\bigg)^2 (\eps^2 + 2(C'+\eps)\eps)^2 \\
		&= (\|\bx_s\|_F^2)^2 (\eps^2 + 2(C'+\eps)\eps)^2 \\
    \label{kytric20}
    &\leq  (C^2 \eps^2 + 2 C^2 (C'+\eps) \eps)^2.
\end{align}
We now use \eqref{kytric1} to bound a related term containing the observations: 
\begin{align}
    & \bigg|\sum_{i=1}^n(y^{(i)}-|\langle x_s^{(i)}, b_s\rangle|^2)^2- \sum_{i=1}^n(y^{(i)}-|\langle x_s^{(i)}, \tilde{b}_s\rangle|^2)^2\bigg|\nonumber\\
    &\quad=\bigg| \sum_{i=1}^n\Big(-|\langle x_s^{(i)}, b_s\rangle|^2+|\langle x_s^{(i)}, \tilde{b}_s\rangle|^2\Big) \Big(y^{(i)}-|\langle x_s^{(i)}, b_s\rangle|^2+ y^{(i)}-|\langle x_s^{(i)}, \tilde{b}_s\rangle|^2 \Big)\bigg| \label{beauty1}\\
	&\quad=\bigg| \sum_{i=1}^n\Big(-|\langle x_s^{(i)}, b_s\rangle|^2+|\langle x_s^{(i)}, \tilde{b}_s\rangle|^2\Big) \Big(2(y^{(i)}-|\langle x_s^{(i)}, b_s\rangle|^2)+ |\langle x_s^{(i)}, b_s\rangle|^2-|\langle x_s^{(i)}, \tilde{b}_s\rangle|^2 \Big)\bigg| \label{beauty1a}\\
    &\quad\leq \bigg| \sum_{i=1}^n\Big(-|\langle x_s^{(i)}, b_s\rangle|^2+|\langle x_s^{(i)}, \tilde{b}_s\rangle|^2\Big) \Big(2(y^{(i)}-|\langle x_s^{(i)}, b_s\rangle|^2)\Big)\bigg| + \bigg| \sum_{i=1}^n\Big(|\langle x_s^{(i)}, b_s\rangle|^2-|\langle x_s^{(i)}, \tilde{b}_s\rangle|^2\Big)^2 \bigg|	\label{beauty1b}\\
    &\quad=2\bigg|\sum_{i=1}^n\Big(-|\langle x_s^{(i)}, b_s\rangle|^2+|\langle x_s^{(i)}, \tilde{b}_s\rangle|^2\Big) z_b^{(i)}\bigg| + \bigg| \sum_{i=1}^n\Big(|\langle x_s^{(i)}, b_s\rangle|^2-|\langle x_s^{(i)}, \tilde{b}_s\rangle|^2\Big)^2 \bigg| \label{beauty2}\\
    &\quad \leq 2\sum_{i=1}^n\Big|-|\langle x_s^{(i)}, b_s\rangle|^2+|\langle x_s^{(i)}, \tilde{b}_s\rangle|^2\Big| \cdot \big|z_b^{(i)}\big| + \bigg| \sum_{i=1}^n\Big(|\langle x_s^{(i)}, b_s\rangle|^2-|\langle x_s^{(i)}, \tilde{b}_s\rangle|^2\Big)^2 \bigg| \\
    &\quad  \leq 2 \sum_{i=1}^n\Big|-|\langle x_s^{(i)}, b_s\rangle|^2+|\langle x_s^{(i)}, \tilde{b}_s\rangle|^2\Big| \cdot \|\bz_b\|_2  + \bigg| \sum_{i=1}^n\Big(|\langle x_s^{(i)}, b_s\rangle|^2-|\langle x_s^{(i)}, \tilde{b}_s\rangle|^2\Big)^2 \bigg| 		\label{beauty5} \\
    \label{eq205beauty}
    &\quad \leq 2 C'' \sum_{i=1}^n\Big|-|\langle x_s^{(i)}, b_s\rangle|^2+|\langle x_s^{(i)}, \tilde{b}_s\rangle|^2\Big|+\bigg| \sum_{i=1}^n\Big(|\langle x_s^{(i)}, b_s\rangle|^2-|\langle x_s^{(i)}, \tilde{b}_s\rangle|^2\Big)^2\bigg|\\
		   \label{eq206beauty}
    &\quad \leq 2 C''( C^2 \eps^2 + 2 C^2 (C'+\eps) \eps)+ ( C^2 \eps^2 + 2 C^2 (C'+\eps) \eps)^2,
\end{align} 
where \eqref{beauty1} uses $a^2 - b^2 = (a-b)(a+b)$, \eqref{beauty2} uses the definition of $\bz_b$ (whose $i$-th entry is denoted by $z_b^{(i)}$),  \eqref{beauty5} uses $|z_b^{(i)}| \le \|\bz_b\|_2$, \eqref{eq205beauty} follows since $\|\bz_b\|_2 \le C''$ within $\calA$, and~\eqref{eq206beauty} follows from~\eqref{kytric1} and~\eqref{kytric20}. 

{\bf Bounding a density ratio.} Let $\delta\{ \cdot \}$ be the Dirac delta function, and observe that
\begin{align}
    f_{\beta_s|\bX_s\bY}(b_s|\bx_s,\by)&=\int_{\bbR^n} f_{\beta_s \bZ|\bX_s\bY}(b_s,\bz|\bx_s,\by) \mu(d\bz)\\
    &=\frac{\int_{\bbR^n} f_{\beta_s\bZ\bY|\bX_s}(b_s,\bz,\by|\bx_s) \mu(d\bz)}{f_{\bY|\bX_s}(\by|\bx_s)}\\
    &=\frac{\int_{\bbR^n} f_{\beta_s}(b_s)\prod_{i=1}^n \big( f_Z(z^{(i)})\delta\{y^{(i)}=|\langle x_s^{(i)}, b_s\rangle|^2+z^{(i)}\} \big) \mu(d\bz)}{f_{\bY|\bX_s}(\by|\bx_s)}\\
    &=\frac{ f_{\beta_s}(b_s)\prod_{i=1}^n \int_{\bbR} f_Z(z^{(i)})\delta\{y^{(i)}=|\langle x_s^{(i)}, b_s\rangle|^2+z^{(i)}\} \mu(dz^{(i)})}{f_{\bY|\bX_s}(\by|\bx_s)}\\
    \label{etric1}
    &=\frac{f_{\beta_s}(b_s)\prod_{i=1}^n f_Z(y^{(i)}-|\langle x_s^{(i)}, b_s\rangle|^2)}{f_{\bY|\bX_s}(\by|\bx_s)}.
\end{align}
Recalling the distributions $Z \sim \calN(0,\sigma^2)$ and $\beta_s \sim \mathcal{CN}(0, \bI_k \sigma_{\beta}^2)$, it follows from \eqref{etric1} that
\begin{align}
    \frac{f_{\beta_s|\bX_s\bY}(b_s|\bx_s,\by)}{f_{\beta_s|\bX_s\bY}(\tilde{b}_s|\bx_s,\by)}&=\frac{f_{\beta_s}(b_s)\prod_{i=1}^n f_Z(y^{(i)}-|\langle x_s^{(i)}, b_s\rangle|^2)}{f_{\beta_s}(\tilde{b}_s)\prod_{i=1}^n f_Z(y^{(i)}-|\langle x_s^{(i)}, \tilde{b}_s\rangle|^2)}\\
		\label{tictic}
    &=\exp\bigg(\frac{-\|b_s\|_2^2+\|\tilde{b}_s\|_2^2}{\sigma_{\beta}^2}\bigg)\exp\bigg(\frac{-\sum_{i=1}^n (y^{(i)}-|\langle x_s^{(i)}, b_s\rangle|^2)^2+\sum_{i=1}^n (y^{(i)}-|\langle x_s^{(i)}, \tilde{b}_s\rangle|^2)^2}{2\sigma^2}\bigg)\\
    \label{fitrick}
    &\leq \exp\bigg(\frac{\eps^2 + 2 C' \eps}{\sigma_{\beta}^2}\bigg)\exp\bigg(\frac{2 C''( C^2 \eps^2 + 2 C^2 (C'+\eps) \eps)+ ( C^2 \eps^2 + 2 C^2 (C'+\eps) \eps)^2}{2\sigma^2}\bigg)
\end{align}
where \eqref{fitrick} uses \eqref{etric5} and \eqref{eq206beauty}.  Now, since $C=\sqrt{2kn}$, $C'=\sqrt{k \sigma_{\beta}^2 \log n}$, $C''=\sqrt{2\sigma^2 n}$, and $\sigma_{\beta}^2=\Theta(\frac{1}{k})$, we see from~\eqref{fitrick} that if we choose
\begin{align}
\label{choiceeps}
\eps=\frac{1}{ n^2 k^2\log n},
\end{align} 
then we obtain
\begin{align}
    \label{beaumod}
    \limsup_{k \to \infty} \frac{f_{\beta_s|\bX_s\bY}(b_s|\bx_s,\by)}{f_{\beta_s|\bX_s\bY}(\tilde{b}_s|\bx_s,\by)}\leq 1,
\end{align} 
whenever $(b_s,\bx_s,\by) \in \calA$ and $\|\tilde{b}_s-b_s\|_2 \leq \eps$.

{\bf Bounding an average log-density.} Let $(b_s^*,\bx_s^*,\by^*)$ be an arbitrary point in $\calA$, and define
\begin{align}
    \tilf_{\rm{\min}}(\calA)=\min_{\tilde{b}_s \,:\, \|\tilde{b}_s-b_s^*\|_2 \leq \eps}f_{\beta_s|\bX_s\bY}(\tilde{b}_s|\bx_s^*,\by^*).
\end{align} 
From~\eqref{beaumod}, we have $f_{\beta_s|\bX_s\bY}(\tilde{b}_s|\bx_s^*,\by^*) \geq f_{\beta_s|\bX_s\bY}(b_s^*|\bx_s^*,\by^*)(1+o(1))$ whenever $\|\tilde{b}_s-b_s^*\|_2 \leq \eps$. Hence, we have
\begin{align}
    \label{eq242note2a}
    \tilf_{\rm{\min}}(\calA)\geq f_{\beta_s|\bX_s\bY}(b_s^*|\bx_s^*,\by^*)(1+o(1)),
\end{align} 
where $o(1)$ is vanishing as $k\to \infty$. On the other hand, we trivially have
$\tilf_{\rm{\min}}(\calA)\leq f_{\beta_s|\bX_s\bY}(b_s^*|\bx_s^*,\by^*)$, and hence
\begin{align}
\label{eq242note2}
    \tilf_{\rm{\min}}(\calA)= f_{\beta_s|\bX_s\bY}(b_s^*|\bx_s^*,\by^*)(1+o(1)).
\end{align}
Now, defining the ball $B_{\eps}(b_s^*):=\{\tilde{b}_s \in \bbC^k: \|\tilde{b}_s-b_s^*\|_2 \leq \eps\}$, we have
\begin{align}
    1 &\geq \bbP[\beta_s \in B_{\eps}(b_s^*)|\bX_s=\bx_s^*,\bY=\by^*]\\
    &\geq  \mbox{vol}(B_{\eps}(b_s^*))\tilf_{\rm{\min}}(\calA)\\
    &=\frac{\pi^k}{k!}\eps^{2k} \tilf_{\rm{\min}}(\calA), 
\end{align}
where $\frac{\pi^k}{k!}\eps^{2k}$ is the  volume of the ball $B_{\eps}(b^*)$~\cite{Abr65}. Therefore, we have
\begin{align}
    \tilf_{\rm{\min}}(\calA)&\leq \frac{k!}{\pi^k} \eps^{-{2k}}\\
    \label{daughter1}
    &=\frac{k!}{\pi^k} \big(n^2 k^2\log n\big)^{2k}
\end{align}
by \eqref{choiceeps}.  Combining \eqref{eq242note2} and~\eqref{daughter1} gives
\begin{align}
    f_{\beta_s|\bX_s\bY}(b_s^*|\bx_s^*,\by^*)
        \leq \frac{k!}{\pi^k} \big(n^2 k^2\log n\big)^{2k} (1+o(1)).  \label{eq248note1}
\end{align}
Since $(b_s^*,\bx_s^*,\by^*)$ can be arbitrarily chosen within $\calA$, we rename it to $(b_s,\bx_s,\by) \in \calA$, and take the logarithm to deduce that
\begin{align}
 \log f_{\beta_s|\bX_s\bY}(b_s|\bx_s,\by) &\leq \log\bigg( \frac{k!}{\pi^k}  \big(n^2 k^2\log n\big)^{2k} (1+o(1))\bigg)\\
&=\log k! -k \log \pi + 2k \log (n^2 k^2\log n) + o(1)\\
    \label{eq252newnew}
    &=\Theta(k \log n),
\end{align} 
by the assumption $n=\Omega(k)$. 

{\bf Bounding a mutual information-like term.} The mutual information is the  average of a log-density ratio, and that ratio may be positive or negative in general.  We will find it more convenient to apply the function $[\cdot]^+$ to the log-density ratio, and proceed as follows:
\begin{align}
    \label{defineI0+}
    I_0^+&:=\bbE\bigg[\bigg[\log\frac{f_{\bY|\bX_s \beta_s}(\bY|\bX_s,\beta)}{f_{\bY|\bX_s}(\bY|\bX_s)}\bigg]^+\,\bigg|\,(\beta_s,\bX_s,\bY) \in \calA \bigg]\\
    &=\bbE\bigg[\bigg[\log\frac{f_{\beta_s|\bX_s\bY}(\beta_s|\bX_s,\bY)}{f_{\beta_s}(\beta_s)}\bigg]^+\,\bigg|\,(\beta_s,\bX_s,\bY) \in \calA\bigg] \label{Bayes} \\
    &=\bbE\bigg[\big[-\log f_{\beta_s}(\beta_s)+\log f_{\beta_s|\bX_s\bY}(\beta_s|\bX_s,\bY)\big]^+\,\bigg|\,(\beta_s,\bX_s,\bY) \in \calA\bigg]\\
		\label{noting99}
    &=\bbE\bigg[\Big[k \log (\pi \sigma_{\beta}^2)+\frac{\|\beta_s\|_2^2}{\sigma_{\beta}^2}+\log f_{\beta_s|\bX_s\bY}(\beta_s|\bX_s,\bY)\Big]^+\,\bigg|\,(\beta_s,\bX_s,\bY) \in \calA\bigg]\\
    \label{eqnote}
    &\leq \bbE\bigg[\frac{\|\beta_s\|_2^2}{\sigma_{\beta}^2}+ \big[k \log (\pi \sigma_{\beta}^2)+\log f_{\beta_s|\bX_s\bY}(\beta_s|\bX_s,\bY)\big]^+ \,\Big|\, (\beta_s,\bX_s,\bY) \in \calA\bigg]\\
		\label{eq124newnew}
    &\leq \frac{k \sigma_{\beta}^2}{\sigma_{\beta}^2 \bbP[(\beta_s,\bX_s,\bY) \in \calA]}+ \bbE\Big[\big[k \log (\pi \sigma_{\beta}^2)+\log f_{\beta_s|\bX_s\bY}(\beta_s|\bX_s,\bY) \big) \big]^+ \,\Big|\, (\beta_s,\bX_s,\bY) \in \calA\Big]\\
    \label{eqnote10}
    &=\frac{k}{\bbP[(\beta_s,\bX_s,\bY) \in \calA]} +O(k\log k) + O(k\log n) \\
     \label{eq260new}
    &=O(k\log n),
\end{align} 
where \eqref{Bayes} follows from Bayes' rule,~\eqref{noting99} follows from the fact that $f_{\beta_s}(b_s)= \frac{1}{(\pi \sigma_{\beta}^2)^k}\exp\big(-\frac{\|b_s\|_2^2}{\sigma_{\beta}^2}\big)$ for all $b_s\in \bbC^k$, \eqref{eqnote} applies $[a+b]^+ \le a + [b]^+$ for $a\geq 0$,~\eqref{eq124newnew} uses~$\bbE[\|\beta_s\|_2^2] \geq \bbP[(\beta_s,\bX_s,\bY) \in \calA]\cdot \bbE[\|\beta_s\|_2^2 \,|\, (\beta_s,\bX_s,\bY) \in \calA]$, \eqref{eqnote10} follows from~\eqref{eq252newnew} and the assumption $\sigma_{\beta}^2 = \Theta\big(\frac{1}{k}\big)$, and \eqref{eq260new} uses $\bbP[\calA] \to 1$ and the assumption $n=\Omega(k)$.

{\bf Wrapping up.} It follows from~\eqref{eq260new} and Markov's inequality that for any $\gamma>0$,
\begin{align}
    \bbP\bigg[\log\frac{f_{\bY|\bX_s \beta_s}(\bY|\bX_s,\beta_s)}{f_{\bY|\bX_s}(\bY|\bX_s)}>\gamma \,\bigg|\, (\beta_s,\bX_s,\bY) \in \calA\bigg]&\leq \frac{I_0^+}{\gamma}\\
    &=\frac{O(k \log n)}{\gamma}.
\end{align}
Hence, we have
\begin{align}
    P_0(\gamma)
    &=\bbP\bigg[\log\frac{f_{\bY|\bX_s \beta_s}(\bY|\bX_s,\beta_s)}{f_{\bY|\bX_s}(\bY|\bX_s)}>\gamma\bigg]\\
    &=\bbP\bigg[\log\frac{f_{\bY|\bX_s \beta_s}(\bY|\bX_s,\beta_s)}{f_{\bY|\bX_s}(\bY|\bX_s)}>\gamma \,\bigg|\,(\beta_s,\bX_s,\bY) \in \calA\bigg]\bbP\big[(\beta_s,\bX_s,\bY) \in \calA\big]\nonumber\\
    &\quad+ \bbP\bigg[\log\frac{f_{\bY|\bX_s \beta_s}(\bY|\bX_s,\beta_s)}{f_{\bY|\bX_s}(\bY|\bX_s)}>\gamma \,\bigg|\,(\beta_s,\bX_s,\bY) \notin \calA\bigg]\bbP\big[(\beta_s,\bX_s,\bY) \notin \calA\big]\\
    &\leq \frac{O(k \log n)}{\gamma} +\bbP\big[(\beta_s,\bX_s,\bY) \notin \calA\big]\\
    &=\frac{O(k \log n)}{\gamma}+o(1).
\end{align}
This concludes the proof of Proposition~\ref{gaprof}.

%
%
\section{Conclusion}\label{sec:conclu}

We have characterized the information-theoretic limits of approximate support recovery in the complex phase retrieval model with Gaussian measurements, under both discrete and Gaussian distributions on the unknown non-zero entries.  Along the way, we established novel concentration bounds for conditional information random variables, which may be of independent interest.  Our achievability and converse bounds have matching scaling laws, as well as near-matching constant factors as the SNR increases.
There are numerous potential directions for further work, including (i) handling the exact recovery criterion, (ii) improving our results in the low-SNR regime via tighter mutual information bounds, (iii) extending our achievability results to general scalings $k = o(p)$, (iv) handling the linear sparsity regime $k = \Theta(p)$ without any additional assumptions, (v) performing analogous studies for non-Gaussian measurement matrices, such as Fourier measurements, and (vi) seeking computationally efficient algorithms whose support recovery performance comes close to the fundamental limits.


\appendices 

%
%
\section{Signal-to-Noise Ratio (SNR) Calculations} \label{app:SNR}

{\bf Gaussian $\beta_s$.} For the real Gaussian linear model in~\cite[Corr.~2]{Scarlet2017a}, we have i.i.d.~$\calN(0,1)$ measurements, $\calN\big(0,\frac{c_{\beta}}{\sigma^2}\big)$ entries of $\beta_s$, and $\calN(0,\sigma^2)$ noise, leading to an SNR of $\frac{c_{\beta}}{\sigma^2}$.

The complex Gaussian phase retrieval setting in Section~\ref{model}  with $\mathcal{CN}\big(0,\frac{c_{\beta}}{\sigma^2})$ entries of $\beta_s$ is slightly more complicated. Noting that a standard $\chi_2^2$ random variable has mean $2$, variance $4$, and second moment $8$, we find that the expected SNR for sending a support vector $s \in \calS$ is 
\begin{align}
    \mbox{SNR}&=\frac{\bbE[|\langle X_s,\beta_s\rangle|^4]}{\sigma^2}\\
    &=\frac{\bbE[\bbE[|\langle X_s,\beta_s\rangle|^4]|X_s]}{\sigma^2}\\
    \label{eqnoteh0}
    &=\frac{2\sigma_{\beta}^4 \bbE[\|X_s\|_2^4]}{\sigma^2}\\
    \label{eqnoteg}
    &=\frac{ 2 \sigma_{\beta}^4 k(k+1)}{\sigma^2} \\
    \label{eqnoteg2}
    &=\frac{ 2 c_{\beta}^2(1+\frac{1}{k})}{\sigma^2},
\end{align} 
where~\eqref{eqnoteh0} follows from the fact that given $X_s$, $\langle X_s,\beta_s\rangle \sim \mathcal{CN}(0, \sigma_{\beta}^2 \|X_s\|_2^2)$ so $\frac{2}{\sigma_{\beta}^2 \|X_s\|_2^2} |\langle X_s,\beta_s\rangle|^2$ has a $\chi_2^2$ distribution,~\eqref{eqnoteg} follows from the fact that $2\|X_s\|_2^2$ has a $\chi_{2k}^2$ distribution so $\bbE[(2\|X_s\|_2^2)^2]=\big(\bbE[2\|X_s\|_2^2]\big)^2+ \var[2\|X_s\|_2^2]=(2k)^2+ 4k=4k(k+1)$, and \eqref{eqnoteg2} uses $\sigma_{\beta}^2 = \frac{c_{\beta}}{k}$. Since we only consider scaling regimes where $k \to \infty$, the term $\frac{1}{k}$ is negligible.

{\bf Discrete $\beta_s$.} For the real discrete linear model in~\cite[Sec. IV-A]{Scarlet2017a}, we have i.i.d.~$\calN(0,1)$ measurements, a $k$-sparse random vector $\beta_s$ which is a uniformly random permutation of $b_s=(b_1,b_2,\cdots,b_k)$, and $\calN(0,\sigma^2)$ noise, leading to an SNR of $\frac{\|b_s\|_2^2}{\sigma^2}$. In particular, when $|b_1|=|b_2|=\cdots=|b_k|=\sqrt{\frac{c_{\beta}}{k}}$, the SNR is equal to $\frac{c_{\beta}}{\sigma^2}$. 

For the complex discrete phase retrieval setting in Section~\ref{model} with $\beta_s$ being a uniformly random permutation of $b_s=(b_1,b_2,\cdots,b_k)$ and with $\mathcal{CN}(0,\sigma^2)$ noise, we can use similar arguments as in the Gaussian case to show that
\begin{align}
    \mbox{SNR}=\frac{2\|b_s\|^4}{\sigma^2}.
\end{align}
In particular, for the case $|b_1|=|b_2|=\cdots=|b_k|=\sqrt{\frac{c_{\beta}}{k}}$, we have
\begin{align}
\mbox{SNR}=\frac{2c_{\beta}^2}{\sigma^2}.
\end{align}
In addition, since the ``sorted'' vector $b'_s$ satisfies $\sum_{i=1}^{\lfloor \alpha k \rfloor} |b_i'|^2 =\frac{\lfloor \alpha k \rfloor}{k} c_{\beta}\to \alpha c_{\beta}$ (as $k \to \infty$) and similarly $\sum_{i=\lfloor \alpha k \rfloor+1}^{k} |b_i'|^2 \to  (1-\alpha) c_{\beta}$,  the mutual information terms \eqref{defI1} and~\eqref{defI2} simplify to
\begin{gather}
\label{tolim1}
I_1(\alpha,k)\to \frac{1}{2} \log\bigg[\frac{4}{\exp(2h(Z))}(\alpha c_{\beta})^2+1\bigg],\\
\label{tolim2}
I_2(\alpha,k)\to \frac{1}{2} \log\bigg[\frac{2\pi e}{\exp(2h(Z))}(\alpha c_{\beta})^2+1\bigg]+\frac{1}{2}\log \bigg[1+\frac{\alpha (1-\alpha) c_{\beta}^2 }{(\alpha c_{\beta})^2 +\frac{\exp(2h(Z))}{2\pi e}}\bigg]+\frac{1}{2}\log\bigg(\frac{\pi e}{2}\bigg).
\end{gather}
These simplifications readily permit the numerical evaluation of \eqref{eq3main}--\eqref{eq4main} in Theorem~\ref{cormain1} as $k \to \infty$.

{\bf Matching the linear and phase retrieval models.} In light of the above calculations, in Figure~\ref{fig:EvalPic0} and Figure \ref{fig:EvalPic1}, we match the SNR of the two models (real linear and complex phase retrieval) by taking $c_{\beta}$ from the phase retrieval model and squaring it and then multiplying it by $2$ to get the value for the linear model. 

%
%
\section{Proof of Theorem \ref{mutulthm} (Mutual Information Bounds)} \label{app:mutualinfo}

First, for a fixed partition $(s_{\rm eq},s_{\rm dif})$ of the support set $s$, we rewrite the acquisition model in~\eqref{acqui} as
\begin{align}
    Y=| \langle X_{s_{\rm{eq}}},\beta_s \rangle+ \langle X_{s_{\rm{dif}}},\beta_s \rangle|^2 +Z.
\end{align}
Conditioned on $\beta_s=b_s$, this gives
\begin{align}
    \label{keymodel}
    Y=|\sqrt{v_{\rm{eq}}}W_{\rm{eq}}+\sqrt{v_{\rm{dif}}}W_{\rm{dif}}|^2+Z,
\end{align}
where $v_{\rm{eq}}=\sum_{i \in s_{\rm{eq}}} |b_i|^2$, $v_{\rm{dif}}=\sum_{i \in s_{\rm{dif}}}|b_i|^2$, and $W_{\rm{eq}}, W_{\rm{dif}}$ are independent $\mathcal{CN}(0,1)$ random variables (recall that $X_s$ has i.i.d.~$\mathcal{CN}(0,1)$ entries).


Next, given $\beta_s=b_s$ and $W_{s_{\rm{eq}}}=w_{s_{\rm{eq}}}$, we write $Y=U_{w_{\rm{eq}}}+Z$, where
\begin{equation}    
    U_{w_{\rm{eq}}}=|\sqrt{v_{\rm{eq}}}w_{\rm{eq}}+\sqrt{v_{\rm{dif}}}W_{\rm{dif}}|^2 \label{eq:Uweq}
\end{equation}
follows a non-central $\chi^2$ distribution with two degrees of freedom, which is log-concave~\cite{Yu2011}.  
Observe that
\begin{align}
    I_{s_{\rm{dif}},s_{\rm{eq}}}(b_s) &=I(X_{s_{\rm{dif}}};Y|X_{s_{\rm{eq}}},\beta_s=b_s)\\
    &=h(Y|X_{s_{\rm{eq}}},\beta_s=b_s)-h(Z)\\
    &=\bbE_{\tilde{X}_{s_{\rm{eq}}}}[h(Y|X_{s_{\rm{eq}}}=\tilde{X}_{s_{\rm{eq}}}, \beta_s=b_s)]-h(Z)\\
    \label{funy7}
    &=\bbE_{W_{\rm{eq}}}[h(U_{W_{\rm{eq}}}+Z)]-h(Z), 
    \end{align}
    where $\tilde{X}_{s_{\rm{eq}}} \sim f_X^{|s_{\rm{eq}}|}$.
    The entropy of $U_{w_{\rm{eq}}}+Z$ can be lower bounded using the entropy power inequality as $\exp(2 h(U_{w_{\rm{eq}}}+Z))\geq \exp(2 h(U_{w_{\rm{eq}}}))+\exp(2 h(Z))$~\cite{elgamal}, or equivalently
    \begin{align}
    \label{powerin}
    h(U_{w_{\rm{eq}}}+Z) \geq \frac{1}{2}\log\big(\exp(2h(U_{w_{\rm{eq}}}))+\exp(2h(Z))\big).
\end{align}
To find an upper bound on the entropy of $U_{w_{\rm{eq}}}+Z$, we use the reverse entropy power inequality~\cite[Theorem.~7]{Marsiglietti2018} for two uncorrelated log-concave random variables $U_{w_{\rm{eq}}}$ and $Z$ to obtain $\exp(2h(U_{w_{\rm{eq}}}+Z)) \leq \frac{\pi e}{2}\big(\exp(2h(U_{w_{\rm{eq}}}))+\exp(2h(Z))\big)$, 
or equivalently,
\begin{align}
    \label{repowerin}
    h(U_{w_{\rm{eq}}}+Z)\leq \frac{1}{2}\log \Big(\frac{\pi e}{2}\Big)+\frac{1}{2}\log\big(\exp(2h(U_{w_{\rm{eq}}}))+\exp(2h(Z))\big).
\end{align}

We now consider upper and lower bounding the entropy of $U_{w_{\rm{eq}}}$.  For the upper bound, we simply use that the Gaussian distribution maximizes entropy for a given variance:
\begin{align}
    \label{10eq}
    h(U_{w_{\rm{eq}}})\leq \frac{1}{2}\log (2\pi e \var(U_{w_{\rm{eq}}})).
\end{align}
Moreover, the result of \cite[Theorem~3]{Marsiglietti2018} states that this upper bound is nearly tight for log-concave random variables:
\begin{align}
    \label{9eq}
    h(U_{w_{\rm{eq}}})\geq \frac{1}{2}\log (4 \var(U_{w_{\rm{eq}}})).
\end{align}
Indeed, $U_{w_{\rm{eq}}}=v_{\rm{dif}}\big|\mathcal{CN}\big(\sqrt{\frac{v_{\rm{eq}}}{v_{\rm{dif}}}}w_{\rm{eq}},1\big)\big|^2$ ({\em cf.}, \eqref{eq:Uweq}) has a non-central $\chi^2$ distribution with two degrees of freedom, which is log-concave \cite{Yu2011}.  In addition, the variance is given by~\cite[p.~45]{Proakis} 
\begin{align}
    \label{11eq}
    \var(U_{w_{\rm{eq}}})=v_{\rm{dif}}^2 \bigg(1+ \frac{v_{\rm{eq}}}{v_{\rm{dif}}}|w_{\rm{eq}}|^2\bigg).
\end{align} 
Hence, from~\eqref{10eq} and~\eqref{11eq}, we obtain
\begin{align}
    h(U_{w_{\rm{eq}}}) &\leq \frac{1}{2}\log\Bigg[ 2\pi e\, v_{\rm{dif}}^2 \bigg(1+ \frac{v_{\rm{eq}}}{v_{\rm{dif}}}|w_{\rm{eq}}|^2\bigg)  \Bigg], \label{eq18note1}
\end{align}
and from from~\eqref{9eq}, we obtain
\begin{align}
    \label{eq19note1}
    h(U_{w_{\rm{eq}}})&\geq \frac{1}{2}\log \bigg[ 4 v_{\rm{dif}}^2 \bigg(1+ \frac{v_{\rm{eq}}}{v_{\rm{dif}}}|w_{\rm{eq}}|^2\bigg)  \Bigg].
\end{align}
It follows from~\eqref{repowerin} and~\eqref{eq18note1} that
\begin{align}
    h(U_{w_{\rm{eq}}}+Z) &\leq \frac{1}{2}\log \bigg(\frac{\pi e}{2}\bigg)+\frac{1}{2}\log\bigg(2\pi e \big(v_{\rm{dif}}^2 + v_{\rm{dif}}v_{\rm{eq}}|w_{\rm{eq}}|^2\big)+\exp(2h(Z))\bigg)\\
    \label{eq19ty1}
    &=\frac{1}{2}\log \bigg(\frac{\pi e}{2}\bigg)+h(Z)+\frac{1}{2}\log\bigg(\frac{2\pi e}{\exp(2h(Z))} \big(v_{\rm{dif}}^2 + v_{\rm{dif}} v_{\rm{eq}}|w_{\rm{eq}}|^2\big)+1\bigg)\\
    \label{tem143}
    &= \frac{1}{2}\log \bigg(\frac{\pi e}{2}\bigg)+h(Z)+\frac{1}{2}\log\bigg[\bigg(\frac{2\pi e}{\exp(2h(Z))}\bigg)v_{\rm{dif}}^2+1\bigg]+\frac{1}{2}\log\bigg(1+\frac{v_{\rm{dif}}v_{\rm{eq}}}{v_{\rm{dif}}^2+\frac{\exp(2h(Z))}{2\pi e}}|w_{\rm{eq}}|^2\bigg),
\end{align} 
where the two equalities are simple algebraic manipulations.  Similarly, it follows from~\eqref{powerin} and~\eqref{eq19note1} that
\begin{align}
    h(U_{w_{\rm{eq}}}+Z)&\geq \frac{1}{2}\log\Big(4 \big(v_{\rm{dif}}^2 + v_{\rm{dif}}v_{\rm{eq}}|w_{\rm{eq}}|^2\big)+\exp(2h(Z))\Big)\\
    \label{terma150}
    &\geq \frac{1}{2}\log\big(4 v_{\rm{dif}}^2 +\exp(2h(Z))\big)\\
    \label{eq149terma}
    &=\frac{1}{2}\log \bigg[\bigg(\frac{4}{\exp(2h(Z))}\bigg)v_{\rm{dif}}^2+1\bigg]+h(Z).
\end{align}
Returning to \eqref{funy7}, we have 
\begin{align}
    I_{s_{\rm{dif}},s_{\rm{eq}}}(b) 
    &=\bbE_{W_{\rm{eq}}}[h(U_{W_{\rm{eq}}}+Z)]-h(Z)\\
    \label{taug1a}
    &\leq \frac{1}{2}\log \bigg(\frac{\pi e}{2}\bigg)+\frac{1}{2}\log\bigg[\bigg(\frac{2\pi e}{\exp(2h(Z))}\bigg)v_{\rm{dif}}^2+1\bigg]+\frac{1}{2}\log\bigg(1+\frac{v_{\rm{dif}}v_{\rm{eq}}}{v_{\rm{dif}}^2+\frac{\exp(2h(Z))}{2\pi e}}\bbE[|W_{\rm{eq}}|^2]\bigg)\\
    \label{taug1}
    &=\frac{1}{2}\log \bigg(\frac{\pi e}{2}\bigg)+\frac{1}{2}\log\bigg[\bigg(\frac{2\pi e}{\exp(2h(Z))}\bigg)v_{\rm{dif}}^2+1\bigg]+\frac{1}{2}\bbE\bigg[\log\bigg(1+\frac{v_{\rm{dif}}v_{\rm{eq}}}{v_{\rm{dif}}^2+\frac{\exp(2h(Z))}{2\pi e}}\bigg)\bigg],
\end{align} 
where~\eqref{taug1a} follows from \eqref{tem143} and the concavity of the function $\log(1+x)$ for $x>-1$,~\eqref{taug1} follows from the fact that $W_{\rm{eq}}\sim \mathcal{CN}(0,1)$.

Finally, from~\eqref{funy7} and~\eqref{eq149terma}, we have
\begin{align}
    I_{s_{\rm{dif}},s_{\rm{eq}}}(b_s) \geq \frac{1}{2}\log \bigg[\bigg(\frac{4}{\exp(2h(Z))}\bigg)v_{\rm{dif}}^2+1\bigg], \label{taug2}
\end{align}
and \eqref{boundmut} follows from~\eqref{taug1}--\eqref{taug2}. 

%
%
\section{Proof of Proposition \ref{keyext} and Corollary \ref{verykey} (General Concentration of Conditional Information)} \label{app:conc_general}

Before proceeding, we briefly explain the notation used throughout this appendix.  The first two lemmas below concern generic vectors $\bx \in \bbR^n$, and the remainder of the appendix concerns joint density functions on $(X,Y)$ with $X \in \bbR^{2k}$ and $Y \in \bbR$, and more generally on $(\bX,\bY)$ with $\bX \in \bbR^{2kn}$ and $\bY \in \bbR^n$.  Initially, this should be viewed as generic notation; in Appendix \ref{app:conc_phase}, we will specialize to the phase retrieval setting by interpreting complex vectors in $\bbC^k$ as equivalently being in $\bbR^{2k}$.

\subsection{Technical Analysis}

The following lemma gives a sufficient condition for interchanging certain derivatives and integrals, and perhaps more importantly, establishes bounds on certain first and second derivatives that will eventually be used to bound the key quantity $K_1$ in Proposition \ref{keyext}.  Here and subsequently, $L^1(\bbR^n)$ denotes the set of absolutely integrable functions on $\bbR^n$.

\begin{lemma} \label{bestcup} 
    Fix $n \in \bbZ^+$, and let $g:\bbR^n \times \bbC \to \bbC$. Assume that $g(\bx,u)$ is a real entire function\footnote{A {\em real entire function} is a function on $\bbC$ which is analytic (complex differentiable or holomorphic) on the entire complex plane and assumes real values on the real axis. For our purposes, it suffices to understand that the exponential function $g(t) = e^{ct}$ falls in this class, and that any function in this class restricted to the real line is always equal to its infinite Taylor expansion \cite[Sec. 2.3]{Steinbook}. \label{foot:entire} } 
in $u$ for each fixed $\bx \in \bbR^n$ such that $g(\bx,u) \geq 0$ for all $(\bx,u) \in \bbR^n \times \bbR^+$. In addition, assume that either $(-1)^l \frac{\partial^l g}{\partial u^l}(\bx,t) \geq 0$ for all pairs $(l,t) \in \bbN \times \bbR^+$ or 
    $\frac{\partial^l g}{\partial u^l}(\bx,t) \geq 0$ for all pairs $(l,t) \in \bbN \times\bbR^+$.  For $t \in \bbR^+$, define
    \begin{align}
    \label{mor1}
    T(t):=\int_{\bbR^n} g(\bx,t)\mu(d\bx).
    \end{align}
    
    (i) If $g(\bx,u) \in L^1(\bbR^n) $ for all $u\in \bbR^+$, we have that $T(t)$ is twice differentiable and that
    \begin{align}
        \frac{d T^l(t)}{d t^l}&=\int_{\bbR^n} \frac{\partial g^l}{\partial u^l}(\bx,t) \mu(d\bx), \label{eq:T_deriv}
    \end{align} 
    for $l\in \{1,2\}$. \\
    
    (ii) Let $\calG$ be a subset of $(0,\infty)$.  Under the condition
    \begin{align}
        \label{condequip1}
     \sup_{u\in \calG} T(u)\leq T^*
    \end{align} 
    for some constant $T^*$, we have
    \begin{align}
        \label{crazy1}
       T''(t) \leq  \frac{75 T^*}{t^2}
    \end{align} 
    for any $t \in \calG^o$, where $\calG^o$ is an interior of the set $\calG$.
		
    (iii) Let $\calG$ be a subset of $(0,\infty)$.  Under the condition
    \begin{align}
        \label{condequip11}
     \sup_{u\in \calG} u T(u)\leq T^{\dagger}
    \end{align} 
    for some constant $T^{\dagger}$, we have
    \begin{align}
        \label{crazy11}
       (t T(t))'' \leq \frac{150 T^{\dagger}}{t^2}
    \end{align} 
    for any $t \in \calG^o$.
    \end{lemma}
\begin{IEEEproof}
    See Appendix~\ref{proof:bestcup}.
\end{IEEEproof}


Fradelizi,~Madiman, and~Wang~\cite{Fradelizi2016} state that we can exchange analogous integrals and derivatives if the function under the integral is in $L^1$, but we are not aware of a proof. They also noted that $f_{\bX}^t(\cdot)$ satisfies this property for any $t>0$ when $f_{\bX}(\cdot)$ is log-concave.  However, we cannot use such results directly, because we will be considering joint distributions that fail to be {\em jointly} log-concave.

The following lemma formally states that the integral of any power of a log-concave random vector is in $L^1$,
and provides an explicit upper bound on such an integral (to be used in Corollary~\ref{immedi} below).

\begin{lemma} \label{logconcave} 
    Fix $n\in \bbZ^+$, and let $f: \bbR^n\to \bbR^+$ be a log-concave function such that $\|f\|_1 <\infty$ and $\|f\|_{\infty}<\infty$.\footnote{Here $\|f\|_1$ denotes the integral of the absolute value, and $\|f\|_{\infty}$ denotes the maximum absolute value. } Then, for all $t>0$, the following holds:
    \begin{align}
        \label{eq121neweq}
        \int_{\bbR^n} f^t(\bx) \mu(d\bx) \leq D \frac{ (\|f\|_{\infty}+1)^t}{t^n},
    \end{align}  
    where $D$ is finite and is defined as
		\begin{align}
		\label{defD}
		D:=\sup_{t>0} (\|f\|_{\infty}+1)^{-t}\int_{\bbR^n} t^n f^t(\bx) \mu(d\bx).
		\end{align}
\end{lemma}
\begin{IEEEproof}
    Observe that
    \begin{align}
        \int_{\bbR^n} t^n f^t(\bx) \mu(d\bx)&=\int_{\bbR^n} t^n \exp(t \log f(\bx)) \mu(d\bx)\\
        \label{eqnewy}
        &= \int_{\bbR^n} \exp(t \log f(\bz/t)) \mu(d\bz),
    \end{align} 
    where~\eqref{eqnewy} follows from a change of variable $\bz=t \bx$. 
    Noting that $t \log f(\bz/t)$ is jointly concave as a function of $(\bz,t) \in \bbR^n \times (0,\infty)$ \cite[Lemma 2.8]{Fradelizi2016}, we find that $\int_{\bbR^n} \exp(t \log f(\bz/t)) \mu(d\bz)$ is log-concave in $t$ by Pr\'{e}kepa's theorem~\cite{Prekopa1973}, which states that the marginal function of a jointly log-concave function is log-concave. Since the product of two log-concave functions is log-concave, we deduce from \eqref{eqnewy} that the function $(\|f\|_{\infty}+1)^{-t}\int_{\bbR^n} t^n f^t(\bx) \mu(d\bx)$ is also log-concave in $t$. 
    
    To establish that the supremum over $t > 0$ in \eqref{defD} is bounded, we will combine the log-concavity property with the limiting behavior as $t \to \infty$.  We write 
    \begin{align}
        (\|f\|_{\infty}+1)^{-t}\int_{\bbR^n} t^n f^t(\bx) \mu(d\bx)&=\int_{\bbR^n} t^n \bigg(\frac{f(\bx)}{\|f\|_{\infty}+1}\bigg)^t\mu(d\bx) \label{f_infty}
    \end{align}
    and consider taking the limit $t \to \infty$ on both sides.  For this purpose, we need to establish some technical conditions for applying the monotone convergence theorem~\cite[Ch.~18]{Royden}:
    \begin{enumerate}
        \item For fixed $\bx\in \bbR^n$, the function 
        $t^n \big(\frac{f(\bx)}{\|f\|_{\infty}+1}\big)^t$ is non-increasing for $t$ sufficiently large, since $0 \le \frac{f(\bx)}{\|f\|_{\infty}+1} < 1$ by the definition of $\|f\|_{\infty}$.
        \item For each fixed $\bx\in \bbR^n$, we have 
        \begin{align}
            \label{eq:convzero}
            \lim_{t\to \infty} t^n \bigg(\frac{f(\bx)}{\|f\|_{\infty}+1}\bigg)^t, &=0
        \end{align} 
        again using $0 \le \frac{f(\bx)}{\|f\|_{\infty}+1} < 1$.
        \item The function $t^n \big(\frac{f(\bx)}{\|f\|_{\infty}+1}\big)^t$ is integrable with respect to $\bx$ for any fixed $t \ge 1$; this is because $\big(\frac{f(\bx)}{\|f\|_{\infty}+1}\big)^t \le \frac{f(\bx)}{\|f\|_{\infty}+1}$ for $t \ge 1$, and $\|f\|_1 \le \infty$.
    \end{enumerate}
    Taking limits in \eqref{f_infty} and applying \eqref{eq:convzero} and the monotone convergence theorem~\cite[Ch.~18]{Royden}, we obtain
    \begin{align}
        \label{contype1}
        \lim_{t\to \infty} (\|f\|_{\infty}+1)^{-t}\int_{\bbR^n} t^n f^t(\bx) \mu(d\bx)&=0.
    \end{align}
    
    Summarizing the above findings, we have shown that the function $\kappa(t) := \log \big( (\|f\|_{\infty}+1)^{-t} \int_{\bbR^n} t^n f^t(\bx) \mu(d\bx)\big)$ is concave in $t$ (and is therefore continuous wherever it takes finite values), is bounded from above for any fixed $t \ge 1$, and tends to $-\infty$ as $t \to \infty$.  These properties immediately imply that $\sup_{t \ge 1} \kappa(t) < \infty$, so to establish $D < \infty$ in \eqref{defD}, it only remains to show that $\sup_{t \in (0,1)} \kappa(t) < \infty$. 
    
    If $\|f\|_1 = 0$, then $f(\bx)$ is zero almost everywhere, and the claim $D < \infty$ is trivial, so we proceed assuming that $\|f\|_1 > 0$.  In this case, $\int_{\bbR^n} f^t(\bx) \mu(d\bx) > 0$ for any fixed $t > 0$, which implies that $\inf_{t \in [1,2]} \kappa(t) > -\infty$ (again, a concave function is continuous wherever it takes finite values).  By concavity, we have for $t \in (0,1)$ that $\kappa(1) \ge \frac{1}{2}\big( \kappa(t) + \kappa(2-t)\big)$, or equivalently 
    \begin{align}
        \kappa(t) 
            &\le 2\kappa(1) - \kappa(2-t) \\
            &\le 2 \sup_{u \ge 1} \kappa(u) - \inf_{u \in [1,2]} \kappa(u).
    \end{align}
    Hence, having already shown that $\sup_{u \ge 1} \kappa(u) < \infty$ and $\inf_{u \in [1,2]} \kappa(u) > -\infty$, we deduce that $\sup_{t \in (0,1)} \kappa(t) < \infty$ and hence $D < \infty$.
    
    
\end{IEEEproof}

We note that the preceding lemmas concern general vectors $\bx$ that need not be related to the matrix $\bX$ in the phase retrieval setting.  Henceforth, we gives results concerning {\em pairs} $(\bx,\by)$, which will later be directly equated with the relevant quantities in the phase retrieval problem. 

In the following lemma, we specialize the first part of Lemma~\ref{bestcup} to functions of $(\bx,\by) \in \bbR^{2kn}$ under the condition of a certain integral being finite.  This condition is explored further below.

\begin{corollary}\label{simlem} 
   Fix $n,k \in \bbZ^+$, and let $(\bX,\bY) \in \bbR^{2kn} \times \bbR^n$ be random vectors with joint distribution $f_{\bX\bY}$. For each $t \in \bbR^+$, define
    \begin{align}
        \label{defLn}
        L_n(t):=\int_{\bbR^{2kn} \times \bbR^n} f_{\bX}(\bx) f_{\bY|\bX}^t(\by|\bx) \mu(d\bx \times d\by).
    \end{align}  
    Then, under the condition that
		\begin{align}
		\label{condB1}
        L_n(t) &<\infty
		\end{align} holds for all $t \in \bbR^+$, we have that $L_n(t)$ is twice differentiable and
    \begin{align}
        \label{estL1}
        L_n'(t)&=\int_{\bbR^{2kn}\times \bbR^n} f_{\bX}(\bx)  f_{\bY|\bX}^t (\by|\bx) \log f_{\bY|\bX}(\by|\bx) \mu(d\bx \times d\by),\\
        \label{estL2}
        L_n''(t)&=\int_{\bbR^{2kn} \times \bbR^n} f_{\bX}(\bx) f_{\bY|\bX}^t (\by|\bx) (\log f_{\bY|\bX}(\by|\bx))^2 \mu(d\bx\times d\by). 
    \end{align}
\end{corollary}
\begin{IEEEproof} 
    We use the first part of Lemma~\ref{bestcup} with $(\bx,\by)$ playing the role of $\bx$ therein, and $f_{\bX}f_{\bY|\bX}^t$ playing the role of $g$.  Note that for each fixed $(\bx,\by) \in \bbR^{2kn}\times \bbR^n$, $f_{\bY|\bX}^t(\by|\bx) f_{\bX}(\bx)=\exp(t \log f_{\bY|\bX}(\by|\bx))$ is an entire function in $t \in \bbC$~\cite[Sec. 2.3]{Steinbook} and that $f_{\bY|\bX}^t(\by|\bx) f_{\bX}(\bx) \in \bbR_+$ for all $t \in \bbR^+$. In addition, for each fixed $(\bx,\by) \in \bbR^{kn} \times \bbR^n$, we have
    \begin{align}
        \label{ddt1}
        \frac{\partial^l f_{\bY|\bX}^t(\by|\bx) f_{\bX}(\bx)}{\partial t^l}=f_{\bX}(\bx) f_{\bY|\bX}^t (\by|\bx) \big(\log f_{\bY|\bX}(\by|\bx)\big)^l,\quad \forall l\in \bbZ^+, t\in \bbR^+,
    \end{align}
    or equivalently,
    \begin{align}
        \label{ddt2}
        (-1)^l \frac{\partial^l f_{\bY|\bX}^t(\by|\bx) f_{\bX}(\bx)}{\partial t^l}=f_{\bX}(\bx)  f_{\bY|\bX}^t (\by|\bx) \big(-\log f_{\bY|\bX}(\by|\bx)\big)^l \geq 0, \quad \forall l \in \bbZ^+, t\in \bbR^+.
    \end{align} 
    Hence, for each fixed $(\bx,\by)$ we have that $\frac{\partial^l f_{\bY|\bX}^t(\by|\bx) f_{\bX}(\bx)}{\partial t^l}\geq 0$ for all pairs $(l,t) \in \bbN \times\bbR^+$ if $f_{\bY|\bX}(\by|\bx)>1$, and that $(-1)^l\frac{\partial^l f_{\bY|\bX}^t(\by|\bx) f_{\bX}(\bx)}{\partial t^l}\geq 0$ for all pairs $(l,t) \in \bbN \times\bbR^+$ if $f_{\bY|\bX}(\by|\bx)\leq 1$, so that the assumption of Lemma \ref{bestcup} is satisfied in both cases. 
\end{IEEEproof}

The following lemma provides sufficient conditions under which \eqref{condB1} holds.

\begin{lemma}\label{keynotelem}  
    Fix $n,k \in \bbZ^+$, and let $(\bX,\bY) \sim f_{\bX\bY}$.  Under the conditions
		\begin{gather}
		\label{cond1C1}
		\int_{\bbR^n} f_{\bY}^{t}(\by)\mu(d\by) <\infty,\quad \forall 0<t\leq 1,\\
		\label{cond1C2}
		\sup_{\bx \in \bbR^{2kn}, \by \in \bbR^n} f_{\bY|\bX}(\by|\bx) <\infty,
		\end{gather} 
        we have that \eqref{condB1} of Corollary~\ref{simlem} holds for all $t \in \bbR^+$, i.e., $L_n(t) = \int_{\bbR^{2kn}} f_{\bX}(\bx) \int_{\bbR^n} f_{\bY|\bX}^{t} (\by|\bx)\mu(d\by) \mu(d\bx)<\infty$.  More specifically, we have
    \begin{align}
        \label{ladyfinal}
        L_n(t) \leq 2 \int_{\bbR} f_{\bY}^{t}(\by)\mu(d\by)
    \end{align} 
    for all $0<t\leq 1$, and
    \begin{align}
        \label{manfinal}
        L_n(t) \leq \bigg( \sup_{\bx \in \bbR^{2kn}, \by \in \bbR^n} f_{\bY|\bX}(\by|\bx) \bigg)^{t-1} 
    \end{align} 
    for all $t>1$.
\end{lemma}
\begin{IEEEproof}
    See  Appendix~\ref{proof:keynotelem}. 
\end{IEEEproof}

The following corollary shows that the sufficient conditions of Lemma \ref{keynotelem} are satisfied when $(\bX,\bY)$ are i.i.d.~according to a joint distribution on $(X,Y)$ corresponding to an additive noise model with a log-concave marginal $f_Y$.  The latter condition can be interpreted as stating that $f_{Y|X}(\cdot|x)$ is log-concave ``on average".  In addition, explicit upper bounds on \eqref{defLn} are given that will be useful later.

\begin{corollary}\label{rmk20}  
    Fix $n,k \in \bbZ^+$, and let $(\bX,\bY)\sim f_{XY}^n$ be i.i.d.~on $f_{XY}$ with $x \in \bbR^{2k}$ and $Y \in \bbR$.  Assume that $f_Y$ is log-concave, and that given $X=x$, we have $Y=U_x+ Z$, where $U_x$ and $Z$ are independent random variables and $\|f_Z\|_{\infty}<\infty$. Then conditions~\eqref{cond1C1}--\eqref{cond1C2} of Lemma \ref{keynotelem} hold, and in addition, we have
    \begin{gather}
        \sup_{\bx \in \bbR^{2kn}, \by \in \bbR^n} f_{\bY|\bX}(\by|\bx)\leq  \|f_Z\|_{\infty}^n, \label{supbound1} \\
        \|f_{\bY}\|_{\infty} \leq \|f_Z\|_{\infty}^n. \label{supbound2}
    \end{gather}
\end{corollary}
\begin{IEEEproof} 
First, for all $(x,y) \in \bbR^{2k} \times \bbR$, we have
    \begin{align}
    f_{Y|X}(y|x)=f_{U_x}(y)* f_Z(y)=\int_{-\infty}^{\infty} f_{U_x}(t) f_Z(y-t) dt\leq \|f_Z\|_{\infty} \int_{-\infty}^{\infty} f_{U_x}(t) dt= \|f_Z\|_{\infty},
	  \end{align}  
    and hence
    \begin{align}
    \label{todayspecial}
    f_{\bY|\bX}(\by|\bx)=\prod_{i=1}^n f_{Y|X}(y^{(i)}|x^{(i)}) \leq  \|f_Z\|_{\infty}^n
    \end{align}
   or equivalently $\sup_{\bx \in \bbR^{2kn}, \by \in \bbR^n} f_{\bY|\bX}(\by|\bx) \leq \|f_Z\|_{\infty}^n<\infty$. This means that condition~\eqref{cond1C2} of Lemma~\ref{keynotelem} holds. Moreover, we have from~\eqref{todayspecial} that
	\begin{align}
    	f_{\bY}(\by)&=\int_{\bbR^{2kn}} f_{\bX}(\bx) f_{\bY|\bX}(\by|\bx)\mu(d\bx)\\
    	&\leq \int_{\bbR^{2kn}} f_{\bX}(\bx) \|f_Z\|_{\infty}^n \mu(d\bx)\\
    	&= \|f_Z\|_{\infty}^n.
	\end{align} 
	Combining this with the log-concavity of $Y$ (and hence $\bY$) and applying Lemma~\ref{logconcave}, we deduce that condition~\eqref{cond1C1} of Lemma~\ref{keynotelem} holds.  
\end{IEEEproof}

The preceding results will be used in conjunction with the following lemma in order to bound the key quantity $K_1$ appearing in Proposition \ref{keyext}.  This result is a counterpart to part of the analysis in \cite[proof of Theorem~2.3]{Fradelizi2016}, but it is proved using different methods.\footnote{The analysis in \cite[proof of Theorem~2.3]{Fradelizi2016} does not seem to be feasible for our purposes unless $(X,Y)$ are jointly log-concave, since otherwise we cannot confirm that $\barG(t):=n \log t+ \log \int_{\bbR^{2kn}} f_{\bX}(\bx) \int_{\bbR^n} f_{\bY|\bX}(\by|\bx) \mu(d\by) \mu(d\bx)$ is concave.}


\begin{lemma}\label{tonightbest} 
    Fix $k \in \bbZ^+$, and let $(X,Y)\sim f_{XY}$ such that $X \in \bbR^{2k}$, $Y\in \bbR$, and the distribution of $Y$ is log-concave.  Define $L_1(t):=\int_{\bbR^{2k}} f_{X}(x) \int_{\bbR} f_{Y|X}^t(y|x)\mu(dy)\mu(dx)$, and suppose that
    \begin{gather}
    \label{ladynight}
    \sup_{t \in (0,1]}tL_1(t) \leq P_1 \\
    \label{ladynight2}
    \sup_{t>1} \barQ_1^{1-t} L_1(t)\leq P_2,
    \end{gather} 
    for some positive constants $P_1$, $P_2$, and $\barQ_1$.  Then, defining $C=150\max\{P_1,P_2\}$, we have
    \begin{align}
    (tL_1(t))'' \leq \frac{C }{t^2}
    \end{align} 
    for all $t \in \big(0,1\big]$, and
    \begin{align}
    (\barQ_1^{1-t} L_1(t))'' \leq \frac{C}{t^2}
    \end{align} 
    for all $t>1$.
\end{lemma}
\begin{IEEEproof} 
    This result follows from Lemma~\ref{bestcup} (with $n=2k+1$, since we consider $(X,Y)$ jointly) applied separately for the following two cases:
    \begin{itemize}
        \item For $0<t\leq 1$, set $g(x,y,t):= f_{X}(x) f_{Y|X}^t (y|x)$ and use the third part of the lemma with $\calG=(0,1]$;
        \item For $t>1$, set $g(x,y,t):=\barQ_1^{1-t} f_{X}(x) f_{Y|X}^t (y|x)$ and use the second part of the lemma with $\calG=[1,\infty)$.
    \end{itemize}
    Note that $f_{X}(x) f_{Y|X}^t (y|x)$ and $\barQ_1^{1-t} f_{X}(x) f_{Y|X}^t (y|x)=f_X(x) \big(\frac{f_{Y|X}(y|x)}{\barQ_1}\big)^t$ are both real entire functions in $t \in \bbC$ for each fixed $(x,y)\in \bbR^{2k} \times \bbR$ (see Footnote \ref{foot:entire} on page \pageref{foot:entire}).
    In addition, both functions are non-negative valued, and the required conditions on the derivatives hold by the same argument as \eqref{ddt1}--\eqref{ddt2}.
\end{IEEEproof}

\subsection{Proof of Proposition~\ref{keyext} (General Exponential Bound)} \label{proof:keyext}

Recall the notation $f_{XY}$, $\barQ$ and $K_1$ as per the proposition statement, and define
\begin{align}
    \label{tesby}
    F(t)&:=\log (L(t)),
\end{align} 
where $L(t):=\int_{\bbR^{2k} \times \bbR} f_{Y|X}^t(y|x) f_{X}(x)\mu(dx \times dy)$ as stated in~\eqref{defL}. 
From Corollary~\ref{simlem} with $n=1$, we have
\begin{align}
    \label{test1}
    L(1)&=\int_{\bbR^{2k} \times \bbR} f_{XY}(x,y) \mu(dx\times dy)=1,\\
    L'(1)&=\int_{\bbR^{2k} \times \bbR} f_{XY}(x,y) \log f_{Y|X}(y|x)\mu(dx\times dy)\\
    \label{test2}
    &=-h(Y|X),
\end{align}
and in addition, the definition of $K_1$ in \eqref{defKK} immediately implies 
\begin{align}
    \label{test3}
    (uL(u))'' &\leq \frac{K_1}{u^2}, \quad u\in (0,1],\\
		\label{test4pu}
	 (\barQ^{1-u} L(u))'' &\leq \frac{K_1}{u^2}, \quad u\in (1,\infty).
\end{align}
Now, from the Taylor-Lagrange formula~(e.g., see \cite[proof of Theorem 3.1]{Fradelizi2016}) for the function $t L(t)$, for every $t \in \big(0,1]$, we have
\begin{align}
    t L(t) &=L(1)+(t-1)\frac{d (t L(t))}{d t}(1)+ \int_1^{t} (t-u) \frac{d^2 (t L(t))}{d t^2}(u) du\\
    \label{eqsky1}
    &\leq L(1)+(t-1)[L'(1)+L(1)]+ K_1\int_t^1 \frac{u-t}{u^2}  du\\
    \label{key5a}
    &=1+ (t-1)[-h(Y|X)+1]+ K_1(t-1-\log t),
\end{align} 
where~\eqref{eqsky1} follows from~\eqref{test3} along with direct differentiation, and~\eqref{key5a} follows from~\eqref{test1} and~\eqref{test2}. It follows from~\eqref{key5a} that for all $t \in (0,1]$, we have
\begin{align}
F(t)&=\log L(t)\\
&=\log (t L(t))-\log t \\
&\leq \log \big(1+ (t-1)[-h(Y|X)+1]+ K_1(t-1-\log t)\big)-\log t \\
\label{key5}
&\leq (t-1)[-h(Y|X)+1]+ K_1(t-1-\log t) -\log t\\
\label{5keyedu}
&=(1-t) h(Y|X) +(K_1+1)(t-1-\log t),
\end{align} where~\eqref{key5} follows from the fact that $\log(1+x)\leq x$ for all $x>-1$.

In addition, from the Taylor-Lagrange formula for the function $\barQ^{1-t} L(t)$, for every $t \in \big(1,\infty)$, we have
\begin{align}
    \barQ^{1-t} L(t) &=L(1)+(t-1)\frac{d (\barQ^{1-t} L(t))}{d t}(1)+ \int_1^t (t-u) \frac{d^2 (\barQ^{1-t}L(t))}{d t^2}(u) du\\
    \label{eqsky1edu}
    &\leq L(1)+(t-1)[-\barQ^{1-1}(\log \barQ) L(1)+ \barQ^{1-1} L'(1)]+ K_1\int_1^t \frac{t-u}{u^2}  du\\
    \label{key5aedu}
    &=1+ (t-1)[-\log \barQ-h(Y|X)]+ K_1(t-1-\log t),
\end{align} 
where~\eqref{eqsky1edu} follows from~\eqref{test4pu} along with direct differentiation, and~\eqref{key5aedu} follows from~\eqref{test1} and~\eqref{test2}. It follows from~\eqref{key5aedu} that for all $t \in (1,\infty)$, we have
\begin{align}
F(t)&=\log L(t)\\
&=\log (\barQ^{1-t} L(t))-\log \barQ^{1-t} \\
&\leq \log \big(1+ (t-1)[-\log \barQ-h(Y|X)]+ K_1(t-1-\log t)\big)-(1-t) \log \barQ \\
\label{key5edu}
&\leq (t-1)[-\log \barQ-h(Y|X)]+ K_1(t-1-\log t)-(1-t) \log \barQ \\
&=(1-t) h(Y|X) +K_1 (t-1-\log t)\\
\label{key6edu}
&\leq (1-t) h(Y|X) +(K_1+1) (t-1-\log t),
\end{align} where~\eqref{key5edu} follows from the fact that $\log(1+x)\leq x$ for all $x>-1$, and~\eqref{key6edu} follows from the fact that $t-1-\log t \geq 0$ for all $t>0$.

Combining the cases in~\eqref{5keyedu} and~\eqref{key6edu}, we have
\begin{align}
\label{magnificient}
F(t) \leq  (1-t) h(Y|X) +(K_1+1) (t-1-\log t)
\end{align} 
for all $t>0$.  On the other hand, since $F(t) = \log L(t)$, we also have
\begin{align}
    \exp(F(t))&=\int_{\bbR^{2k} \times \bbR} \exp((t-1)\log f_{Y|X}(y|x)) f_{XY}(x,y) \mu(dx\times dy)\\
    &=\bbE[\exp((t-1)\log f_{Y|X}(Y|X))]\\
    \label{key6}
    &=\bbE[\exp((1-t)\tilh(Y|X))]
\end{align}
It follows from~\eqref{magnificient} and~\eqref{key6} that
\begin{align}
    \bbE[\exp((1-t)\tilh(Y|X))] \leq \exp\big( (1-t) h(Y|X) +(K_1+1) (t-1-\log t) \big),
\end{align}
or equivalently
\begin{align}
    \label{newsupcl}
    \bbE[\exp((1-t)\tilh(Y|X)-h(Y|X))] \leq \exp\big((K_1+1) (t-1-\log t) \big)
\end{align} 
for all $t>0$. 
By setting $\mu=1-t$, we obtain~\eqref{keyfact} from~\eqref{newsupcl}, recalling from the definition of $r(\cdot)$ in \eqref{defru} that $r(-\mu)=-\mu-\log(1-\mu)$ for $\mu<1$.  The remaining case $\mu\geq 1$ is trivial, since the right-hand side of~\eqref{keyfact} evaluates to $+\infty$ by the definition $r(-\mu)=+\infty$ for $\mu\geq 1$.

\subsection{Proof of Corollary~\ref{verykey} (General Concentration Corollary)} \label{proof:corkey}

The proof is very similar to that of~\cite[Corollary 3.4]{Fradelizi2016}, with the main idea being to use the Chernoff bound and optimize the exponent. 

By the Chernoff bound, we have for any $\beta>0$ and $\mu>0$ that
\begin{align}
    \label{type3}
    \bbP\big[\tilh(\bY|\bX)-h(\bY|\bX)\leq -\mu\big] &\leq \bbE\big[\exp\big(-\beta \big(\tilh(\bY|\bX)-h(\bY|\bX)\big)\big)\big]	\exp(-\beta \mu),\\
    \label{type4}
    \bbP\big[\tilh(\bY|\bX)-h(\bY|\bX)\geq \mu\big] &\leq \bbE\big[\exp\big(\beta \big(\tilh(\bY|\bX)-h(\bY|\bX)\big)\big)\big]	\exp(-\beta \mu).
\end{align}
Combining these bounds with Proposition~\ref{keyext} (with $\beta = \mu$ in the first case and $\beta = -\mu$ in the second case), we obtain
\begin{align}
    \label{type7}
    \bbP\big[\tilh(\bY|\bX)-h(\bY|\bX)\leq -\mu\big] &\leq \exp\bigg(n(K_1+1)\bigg(r(\beta)-\frac{\beta \mu}{n(K_1+1)}\bigg)\bigg),\\
    \label{type6}
    \bbP\big[\tilh(\bY|\bX)-h(\bY|\bX)\geq \mu\big] &\leq \exp\bigg(n(K_1+1)\bigg(r(-\beta)-\frac{\beta \mu}{n(K_1+1)}\bigg)\bigg),
\end{align}
where  $r(u)$ is defined in \eqref{defru}.  Now, define
\begin{align}
    r^*(t)&=\sup_{u>0} (t u-r(u))\\
    &=\sup_{u>0}(t u-u+\log(1+u)).
\end{align} 
It is easy to see that $r^*(t)=+\infty$ for $t\geq 1$. For $0<t<1$, by differentiating, the supremum is reached at $u=\frac{t}{1-t}>0$ and the maximum value is 
\begin{align}
    \label{eq111type}
    r^*(t)=-t-\log(1-t)=r(-t).
\end{align} 
In fact, $r^*(t)=r(-t)$ holds for all $t > 0$, since $r(-t)$ has value $+\infty$ for $t \ge 1$ by definition.

From~\eqref{type7} and~\eqref{eq111type}, for $\mu>0$, we have
\begin{align}
    \bbP\big[\tilh(\bY|\bX)-h(\bY|\bX)\leq -\mu\big] &\leq \exp\bigg(-n(K_1+1) \sup_{\beta>0}\bigg(\frac{\beta \mu}{n(K_1+1)}-r(\beta)\bigg)\bigg)\\
    &\leq \exp\bigg(-n(K_1+1) r^*\bigg(\frac{\mu}{n(K_1+1)}\bigg)\bigg)\bigg),\\
    &=\exp\bigg(-n(K_1+1) r\bigg(-\frac{\mu}{n(K_1+1)}\bigg)\bigg)\bigg). \label{eq:replacemu1}
\end{align}
Similarly, we can define
\begin{align}
    \tilr^*(t)&:=\sup_{0<u<1} (t u-r(-u))\\
    &=\sup_{0<u<1}(t u+u+\log(1-u)).
\end{align} 
By differentiating, the supremum is reached at $u=\frac{t}{1+t}\in (0,1)$ and the maximum value is 
\begin{align}
    \label{eq111typeb}
    \tilr^*(t)=t-\log(1+t)=r(t)
\end{align} 
for any $t>0$ (here there is no $+\infty$ case).
From~\eqref{type6} and~\eqref{eq111typeb}, for $\mu>0$, we have
\begin{align}
    \bbP\big[\tilh(\bY|\bX)-h(\bY|\bX)\geq \mu\big] &\leq \exp\bigg(-n(K_1+1) \sup_{0<\beta<1}\bigg(\frac{\beta \mu}{n(K_1+1)}-r(-\beta)\bigg)\bigg)\\
    &= \exp\bigg(-n(K_1+1) \tilr^*\bigg(\frac{\mu}{n(K_1+1)}\bigg)\bigg)\bigg),\\
    &=\exp\bigg(-n(K_1+1) r\bigg(\frac{\mu}{n(K_1+1)}\bigg)\bigg)\bigg). \label{eq:replacemu2}
\end{align} 
The proof is completed by replacing $\mu$ by $n(K_1+1)\mu$ in \eqref{eq:replacemu1} and \eqref{eq:replacemu2}, and noting that $h(\bY|\bX) = nh(Y|X)$ for $(\bX,\bY) \sim f_{XY}^n$.

%
%
\section{Proof of Theorem \ref{thre1} (Concentration of Information Density for Phase Retrieval)} \label{app:conc_phase}

The following corollary shows that for the phase retrieval setting, $f_{Y|X_{s_{\rm{eq}}}\beta_s}(\cdot|x_{s_{\rm{eq}}},b_s)$ and $f_{Y|X_s\beta_s}(\cdot|x_s,b_s)$ have the boundedness properties required to apply Lemma \ref{tonightbest}. 

\begin{corollary}\label{immedi} 
    For the phase retrieval model in~\eqref{acqui}, we have for fixed $s$ and $s_{\rm{eq}} \subset s$ that
    \begin{gather}
        \sup_{t\in (0,1]} t \int_{\bbC^{k-\ell}} f_{X_{s_{\rm{eq}}}}(x_{s_{\rm{eq}}}) \int_{\bbR} f_{Y|X_{s_{\rm{eq}}}\beta_s}^{t}(y|x_{s_{\rm{eq}}},b_s) \mu(dy) \mu(dx_{s_{\rm{eq}}})\leq 2D(b_s) (\|f_Z\|_{\infty}+1),\\
        \sup_{t\in (0,1]} t  \int_{\bbC^{k}} f_{X_s}(x_s) \int_{\bbR} f_{Y|X_s\beta_s}^{t}(y|x_s,b_s) \mu(dy) \mu(dx_s) \leq 2D(b_s) (\|f_Z\|_{\infty}+1),
    \end{gather} 
    and
    \begin{gather}
        \label{equip10}
        \sup_{t>1}\|f_Z\|_{\infty}^{1-t} \int_{\bbC^{k-\ell}} f_{X_{s_{\rm{eq}}}}(x_{s_{\rm{eq}}}) \int_{\bbR} f_{Y|X_{s_{\rm{eq}}}\beta_s}^{t}(y|x_{s_{\rm{eq}}},b_s) \mu(dy) \mu(dx_{s_{\rm{eq}}})\leq 1,\\
        \label{equip11}
        \sup_{t>1} \|f_Z\|_{\infty}^{1-t} \int_{\bbC^{k}} f_{X_s}(x_s) \int_{\bbR} f_{Y|X_s\beta_s}^{t}(y|x_s,b_s) \mu(dy) \mu(dx_s) \leq 1
    \end{gather} 
    for all $b_s\in \bbC^k$, where $\ell:=k-|s_{\rm{eq}}|$, and
    \begin{equation}
        \label{defDb}
        D(b_s):=\sup_{t>0} (\|f_{Y|\beta_s=b_s}\|_{\infty}+1)^{-t}\int_{\bbR} t f_{Y|\beta_s}^t(y|b_s) \mu(dy)<\infty.
    \end{equation}
\end{corollary}
\begin{IEEEproof} 
     We condition on $\beta_s = b_s$ (and $S=s$), and consider the resulting joint distributions $(X_s,Y)$ and $(X_{s_{\rm eq}},Y)$ for a single measurement.  The log-concavity properties in Lemma \ref{lem:log_conc} allow us to apply Corollary~\ref{rmk20} (with $n=1$) and subsequently Lemma \ref{keynotelem}.  Substituting \eqref{supbound1} into \eqref{manfinal} yields the following for $t>1$:
    \begin{gather}
        \label{type1}
        \int_{\bbC^{k-\ell}} f_{X_{s_{\rm{eq}}}}(x_{s_{\rm{eq}}}) \int_{\bbR} f_{Y|X_{s_{\rm{eq}}}\beta_s}^{t}(y|x_{s_{\rm{eq}}},b_s) \mu(dy) \mu(dx_{s_{\rm{eq}}}) \leq \|f_Z\|_{\infty}^{t-1},\\
        \label{type2}
        \int_{\bbC^k} f_{X_s}(x_s) \int_{\bbR} f_{Y|X_s\beta_s}^{t}(y|x_s,b_s) \mu(dy) \mu(dx_s) \leq \|f_Z\|_{\infty}^{t-1}.
    \end{gather}
    These equations are equivalent to \eqref{equip10} and~\eqref{equip11}.
    
    For the case $t\in (0,1]$, we first apply Lemma~\ref{logconcave} and Corollary \ref{rmk20} (with $n=1$); the latter implies $\|f_{Y|\beta_s=b_s}\|_{\infty} \le \|f_Z\|_{\infty}<\infty$ via \eqref{supbound2}, which we combine with the former to obtain
    \begin{align}
        \label{verycrazy1}
        \int_{\bbR} f_{Y|\beta_s}^t(y|b_s)\mu(dy) &\leq D(b_s) \frac{(\|f_{Y|\beta_s=b_s}\|_{\infty}+1)^t}{t}\\
				\label{newpo1}
				&\leq D(b_s) \frac{(\|f_Z\|_{\infty}+1)^t}{t},
    \end{align} 
    where $D(b_s)$ is defined in \eqref{defDb}.	Since $t \le 1$ and $\|f_Z\|_{\infty}+1 \ge 1$, we can weaken \eqref{newpo1} to
    \begin{align}
        \label{crazynight2}
        \sup_{t\in (0,1]}t  \int_{\bbR}  f_{Y|\beta}^t(y|b_s)\mu(dy) \leq D(b_s) (\|f_Z\|_{\infty}+1).
    \end{align}
    Now, by applying~\eqref{ladyfinal} of Lemma~\ref{keynotelem} (with $n=1$ and $X_{s_{\rm eq}}$ playing the role of $X$) together with~\eqref{crazynight2}, we have
    \begin{align}
        \sup_{t \in (0,1]} t \int_{\bbC^{k-\ell}} f_{X_{s_{\rm{eq}}}}(x_{s_{\rm{eq}}}) \int_{\bbR} f_{Y|X_{s_{\rm{eq}}}\beta_s}^{t}(y|x_{s_{\rm{eq}}},b_s)\mu(dy) \mu(dx_{s_{\rm{eq}}})
        &\leq 2 \sup_{t \in (0,1]} t  \int_{\bbR} f_{Y|\beta_s}^t(y|b_s)\mu(dy) \\
        &\leq 2D(b_s) (\|f_Z\|_{\infty}+1),
    \end{align} 
    and similarly
    \begin{align}
        \sup_{t \in (0,1]} t \int_{\bbC^k} f_{X_s}(x_s) \int_{\bbR} f_{Y|X_s\beta_s}^{t}(y|x_s,b_s) (y|x_s)\mu(dy) \mu(dx_s)\leq 2D(b_s) (\|f_Z\|_{\infty}+1)
    \end{align} 
    by replacing $X_{s_{\rm{eq}}}$ by $X_s$. 
\end{IEEEproof}

With Corollary \ref{immedi} in place, we are able to use Lemma \ref{tonightbest} to deduce the following result for bounding the crucial quantity $K_1$ in the concentration bounds (first appearing in Proposition \ref{keyext}, and leading to Corollary \ref{verykey} being the form that we will apply).  Note that below, $\barL_{1,b_s}(t)$ and $\barL^{(s_{\rm{eq}})}_{1,b_s}(t)$ are instances of $L_1(t)$ in Lemma \ref{tonightbest}, and $\barK_1(b_s)$ and $\barK^{(s_{\rm{eq}})}_1(b_s)$ are instances of $K_1$.

\begin{lemma}\label{bigband2018} 
    For the phase retrieval model in Section~\ref{model}, for fixed $s$, $s_{\rm{eq}} \subset s$, and $b_s \in \bbC^k$, define 
    \begin{align}
        \barL_{1,b_s}(t)&:=\int_{\bbC^{k}} f_{X_s}(x_s)\int_{\bbR} f_{Y|X_s\beta_s}^t(y|x_s,b_s) \mu(dy) \mu(dx_s),\\
        \barL^{(s_{\rm{eq}})}_{1,b_s}(t)&:=\int_{\bbC^{k-\ell}} f_{X_{s_{\rm{eq}}}}(x_{s_{\rm{eq}}})\int_{\bbR} f_{Y|X_{s_{\rm{eq}}}\beta_s}^t(y|x_{s_{\rm{eq}}},b_s) \mu(dy) \mu(dx_{s_{\rm{eq}}})
    \end{align}     
	for $t>0$, where $\ell:=|s_{\rm{dif}}|=k-|s_{\rm{eq}}|$. Moreover, define
    \begin{align}
    \barK_1(b_s)&:=\max\bigg\{\sup_{t \in (0,1]} t^2 (t \barL^{(s_{\rm{eq}})}_{1,b_s}(t))'',\sup_{t \in (1,\infty)}t^2 (\|f_Z\|_{\infty}^{1-t} \barL_{1,b_s}(t))''\bigg\},\\
    \barK^{(s_{\rm{eq}})}_1(b_s)&:=\max\bigg\{\sup_{t \in (0,1]} t^2 (t \barL^{(s_{\rm{eq}})}_{1,b_s}(t))'',\sup_{t \in (1,\infty)}t^2 (\|f_Z\|_{\infty}^{1-t} \barL^{(s_{\rm{eq}})}_{1,b_s}(t))''\bigg\},
    \end{align} 
    Then, the following bounds hold:
    \begin{align}
    \barK_1(b_s) & \leq C(b_s)-1,\\
    \barK^{(s_{\rm{eq}})}_1(b_s) & \leq C(b_s)-1,
    \end{align}
    where
    \begin{align}
    \label{defCb}
    C(b_s)&:= 150 \max\{2D(b_s)(\|f_Z\|_{\infty}+1),1\},
    \end{align}
    and $D(b_s)$ is defined in \eqref{defDb}.
\end{lemma}
\begin{IEEEproof} 
    This result is obtained by applying Lemma~\ref{tonightbest} (with $(X_s,Y)$ or $(X_{s_{\rm eq}},Y)$ in place of $(X,Y)$, and conditioning on $\beta_s = b_s$), and characterizing the upper bounds $P_1,P_2$ therein using Corollary~\ref{immedi}.  Note that a complex random vector in $\bbC^k$ can be equivalently considered as a real complex vector in $\bbR^{2k}$. 
\end{IEEEproof}

With the above tools in place, we are ready to prove the main result on the concentration of the information density (Theorem \ref{thre1}), and its simplified version (Corollary \ref{trivicor}).

\subsection{Proof of Theorem \ref{thre1}} 

By the assumption of i.i.d.~measurements, we have
\begin{align}
    f_{\bY|\bX_{s_{\rm{eq}}} \bX_{s_{\rm{dif}}} \beta_s}(\by|\bx_{s_{\rm{eq}}},\bx_{s_{\rm{dif}}},b_s)&=\prod_{i=1}^n f_{Y|X_{s_{\rm{eq}}} X_{s_{\rm{dif}}} \beta_s}(y^{(i)}|x_{s_{\rm{eq}}}^{(i)},x_{s_{\rm{dif}}}^{(i)},b_s),\\
    f_{\bY|\bX_{s_{\rm{eq}}} \beta_s}(y|\bx_{s_{\rm{eq}}},b_s)&=\prod_{i=1}^n f_{Y|X_{s_{\rm{eq}}}\beta_s}(y^{(i)}|x_{s_{\rm{eq}}}^{(i)},b_s),
\end{align}
where, recalling $Y = |\langle X_s,\beta_s\rangle| + Z$, the conditional distributions for a single measurement are given by
\begin{align}
    \label{eq208equipA}
    f_{Y|X_{s_{\rm{eq}}} X_{s_{\rm{dif}}} \beta_s}(y|x_{s_{\rm{eq}}},x_{s_{\rm{dif}}},b_s)&:=f_Z(y-|\langle x_{s_{\rm{dif}}}, b_s \rangle + \langle x_{s_{\rm{eq}}}, b_s \rangle|^2),\\
    \label{eq208equipB}
    f_{Y|X_{s_{\rm{eq}}}\beta_s}(y|x_{s_{\rm{eq}}},b_s)&:=(f_{U_{s_{\rm eq}}}* f_Z)(y)
\end{align}
with $U_{s_{\rm eq}}$ being the squared magnitude of a $\mathcal{CN}( \langle x_{s_{\rm{eq}}}, b_s \rangle, \|b_{s_{\rm{dif}}}\|_2^2)$ random variable.  Moreover, we have
\begin{align}
    \label{factor}
    i^n(\bX_{s_{\rm{dif}}};\bY|\bX_{s_{\rm{eq}}},\beta_s=b_s)&=\tilh(\bY|\bX_{s_{\rm{dif}}},\bX_{s_{\rm{eq}}},\beta_s=b_s)-\tilh(\bY|\bX_{s_{\rm{eq}}},\beta_s=b_s),
\end{align}
where $\tilh$ denotes the (conditional) negative log-density ({\em cf.}, \eqref{deftilhnew}).

Recall the log-concavity properties in Lemma \ref{lem:log_conc} for a single measurement, which immediately imply analogous properties for the vector of $n$ independent measurements~\cite[Prop.~3.2]{Saumard2014}. Applying Corollary~\ref{verykey} and bounding $K_1$ therein (as defined in \eqref{defKK}) by $C(b_s)-1$ in accordance with Lemma \ref{bigband2018}, we have for all $\mu>0$ that
\begin{align}
    \label{tr1}
    \bbP\big[\tilh(\bY|\bX_{s_{\rm{dif}}},\bX_{s_{\rm{eq}}},\beta_s=b_s)-n h(Y|X_{s_{\rm{dif}}},X_{s_{\rm{eq}}},\beta_s=b_s)\geq nC(b_s) \mu\big]&\leq \exp(-nC(b_s) r(\mu)),\\
    \label{tr2}
    \bbP\big[\tilh(\bY|\bX_{s_{\rm{dif}}},\bX_{s_{\rm{eq}}},\beta_s=b_s)-n h(Y|X_{s_{\rm{dif}}},X_{s_{\rm{eq}}},\beta_s=b_s)\leq -nC(b_s)\mu\big]&\leq \exp(-nC(b_s)r(-\mu)),
\end{align}
noting the one-to-one correspondence between $\bX_s$ and $(\bX_{s_{\rm{dif}}},\bX_{s_{\rm{eq}}})$. 
Similarly, Corollary~\ref{verykey} and Lemma \ref{bigband2018} also yield 
\begin{align}
    \label{tr3}
    \bbP\big[\tilh(\bY|\bX_{s_{\rm{eq}}},\beta_s=b_s)-n h(Y|X_{s_{\rm{eq}}},\beta_s=b_s)\geq nC(b_s)\mu\big]&\leq \exp(-nC(b_s) r(\mu)),\\
    \label{tr4}
    \bbP\big[\tilh(\bY|\bX_{s_{\rm{eq}}},\beta_s=b_s)-n h(Y|X_{s_{\rm{eq}}},\beta_s=b_s)\geq -n C(b_s) \mu\big] &\leq \exp(-nC(b_s) r(-\mu)).
\end{align}
Finally, observe that for all $\mu>0$, we have
\begin{align}
    \label{facto1}
    &\bbP\big[i^n(\bX_{s_{\rm{dif}}};\bY|\bX_{s_{\rm{eq}}},\beta_s=b_s)-n I(X_{s_{\rm{dif}}};Y|X_{s_{\rm{eq}}},\beta_s=b_s)\leq -2nC(b_s) \mu\big]  \nonumber\\
    &\quad \leq \bbP\big[\tilh(\bY|\bX_{s_{\rm{dif}}},\bX_{s_{\rm{eq}}},\beta_s=b_s)-n h(Y|X_{s_{\rm{dif}}},X_{s_{\rm{eq}}},\beta_s=b_s)\leq -nC(b_s) \mu\big]\nonumber\\
    &\qquad + \bbP\big[\tilh(\bY|\bX_{s_{\rm{eq}}},\beta_s=b_s)-n h(Y|X_{s_{\rm{eq}}},\beta_s=b_s)\geq nC(b_s) \mu\big]\\
    \label{facto2}
    &\quad\leq \exp(-nC(b_s) r(\mu)) + \exp(-n C(b_s)  r(-\mu)),
\end{align} 
where~\eqref{facto1} follows from~\eqref{factor} and the union bound,~\eqref{facto2} follows from~\eqref{tr2}--\eqref{tr3}.  Notice that \eqref{facto2} recovers \eqref{tofact1}, and we similarly obtain \eqref{tofact2} from~\eqref{tr1} and~\eqref{tr4}.  

\subsection{Proof of Corollary \ref{trivicor}} 

Recall that given $\beta_s = b_s$, any given measurement takes the form $Y = |\langle X_s, b_s \rangle|^2 + Z$, where $X_s$ has i.i.d.~$\calN(0,1)$ entries.  Hence, $f_{Y|\beta_s}$ is the convolution of the noisy density $f_Z$ with a $\chi_2^2$ random variable scaled by $\|b_s\|^2$.  This implies that if $\|b_s\|^2$ is a constant (i.e., remains fixed as $k$ increases), then so is $D(b_s)$ (see \eqref{defDb}) and hence also $C(b_s)$ (see \eqref{defCb}).  Equivalently, if $\|b_s\|_2=\Theta(1)$, then $C(b_s)= \Theta(1)$, as required.

%
%
\section{Technical Proofs}

\subsection{Proof of Lemma \ref{bestcup}} \label{proof:bestcup}

{\bf Proof of part (i).} Fix $t \in \bbR^+$. Since $g(\bx,u)$ is a real entire function in $u$ (analytic for all $u \in \bbC$) for each fixed $\bx \in \bbR^n$, by Taylor's expansion~\cite[Theorem~4.4]{Steinbook}, we have 
\begin{align}
    \label{taylorfact}
    g(\bx,u)=g(\bx,t)+\sum_{l=1}^{\infty} \frac{1}{l!}(u-t)^{l}\frac{\partial^l g}{\partial u^l}(\bx,t) 
\end{align} 
for all $u \in \bbR^+$.  Re-arranging, we obtain for $u \ne t$ that
\begin{align}
    \label{bestto}
    \frac{g(\bx,u)-g(\bx,t)}{u-t}=\sum_{l=1}^{\infty} \frac{1}{l!}(u-t)^{l-1}\frac{\partial^l g}{\partial u^l}(\bx,t).
\end{align}
Taking the absolute value, and supposing that  $|u-t|\leq \eps/2$ for some $\eps<2t$, we have
\begin{align}
    \label{ext1}
    \bigg|\frac{g(\bx,u)-g(\bx,t)}{u-t}\bigg| &=  \bigg|\sum_{l=1}^{\infty} \frac{1}{l!} (u-t)^{l-1}\frac{\partial^l g}{\partial u^l}(\bx,t)\bigg|\\
    &\leq \sum_{l=1}^{\infty} \frac{1}{l!}\bigg(\frac{\eps}{2}\bigg)^{l-1}\bigg|\frac{\partial^l g}{\partial u^l}(\bx,t)\bigg|\\
    \label{tofact15}
    &\leq \frac{2}{\eps}\bigg|\sum_{l=1}^{\infty} \frac{1}{l!}(-1)^l\bigg(\frac{\eps}{2}\bigg)^l\frac{\partial^l g}{\partial u^l}(\bx,t)\bigg|+\frac{2}{\eps}\bigg|\sum_{l=1}^{\infty} \frac{1}{l!}\bigg(\frac{\eps}{2}\bigg)^l\frac{\partial^l g}{\partial u^l}(\bx,t)\bigg| \\
    \label{tofact100}
    &=\frac{2}{\eps} \big|g(\bx,t-\eps/2)-g(\bx,t)\big|+\frac{2}{\eps} \big|g(\bx,t+\eps/2)-g(\bx,t)\big| \\
    \label{finto}
    &\leq \frac{2}{\eps} \Big( g(\bx,t-\eps/2)+ 2g(\bx,t)+g(\bx,t+\eps/2)\Big),
\end{align} 
where~\eqref{tofact15} follows from the assumption that either $(-1)^l \frac{\partial^l g}{\partial u^l}(\bx,t)\geq 0$ for all $(l,t)$  or $\frac{\partial^l g}{\partial u^l}(\bx,t)\geq 0$ for all $(l,t)$, \eqref{tofact100} follows from~\eqref{taylorfact} applied twice with $u=t-\eps/2$ and $u=t+\eps/2$, and \eqref{finto} follows from the triangle inequality and the non-negativity of $g$.

We proceed by integrating \eqref{finto} over $\bx$.  By the assumption that $g(\bx,u) \in L^1(\bbR^n)$ for all $u>0$, we have
\begin{align}
    &\int_{\bbR^n} \big(g(\bx,t-\eps/2)+  2g(\bx,t) + g(\bx,t+\eps/2)\big)\mu(d\bx) \\
    \label{lovefact2}
    &\quad =\int_{\bbR^n}  g(\bx,t-\eps/2)\mu(d\bx)+ 2\int_{\bbR^n} g(\bx,t)\mu(d\bx)+\int_{\bbR^n} g(\bx,t+\eps/2)\mu(d\bx)\\
    \label{lovefact3}
    &\quad  =T(t-\eps/2)+ 2 T(t) + T(t+\eps/2)<\infty,
\end{align}
where we applied the definition of $T(t)$ in~\eqref{mor1}, and used the fact that it is finite by the assumption $g(\bx,u) \in L^1(\bbR^n)$ for fixed $u$.  

The definition of $T(t)$ also yields
\begin{align}
    \label{lovefact1}
    \frac{T(u)-T(t)}{u-t}&=\int_{\bbR^n} \frac{g(\bx,u)-g(\bx,t)}{u-t} \mu(d\bx), 
\end{align}
From~\eqref{finto} and~\eqref{lovefact3}, we have that $\big|\frac{g(\bx,u)-g(\bx,t)}{u-t}\big|$ is dominated by the integrable function $\frac{2}{\eps}\big( g(\bx,t-\eps/2)+ 2 g(\bx,t)+g(\bx,t+\eps/2)$, meaning we can apply the dominated convergence theorem~\cite[Ch.~18]{Royden} to obtain
\begin{align}
    \label{eq331equip0}
    T'(t)&=\lim_{u\to t}\frac{T(u)-T(t)}{u-t}\\
    &=\int_{\bbR^n} \lim_{u\to t}\frac{g(\bx,u)-g(\bx,t)}{u-t} \mu(d\bx)\\
    \label{eq331equip}
    &=\int_{\bbR^n} \frac{\partial g}{\partial u}(\bx,t) \mu(d\bx),
\end{align} 
where \eqref{eq331equip0} uses \eqref{lovefact1}, and \eqref{eq331equip} follows from the definition of partial derivative.  We have thus proved \eqref{eq:T_deriv} in the case that $l=1$. 

Similarly to \eqref{eq331equip}, from~\eqref{finto},~\eqref{lovefact3}, and the dominated convergence theorem~\cite[Ch.~18]{Royden}, we have
\begin{align}
    \label{eqtesting}
    \lim_{u\to t}\int_{\bbR^n} \bigg|\frac{g(\bx,u)-g(\bx,t)}{u-t}\bigg| \mu(d\bx)
        &=\int_{\bbR^n} \lim_{u\to t}\bigg|\frac{g(\bx,u)-g(\bx,t)}{u-t}\bigg| \mu(d\bx) \\
        \label{beaumind}
        &=\int_{\bbR^n} \bigg|\frac{\partial g}{\partial u}(\bx,t)\bigg| \mu(d\bx).
\end{align} 
From~\eqref{finto},~\eqref{lovefact3}, and~\eqref{beaumind}, we have
\begin{align}
    \label{eqverygood}
    \barT(t)&:=\int_{\bbR^n} \bigg|\frac{\partial g}{\partial u}(\bx,t)\bigg|\mu(d\bx) \\
		&=\lim_{u\to t}\int_{\bbR^n} \bigg|\frac{g(\bx,u)-g(\bx,t)}{u-t}\bigg| \mu(d\bx)\\
        \label{eqverygoodfinite}
		&\leq \frac{2}{\eps}\big[T(t-\eps/2)+ 2T(t)+ T(t+\eps/2)\big]<\infty.
\end{align} 
Since~\eqref{eqverygood} holds for any $t>0$, 
we similarly have 
\begin{align}
    \label{eqverygood1}
    \int_{\bbR^n} \bigg|\frac{\partial g}{\partial u}(\bx,t-\eps/2)\bigg|\mu(d\bx) &<\infty,\\
    \label{eqverygood2}
    \int_{\bbR^n} \bigg|\frac{\partial g}{\partial u}(\bx,t+\eps/2)\bigg|\mu(d\bx) &<\infty.
\end{align}

Now, following the same steps as those for the first derivative, we have the following analog of \eqref{lovefact1}:
\begin{align}
    \label{love23}
    \frac{T'(u)-T'(t)}{u-t}=\int_{\bbR^n} \frac{\frac{\partial g}{\partial u}(\bx,u)-\frac{\partial g}{\partial u} (\bx,t)}{u-t} \mu(d\bx).
\end{align}
By Taylor's expansion (replacing $g(\bx,u)$ in~\eqref{bestto} by $\frac{\partial g}{\partial u}(\bx,u)$), we also have
\begin{align}
    \frac{\frac{\partial g}{\partial u}(\bx,u)-\frac{\partial g}{\partial u} (\bx,t)}{u-t}=\sum_{l=1}^{\infty} \frac{1}{l!}(u-t)^{l-1}\frac{\partial^{l+1} g}{\partial u^{l+1}}(\bx,t).
\end{align}
Using the same arguments as from~\eqref{ext1} to~\eqref{finto}, we obtain (for $u\neq t$ with $|u-t|\leq \eps/2$ and $t>\eps/2$) that
\begin{align}
    \label{love25}
    \bigg|\frac{\frac{\partial g}{\partial u}(\bx,u)-\frac{\partial g}{\partial u} (\bx,t)}{u-t}\bigg|\leq \frac{2}{\eps}\bigg(\bigg|\frac{\partial g}{\partial u}(\bx,t-\eps/2)\bigg|+2 \bigg|\frac{\partial g}{\partial u}(\bx,t)\bigg|+\bigg|\frac{\partial g}{\partial u}(\bx,t+\eps/2)\bigg|\bigg).
\end{align}
Integrating both sides and applying the definition of $\barT$ in \eqref{eqverygood}, we obtain
\begin{align}
    \label{facto10}
    \int_{\bbR^n} \bigg|\frac{\frac{\partial g}{\partial u}(\bx,u)-\frac{\partial g}{\partial u} (\bx,t)}{u-t}\bigg|\mu(d\bx)
  &\le \frac{2}{\eps} \big[\barT(t-\eps/2)+2\barT(t)+\barT(t+\eps/2)\big]< \infty,
  \end{align}
where the finiteness is by \eqref{eqverygoodfinite}.

From~\eqref{love23},~\eqref{love25}, and~\eqref{facto10}, we have that $\big|\frac{\frac{\partial g}{\partial u}(\bx,u)-\frac{\partial g}{\partial u} (\bx,t)}{u-t}\big|$ is dominated by the integrable function $\frac{2}{\eps}\big(\big|\frac{\partial g}{\partial u}(\bx,t-\eps/2)\big|+2 \big|\frac{\partial g}{\partial u}(\bx,t)\big|+\big|\frac{\partial g}{\partial u}(\bx,t+\eps/2)\big|\big)$. Hence, by the dominated convergence theorem~\cite[Ch.~18]{Royden}, we have
\begin{align}
    T''(t)&=\lim_{u\to t}\frac{T'(u)-T'(t)}{u-t}\\
        \label{l2aproved0}
		&=\lim_{u \to t}\int_{\bbR^n} \frac{\frac{\partial g}{\partial u}(\bx,u)-\frac{\partial g}{\partial u} (\bx,t)}{u-t}\mu(d\bx)\\
		\label{l2aproved}
    &=\int_{\bbR^n} \lim_{u\to t}\frac{\frac{\partial g}{\partial u}(\bx,u)-\frac{\partial g}{\partial u} (\bx,t)}{u-t}\mu(d\bx)  \\
    &=\int_{\bbR^n} \frac{\partial^2 g}{\partial u^2}(\bx,t) \mu(d\bx) \label{l2proved}.
	\end{align} 
This proves \eqref{eq:T_deriv} for $l=2$.

{\bf Proof of part (ii).}   Setting $\eps=t$, we have from \eqref{eq331equip} and \eqref{eqverygoodfinite} that
\begin{align}
T'(t) &\leq \barT(t)\\
    \label{eq339equip0}
    & \leq \frac{2}{t } \big[T(t/2)+ 2T(t)+ T(3 t/2)\big]\\
    \label{eq339equip}
    &\leq \frac{8T^*}{t}
 \end{align} 
by upper bounding each $T(\cdot)$ by $T^*$ in accordance with \eqref{condequip1}.  Moreover, returning to \eqref{l2aproved0}, we have
\begin{align}
    T''(t)
    &=\lim_{u\to t}\int_{\bbR^n} \frac{\frac{\partial g}{\partial u}(\bx,u)-\frac{\partial g}{\partial u} (\bx,t)}{u-t}\mu(d\bx)\\
    \label{A1proof}
    &= \frac{2}{t} \big[\barT(t/2)+2\barT(t)+\barT(3t/2)\big],
\end{align}
where \eqref{A1proof} uses \eqref{facto10} with $\epsilon = t$.  Combining \eqref{A1proof} and~\eqref{eq339equip} gives
\begin{align}
\label{facto11}
		T''(t)&\leq \bigg(\frac{2}{t}\bigg)\bigg[\frac{8T^*}{(t/2)}+ 2\frac{8T^*}{t }+\frac{8T^*}{(3t/2)}\bigg] \\
		\label{toprove}
		&\le \frac{75 T^*}{t^2},
\end{align}
as required.

{\bf Proof of part (iii).} 
We again set $\eps=t$.  The two steps leading to \eqref{eq339equip0} are still valid in this case, but from there we need to proceed differently via the definition of $T^{\dagger}$ in \eqref{condequip11}:
\begin{align}
T'(t) &\leq \barT(t)\\
& \leq \frac{2}{t } \big[T(t/2)+ 2T(t)+ T(3 t/2)\big]\\
    \label{eq339equipsup}
		&\leq \frac{2}{t } \bigg[\frac{T^{\dagger}}{t/2}+ 2 \frac{T^{\dagger}}{t}+ \frac{T^{\dagger}}{3t/2}\bigg]\\
		\label{eq440equiptat}
    &\leq \frac{28 T^{\dagger}}{3t^2}.
 \end{align} 
In addition \eqref{A1proof} is still valid in this case, but is further bounded differently via \eqref{eq440equiptat}:
\begin{align}
    T''(t)
    & \le \frac{2}{t} \big[\barT(t/2)+2 \barT(t)+ \barT(3t/2)\big]\\
    \label{previos}
    &\leq \frac{2}{t}\bigg[\frac{28 T^{\dagger}}{3(t/2)^2}+2\frac{28 T^{\dagger}}{3t^2}+ \frac{28 T^{\dagger}}{3(3t/2)^2}\bigg]\\
    \label{A11proof}
    &=\frac{3248}{27t^3} T^{\dagger}.
\end{align} 
From~\eqref{eq440equiptat} and~\eqref{A11proof}, we have
\begin{align}
    (t T(t))''&=2 T'(t)+ t T''(t)\\
    \label{starcha}
    &\leq 2 \frac{28 T^{\dagger}}{3t^2} + \frac{3248 T^{\dagger}}{27t^2}\\
    \label{lasbo}
    &\le \frac{150 T^{\dagger}}{t^2},
\end{align}
as required.

{\bf Remark.} We could potentially reduce the constant $75$ in~\eqref{toprove} or $150$ in~\eqref{lasbo} by choosing the optimal values of $\eps \in (0, 2t)$. However, for the purposes of this paper, the exact values of these constants are not important.

\subsection{Proof of Lemma \ref{keynotelem}}\label{proof:keynotelem}

{\bf Case 1: $t > 1$.}  For brevity, let $\tilQ:=\sup_{\bx \in \bbR^{2kn}, \by \in \bbR^n} f_{\bY|\bX}(\by|\bx)$.  We have
\begin{align}
    \int_{\bbR^{2kn} \times \bbR^n} f_{\bX}(\bx) f_{\bY|\bX}^{t}(\by|\bx)\mu(d\bx \times d\by)&=\int_{\bbR^{2kn} \times \bbR^n} f_{\bX\bY}(\bx,\by) f_{\bY|\bX}^{t-1}(\by|\bx)\mu(d\bx\times d\by)\\
    \label{nearfinal}
    &\leq \tilQ^{t-1} \int_{\bbR^{2kn} \times \bbR^n} f_{\bX\bY}(\bx,\by)\mu(d\bx \times d\by)\\
    \label{nearfinal2}
    &= \tilQ^{t-1}<\infty.
\end{align}
where~\eqref{nearfinal} follows from the definition of $\tilQ$, and \eqref{nearfinal2} from the assumption $\tilQ < \infty$.

{\bf Case 2: $0<t\le1$.}  In this case, we have
\begin{align}
    \int_{\bbR^{2kn} \times \bbR^n} f_{\bX}(\bx) f_{\bY|\bX}^{t}(\by|\bx)\mu(d\bx\times d\by)&=\int_{\bbR^{2kn} \times \bbR^n} f_{\bX}^{t}(\bx)f_{\bY|\bX}^{t}(\by|\bx) f_{\bX}^{1-t}(\bx)\mu(d\bx \times d\by)\\
    &=\int_{\bbR^{2kn} \times \bbR^n} f_{\bX\bY}^{t}(\bx,\by) f_{\bX}^{1-t}(\bx)\mu(d\bx\times d\by)\\
    &=\int_{\bbR^{2kn} \times \bbR^n} f_{\bY}^{t}(\by) f_{\bX|\bY}^{t}(\bx|\by) f_{\bX}^{1-t}(\bx)\mu(d\bx\times d\by).
\end{align}
Now, for each $(\bx,\by) \in \bbR^{2kn} \times \bbR^n$, if $f_{\bX|\bY}(\bx|\by)\leq f_{\bX}(\bx)$, then we have 
\begin{align}
    \label{modaltest1}
    f_{\bY}^{t}(\by) f_{\bX|\bY}^{t}(\bx|\by) f_{\bX}^{1-t}(\bx) \leq f_{\bY}^{t}(\by) f_{\bX}(\bx),
\end{align}
whereas if $f_{\bX|\bY}(\bx|\by)>f_{\bX}(\bx)$, then we have 
\begin{align}
    \label{modaltest2}
    f_{\bY}^{t}(\by) f_{\bX|\bY}^{t}(\bx|\by) f_{\bX}^{1-t}(\bx) &\leq f_{\bY}^{t}(\by) f_{\bX|\bY}(\bx|\by).
\end{align}
Combining these two cases, we obtain
\begin{align}
    f_{\bY}^{t}(\by) f_{\bX|\bY}^{t}(\bx|\by) f_{\bX}^{1-t}(\bx)\leq f_{\bY}^{t}(\by) f_{\bX}(\bx)+ f_{\bY}^{t}(\by) f_{\bX|\bY}(\bx|\by)
\end{align} 
for all $(\bx,\by) \in \bbR^{2kn} \times \bbR^n$.  Hence,
\begin{align}
    &\int_{\bbR^{2kn} \times \bbR^n}f_{\bY}^{t}(\by) f_{\bX|\bY}^{t}(\bx|\by) f_{\bX}^{1-t}(\bx)\mu(d\bx \times d\by) \nonumber \\
    &\quad \leq \int_{\bbR^{2kn}\times \bbR^n} f_{\bY}^{t}(\by) f_{\bX}(\bx) \mu(d\bx \times d\by) + \int_{\bbR^{2kn} \times \bbR^n} f_{\bY}^{t}(\by) f_{\bX|\bY}(\bx|\by)\mu(d\bx \times d\by)\\
    &\quad=\bigg(\int_{\bbR^{2kn}} f_{\bX}(\bx)\mu(d\bx)\bigg) \cdot \bigg(\int_{\bbR^n} f_{\bY}^{t}(\by)\mu(d\by)\bigg)+\int_{\bbR^n} f_{\bY}^{t}(\by)\bigg[\int_{\bbR^{2kn}} f_{\bX|\bY}(\bx|\by)\mu(d\bx)\bigg] \mu(d\by)\\
    &\quad=\int_{\bbR^n} f_{\bY}^{t}(\by)\mu(d\by)+\int_{\bbR^n} f_{\bY}^{t}(\by)\mu(d\by)\\
    \label{final}
    &\quad= 2\int_{\bbR^n} f_{\bY}^{t}(\by)\mu(d\by)<\infty,
\end{align}
where~\eqref{final} follows from the boundedness assumption in~\eqref{cond1C1}.

\bibliographystyle{IEEEtran}
\bibliography{isitbib}
\end{document}

%% file: preamble.tex
\usepackage[mathscr]{eucal}
\usepackage[cmex10]{amsmath}
\usepackage{epsfig,epsf,psfrag}
\usepackage{amssymb,amsmath,amsthm,amsfonts,latexsym,bm}
\usepackage{amsmath,graphicx,bm,xcolor,url}
\usepackage[caption=false]{subfig} 
\usepackage{fixltx2e}
\usepackage{array}
\usepackage{verbatim}
\usepackage{bm}
\usepackage{algorithmic, cite}
\usepackage{algorithm}
\usepackage{verbatim}
\usepackage{textcomp}
\usepackage{mathrsfs}
\usepackage{multirow}
\usepackage{epstopdf}

\catcode`~=11 \def\UrlSpecials{\do\~{\kern -.15em\lower .7ex\hbox{~}\kern .04em}} \catcode`~=13 

\allowdisplaybreaks[1]

\newcommand{\calA}{\mathcal{A}}

\newcommand{\calC}{\mathcal{C}}

\newcommand{\calG}{\mathcal{G}}

\newcommand{\calN}{\mathcal{N}}

\newcommand{\calS}{\mathcal{S}}
\newcommand{\calT}{\mathcal{T}}

\newcommand{\bI}{\mathbf{I}}

\newcommand{\bx}{\mathbf{x}}
\newcommand{\bX}{\mathbf{X}}
\newcommand{\by}{\mathbf{y}}
\newcommand{\bY}{\mathbf{Y}}
\newcommand{\bz}{\mathbf{z}}
\newcommand{\bZ}{\mathbf{Z}}

\newcommand{\rme}{\mathrm{e}}

\newcommand{\rmF}{\mathrm{F}}

\newcommand{\bbC}{\mathbb{C}}

\newcommand{\bbE}{\mathbb{E}}

\newcommand{\bbN}{\mathbb{N}}

\newcommand{\bbP}{\mathbb{P}}

\newcommand{\bbR}{\mathbb{R}}

\newcommand{\bbZ}{\mathbb{Z}}

\DeclareMathAlphabet{\mathbsf}{OT1}{cmss}{bx}{n}
\DeclareMathAlphabet{\mathssf}{OT1}{cmss}{m}{sl}

\newcommand{\rvP}{\mathsf{P}}

\DeclareSymbolFont{bsfletters}{OT1}{cmss}{bx}{n}  
\DeclareSymbolFont{ssfletters}{OT1}{cmss}{m}{n}
\DeclareMathSymbol{\bsfGamma}{0}{bsfletters}{'000}
\DeclareMathSymbol{\ssfGamma}{0}{ssfletters}{'000}
\DeclareMathSymbol{\bsfDelta}{0}{bsfletters}{'001}
\DeclareMathSymbol{\ssfDelta}{0}{ssfletters}{'001}
\DeclareMathSymbol{\bsfTheta}{0}{bsfletters}{'002}
\DeclareMathSymbol{\ssfTheta}{0}{ssfletters}{'002}
\DeclareMathSymbol{\bsfLambda}{0}{bsfletters}{'003}
\DeclareMathSymbol{\ssfLambda}{0}{ssfletters}{'003}
\DeclareMathSymbol{\bsfXi}{0}{bsfletters}{'004}
\DeclareMathSymbol{\ssfXi}{0}{ssfletters}{'004}
\DeclareMathSymbol{\bsfPi}{0}{bsfletters}{'005}
\DeclareMathSymbol{\ssfPi}{0}{ssfletters}{'005}
\DeclareMathSymbol{\bsfSigma}{0}{bsfletters}{'006}
\DeclareMathSymbol{\ssfSigma}{0}{ssfletters}{'006}
\DeclareMathSymbol{\bsfUpsilon}{0}{bsfletters}{'007}
\DeclareMathSymbol{\ssfUpsilon}{0}{ssfletters}{'007}
\DeclareMathSymbol{\bsfPhi}{0}{bsfletters}{'010}
\DeclareMathSymbol{\ssfPhi}{0}{ssfletters}{'010}
\DeclareMathSymbol{\bsfPsi}{0}{bsfletters}{'011}
\DeclareMathSymbol{\ssfPsi}{0}{ssfletters}{'011}
\DeclareMathSymbol{\bsfOmega}{0}{bsfletters}{'012}
\DeclareMathSymbol{\ssfOmega}{0}{ssfletters}{'012}

\newcommand{\tilf}{\tilde{f}}

\newcommand{\tilh}{\tilde{h}}

\newcommand{\tilQ}{\tilde{Q}}

\newcommand{\tilr}{\tilde{r}}

\newcommand{\hatS}{\hat{S}}

\newcommand{\bars}{\bar{s}}

\newcommand{\barG}{\bar{G}}

\newcommand{\barI}{\bar{I}}

\newcommand{\barK}{\bar{K}}
\newcommand{\barL}{\bar{L}}

\newcommand{\barQ}{\bar{Q}}

\newcommand{\barT}{\bar{T}}

\newcommand{\eps}{\varepsilon}

\DeclareMathOperator{\var}{\mathsf{Var}}

\newtheorem{theorem}{Theorem} 
\newtheorem{lemma}[theorem]{Lemma}

\newtheorem{proposition}[theorem]{Proposition}
\newtheorem{corollary}[theorem]{Corollary}

\newtheorem{remark}[theorem]{Remark}

\newcommand{\qednew}{\nobreak \ifvmode \relax \else
      \ifdim\lastskip<1.5em \hskip-\lastskip
      \hskip1.5em plus0em minus0.5em \fi \nobreak
      \vrule height0.75em width0.5em depth0.25em\fi}